Chapter 13

# Ultrafast Optical Probing of Laser Processing

**Jörn Bonse** [1,*]

[1] Bundesanstalt für Materialforschung und -prüfung (BAM), Unter den Eichen 87, 12205 Berlin, Germany

*: corresponding author
e-mail: joern.bonse@bam.de

## Abstract

Laser treatment has emerged into a cornerstone of industrial manufacturing. This creates the demand to understand and reveal in detail the involved physical processes and their dynamics, to control the process, and to keep guard *in-situ* the laser manufacturing steps. Given the (almost) contactless interaction of light with matter, particularly optical methods appear most suitable for probing laser processing. This chapter reviews the current scientific and technological state of analyzing laser-based material processing by short and ultrashort optical pulses. For that, sequential passive interrogation of material properties (e.g., via reflection, absorption, scattering, diffraction, interference, or specific nonlinear responses) can be employed through so-called *pump-probe* techniques (where the time-resolution is not given by the opto-electronic detector response time but by the duration of an electromagnetic probe pulse that is interrogating the transient scenery at a controlled delay time (typically fs to µs) with respect to the arrival of the pump laser pulse to the excited volume). These approaches are capable to analyze on microscopic to macroscopic scales a plethora of different processes involved in laser processing, such as local heating, ultrafast or thermal melting, material removal via ablation, plasma- and shockwave-formation, optical breakdown in surrounding liquids and gases, shielding and accumulation effects, surface solidification and restructuring, etc. A special focus is set here on the variety of ultrafast optical imaging schemes for visualizing the dynamics of laser processing at the surface and in the bulk of the irradiated materials, including photography and microscopy, holography, tomography, ptychography, and coherent diffractive imaging. The chapter provides a historic survey, highlights important scientific and technological developments and breakthroughs, discusses selected applications, explores the requirements and current limitations, and sheds light on emerging future trends.

# 1 Introduction

### 1.1 Historic Aspects
### 1.2 Laser-Matter Interaction
### 1.3 Pump-Probe Technique

# 2 Spot Probing

### 2.1 Reflectivity
### 2.2 Transmission
### 2.3 Second Harmonic Generation
### 2.4 Diffraction
### 2.5 Scattering

# 3 Microscopic Imaging

### 3.1 Brightfield and Darkfield Microscopy
### 3.2 Shadowgraphy
### 3.3 Interference Microscopy
### 3.4 Phase-Contrast Microscopy
### 3.5 Nomarski Microscopy
### 3.6 Polarization Microscopy
### 3.7 Imaging Reflectometry / Ellipsometry

# 4 Digital Holography

# 5 Ultrafast Pump-Probe Tomography

# 6 Recent Ultrafast Imaging Approaches

### 6.1 Photography
### 6.2 Light Field Tomography
### 6.3 Ptychography
### 6.4 Coherent Diffraction Imaging

# 7 Outlook and Future Developments

### 7.1 Shorter Pulse Durations
### 7.2 Other Wavelength
### 7.3 New Ultrafast Cameras
### 7.4 Extended Pump-Probe Schemes

# 1 Introduction

Laser processing of solids with high intensity optical radiation typically initiates complex and coupled sequences of physical and chemical processes that may start already during the irradiation on fs-timescales and can last into the ms-range, while simultaneously involving various structural, chemical, and topographic material alterations in spatial dimensions ranging from the atomic scale up to the mm-scale (see Chap. 1 (Nolte et al.) and Chap. 3 (Gräf and Müller)). That inherent multi-scale nature of the involved material responses makes it challenging to probe *in-situ* the laser processing and to develop methods that allow a monitoring or even a closed-loop back-feeded control. On the other hand, ultrashort laser pulses with durations down to the fs-range allow to develop extremely fast *in-situ* probing and monitoring methods that enable to track even ultrafast phase transitions, thermal melting, ablation, re-solidification, etc.

In this chapter, I will explore the plethora of possibilities for optically probing laser-matter interaction, specifically during laser-processing. In the introduction (Sect. 1), some general principles and selected historic aspects are discussed. The focus of this chapter is laid on the use of ultrashort pulsed laser radiation that enables ultimate temporal resolution via so-called pump-probe techniques that can be used for localized spot probing (Sect. 2) or for microscopic imaging (Sect. 3). Digital holography methods are explored in Sect. 4, and ultrafast tomography in Sect. 5. Recent ultrafast imaging approaches are reviewed in Sect. 6. Future scientific and technological developments will be outlined in Sect. 7. Having a main focus on ultrafast and microscopic techniques, this chapter aims to complement, update, and extend the already available excellent review articles, book chapters, and books related to this subject [Horn, 2009 / Bäuerle, 2011 / Siegel, 2012 / Mikami, 2016 / Guo, 2019 / Grasso, 2021 / Guo, 2021 / Garcia-Lechuga, 2023 /Stoian, 2023 / Zeng, 2023 / Jürgens, 2024 / Yao, 2024].

## 1.1 Historic Aspects

The fastest intrinsic cognitive capabilities of humans are typically limited at times above a few tenth of a second and are coupled with a (metastable) data storage in different parts of the brain. For getting insights into faster processes, specialized high-speed techniques along with permanent data storage techniques had to be developed, such as photography, cinematography, stroboscopic imaging, etc.

The first *photographic images* were permanently recorded in the 1820s and 1830s by Joseph Nicéphore Niépce and Louis Jacques Mandé Daguerre. At almost the same time the *stroboscopic effect* was discovered in 1825 by Peter Marc Roget, i.e., that a temporally segmented short-time illumination can ”freeze” and combine different states of a dynamically evolving system during a typical exposure time (of the eye or a photographic plate).

Photography early attracted the attention of the military. Already in 1851, a picture of a flying cannon ball was recorded by Thomas Skaife at Plumsted Marshes near Woolwich Arsenal London with a home-made camera [Fuller, 2005]. Due to its high velocity, the cannon ball appears as a strike at an exposure time of 0.02 s and is spatially not resolved in Figure 13.1.

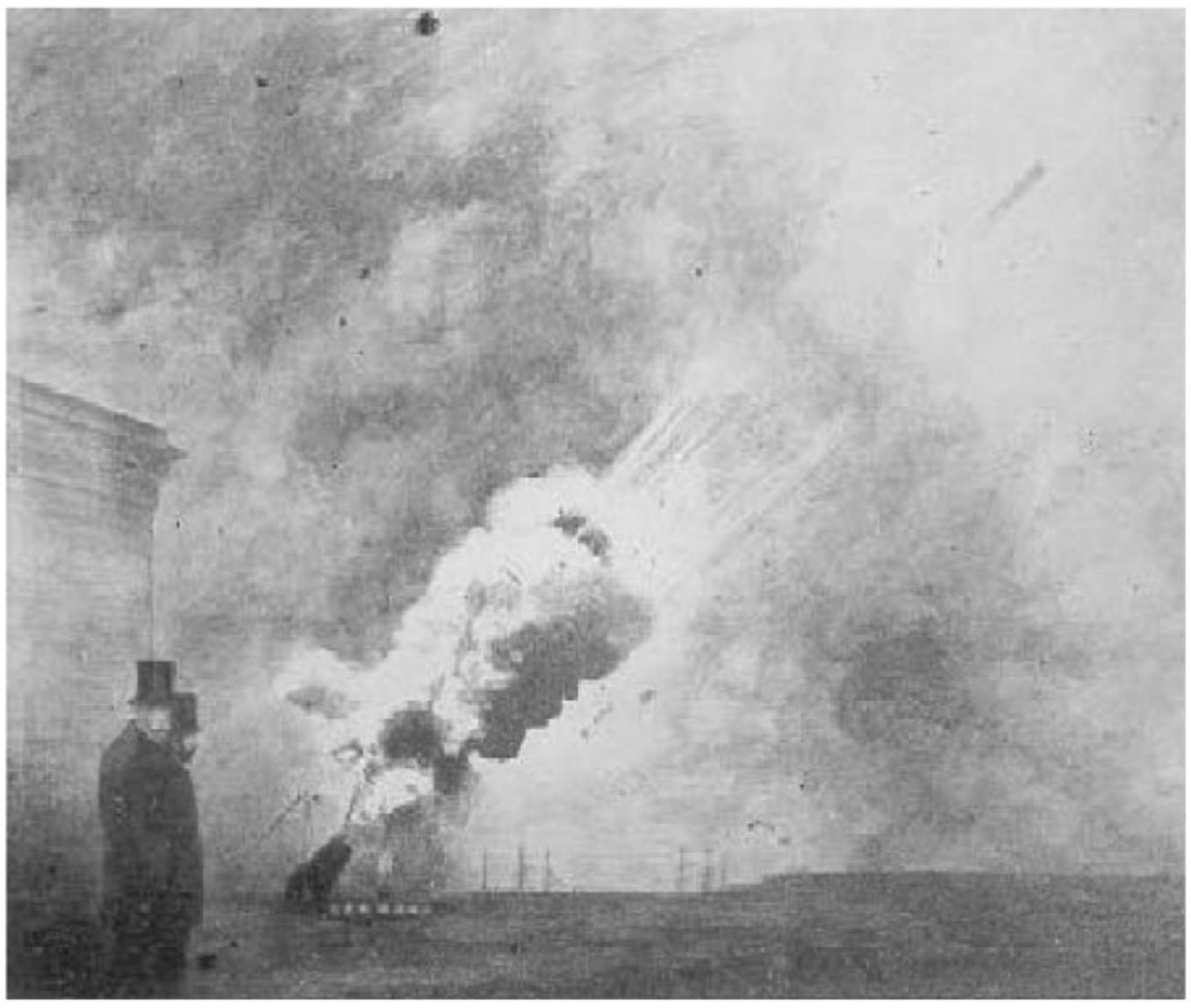

**Fig. 13.1:** Fired cannon with ball in flight. Photograph taken by Thomas Skaife at Plumsted Marshes (Woolwich, UK) in 1851 [Fuller, 2005]. (Picture: Public Domain)

In 1864 August Toepler developed the *Schlieren photography*, that is capable to visualize flows of fluids and gases through local refractive index variations [Krehl, 1995]. Such an approach was used by Peter Salcher and Ernst Mach for taking the first photographic images of supersonic flying gun projectiles in 1886 [Mach, 1887] – the beginning of the field of supersonic aerodynamics, see below.

The practical application of *high-speed photography* in civil life using a battery of 12 different cameras was demonstrated in 1878 by Eadweard Muybridge, answering the question whether horses' feet are actually all off the ground at once during a gallop [Munn, 1878]. They are, see Figure 13.2.

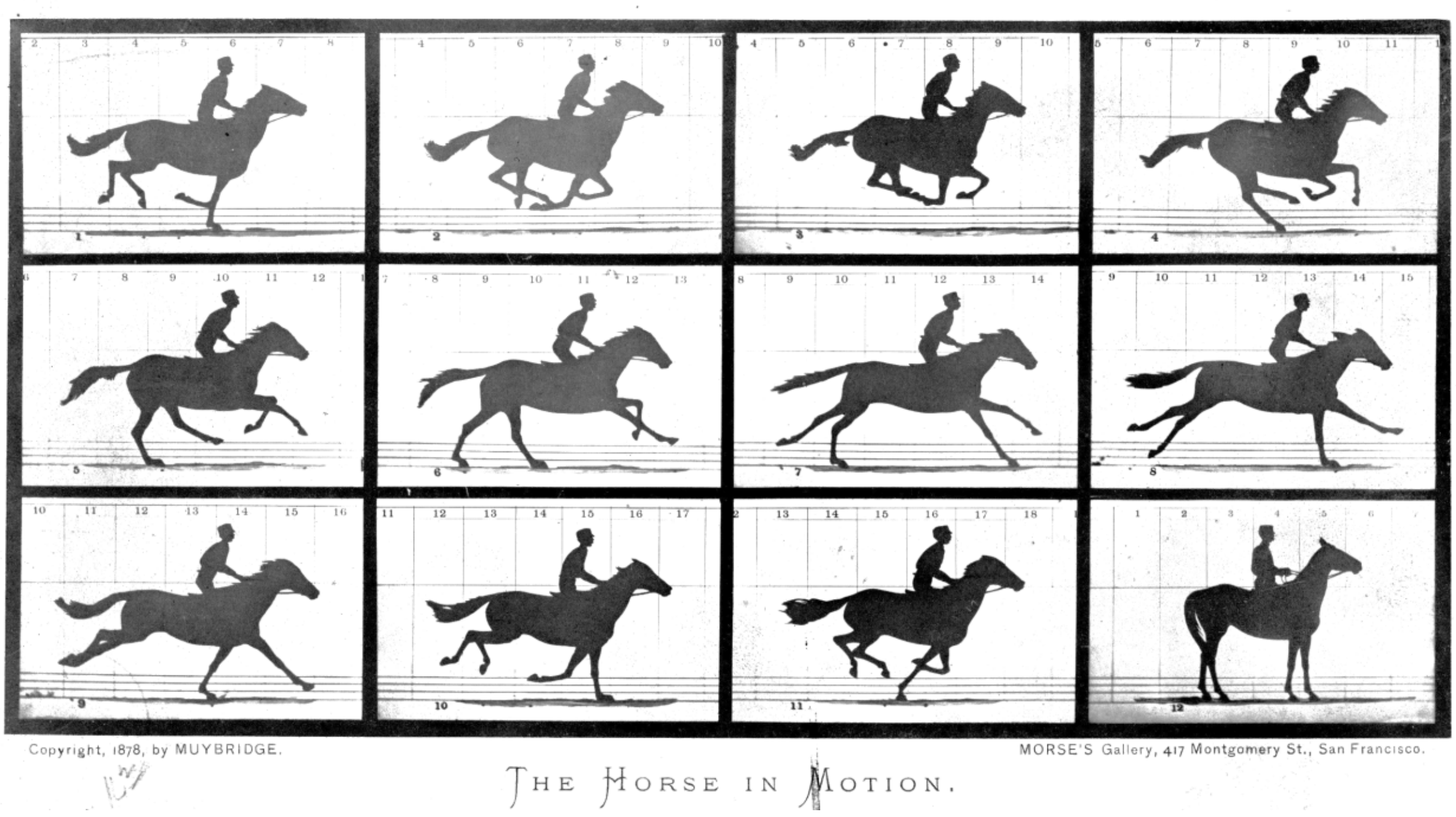

**Fig. 13.2:** Post-illustrated sequence of 12 image frames of the series "The Horse in Motion" (Muybridge, Palo Alto, USA, 1878). The negatives of these photographs were recorded at intervals of ~0.04 s with an exposure time < 0.005 s each. The horizontal lines are separated by four inches each [Wikipedia, Eadweard Muybridge] (Images: Public domain)

The Second World War provided another boost to high-speed photography, particularly by the development of fast streak and framing cameras for high-speed cine of projectiles, studying the dynamics of ignited conventional explosives or even atomic bombs, hypervelocity impact processes, and rockets in flight [Fuller, 2005]. Rotating prism and rotating mirror cameras allowed the photographic film to remain completely stationary, while sweeping the image across it. Figure 13.3 presents a scheme of such a rotating mirror framing camera with a diaphragm. Such cameras are capable of recording several million image frames per second, corresponding to a minimum acquisition time in the sub-µs range.

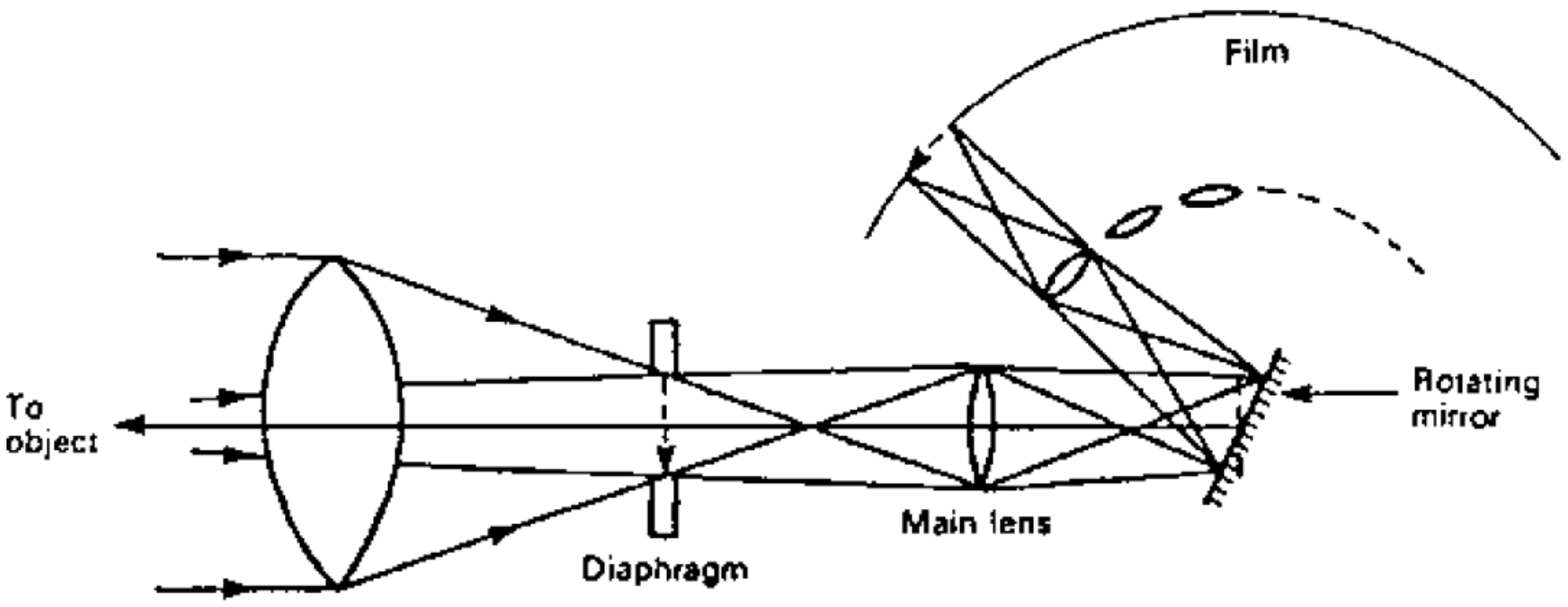

**Fig. 13.3:** Schematic of a rotating mirror framing camera [Fuller, 2005]. The light from the object captured by a lens system is imaged via a rotating mirror to the curved surface of a photographic film. Synchronization of the diaphragm shutter and the mirror rotation with the film geometry allows image recording at rates of up to several million frames/s. (Reprinted from [Fuller, 2005], P.W.W. Fuller in *High-Pressure Shock Compression of Solids VIII – The Science and Technology of High-Velocity Impact*, pp. 251 – 298, 2005, Springer Nature)

Figure 13.4a presents one of the first pictures (negative) of a flying brass projectile, taken by Ernst Mach in Prague in 1888 [Wikipedia, Ernst Mach]. Along with the projectile, it visualizes some bow shockwaves generated by its supersonic propagation, and a whirl zone left behind. The vertical white lines represent the two wire electrodes in the electric spark generator system used for generating a momentary flashlight. Its duration accounts typically to 1/800,000 of a second, i.e., 1.25 µs [Mach, 1897]. Later, Ernst Mach developed together with his son Ludwig an interferometer (nowadays called *Mach-Zehnder interferometer*) to obtain quantitative data on the change in density and pressure of the air behind the bow shockwave.

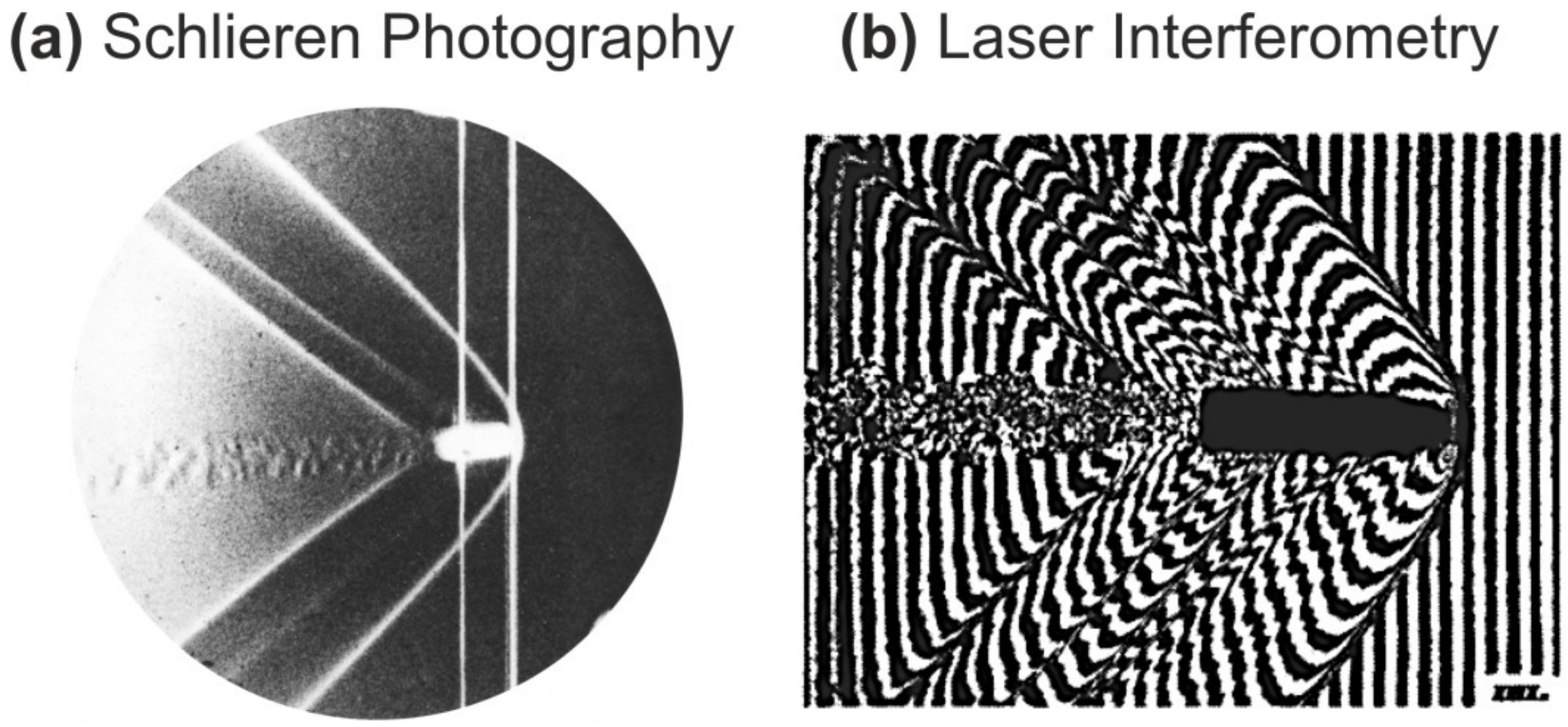

**Fig. 13.4: (a)** Schlieren photograph (negative) of a supersonic flying brass projectile (Ernst Mach, Prague, 1888) [Wikipedia, Ernst Mach] (Image: Public domain). (**b**) Differential interferometry image of a 20 mm caliber projectile flying at 500 m/s, acquired with linearly polarized YAG laser radiation and an exposure time of 15 ns [Fuller, 2005]. (Reprinted from [Fuller, 2005], P.W.W. Fuller in *High-Pressure Shock Compression of Solids VIII – The Science and Technology of High-Velocity Impact*, pp. 251 – 298, 2005, Springer Nature)

After the invention of the laser in 1960 by Theodore Maiman it became obvious that pulsed laser radiation can overcome the limitations of electronically created arc discharge flashlights and can serve as a high brightness source of light that is can be confined to extremely short exposure times. This can be seen in Figure 13.4b, which displays a snapshot of a supersonic flying projectile recorded by laser-interferometry at much higher temporal resolution (exposure time 15 ns) that allows to spatially resolve more details in the image. So far, there are no better alternatives available than probing and monitoring the laser-matter interaction in laser processing with electromagnetic radiation, such as the laser radiation itself.

While the early major developments in time-resolved probing methods were often inspired by military interests, in the second half of the twentieth century also industrial applications have strongly influenced and driven the development of new laser-based time-resolved monitoring approaches. An important example is related to semiconductor technology, where it was observed that a post-treatment by laser radiation can recrystallize semiconductor surfaces that were previously doped (and amorphized) by ion implantation [Shtyrkov, 1976 / Khaibullin, 1978 / Cullis, 1985]. Since defects were removed from the material's surface and its crystal lattice this process is referred to as "*laser annealing*" [Wood PRB-I, 1981 / Wood PRB-II, 1981 / Wood PRB-III, 1982 / Wood, 1984] – a naming in line with the conventional "*thermal annealing*" performed in a furnace. Advanced laser-based time-resolved probing methods were then developed to clarify the underlying mechanisms and to answer the question whether the annealing is achieved via creating a transient melt phase of the semiconductor or whether all-solid-state processes account for the effect [Robinson, 1984 / Lowndes, 1985]. Since then and with the ever-shorter available laser pulse durations, these methods were further improved and applied to many fields ranging from fundamental science to industrial and biomedical applications, nowadays reaching temporal resolutions down to the attosecond (as) range.

A major breakthrough in laser-based high-speed photography was made in 2014 by the development of the *compressed ultrafast photography* (CUP) method that allows a single-pulse recording of a macroscopic transient scenery at a frame rate of one hundred billion frames per second [Gao, 2014]. In analogy to the supersonic propagation of massive projectiles through the air shown above, such a single-shot real-time CUP video recording can even visualize the photonic Mach cone superluminal propagation of light in scattering media, see Figure 13.5 [Liang, 2017].

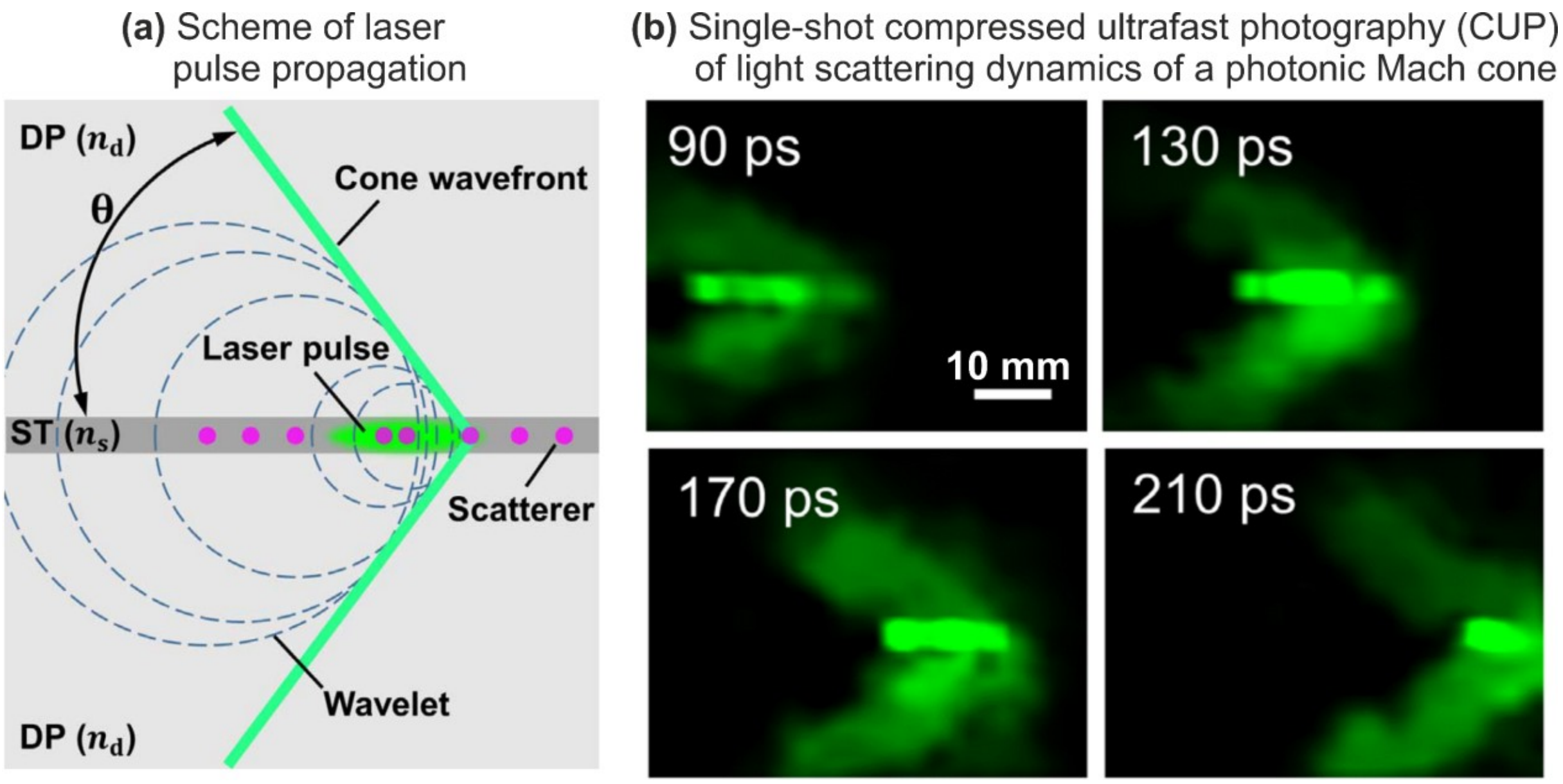

**Fig. 13.5: (a)** Scheme of the thin scattering plate assembly used for visualization of the photonic Mach cone. Abbreviations and symbols: DP: Display tunnel, ST: Source tunnel, $n_d$: refractive index of the DP medium; $n_s$: refractive index of the ST, medium, $\theta$: semi-vertex angle of the photonic Mach cone. (**b**) CUP snapshots of a single green ps-laser pulse propagating through the ST medium, while generating a photonic Mach cone in the DP media [Liang, 2017]. (Reprinted from [Liang, 2017], J. Liang et al., Single-shot real-time video recording of a photonic Mach cone induced by a scattered light pulse. Sci. Adv. **3**, e1601814 (2017), Copyright 2017 under Creative Commons BY-NC 4.0 license, and with permission from the authors of that article. Retrieved from https://doi.org/10.1126/sciadv.1601814)

For that experiment, the authors built a thin scattering plate assembly containing a central "source tunnel" (ST, filled by air mixed with dry ice fog scatterers, refractive index $n_s \sim 1.0$) sandwiched between two "display panels" (DP, consisting of silicone rubber (refractive index $n_d = 1.4$) mixed with scattering aluminum oxide powder), see Figure 13.5a. A collimated green laser pulse (532 nm wavelength, 7 ps pulse duration) propagated through the source tunnel. Since $n_s < n_d$, a photonic Mach cone in is visible in the DPs, accompanying and bound to the laser pulse propagation through the ST, see Figure 13.5b. As an additional test, the replacement of the low refractive index medium in the ST by a high refractive index oil ($n_s = 1.8$) led to the suppression of the Mach cone (image sequence not shown here) [Liang, 2017].

## 1.2 Laser-Matter Interaction

The impact of a single laser pulse or a train of laser pulses at parameters suitable for laser processing of a solid will transiently change the properties of the irradiated material and potentially also its surrounding environment. For analyzing these changes, time-resolved methods are required that can either acquire active emissions *directly* from the laser-irradiated region, or that can probe *indirectly* the properties of the laser-irradiated region, see Figure 13.6.

Figure 13.6a schematically visualizes the interaction of a focused laser beam (incident from the top, red line) with the surface of a solid sample material (gray shaded). From the laser-matter interaction regions different emissions (drawn in blue) can occur. The laser processing may be accompanied by the active emission of matter (e.g., electrons, atoms, ions, clusters, nano- and microparticles, fragments, etc.) emerging from the solid, of electromagnetic radiation such as light or X-rays, or of acoustic emissions, such as sound transmitted through the environment. In some cases, the temporal dynamics of the signal will closely follow that of the laser radiation, e.g., when its reflection or scattering occurs, or when secondary X-rays are generated. In other cases, characteristic material response times will be involved, e.g., when phase transitions such as melting, evaporation, or the formation of a plasma plume are caused by the primary laser radiation. Note that also the presence of a liquid or gaseous environment can significantly change the temporal or spectral emission characteristics and may even attenuate the emitted particles/radiation. Moreover, one must consider that some of the laser-induced processes may occur on ultrashort timescales, i.e., over characteristic times much shorter that common electronics or optoelectronics is capable to detect. Here, a suitable detection technology must be carefully selected, and independent synchronization implemented.

Figure 13.6b sketches the probing of the laser-matter interaction region through passive detection methods. This is typically done by one or more additional probing beams of directed light, X-rays, or electrons. After interaction with the laser-matter interaction zone, e.g., via reflection, transmission, scattering, diffraction, or interference, their signal is recorded, post-processed, and tracked as a function of time. For probing with light, sophisticated methods were already developed. Such all-optical approaches are very appealing since a weak part for probing can be easily split-off from the laser beam used for the laser processing and, thus, has a very precisely defined temporal and phase relation to the main beam. Thus, the probing beam is intrinsically synchronized with the primary laser processing beam. This probe beam is then separately controlled, e.g. via adding additional propagations paths that delay its arrival to the laser processing zone. For ultrashort pulsed laser processing, this idea is particularly used in so-called *pump-probe techniques* which allow to probe the laser-matter interaction with unprecedented accuracy, i.e., with a temporal resolution of the duration of the laser pulses themselves (see Sect. 2). Alternatively synchronized and non-synchronized (free running) *continuous probing methods* may be set up and used for *in-situ* probing the laser-matter interaction on longer times scales.

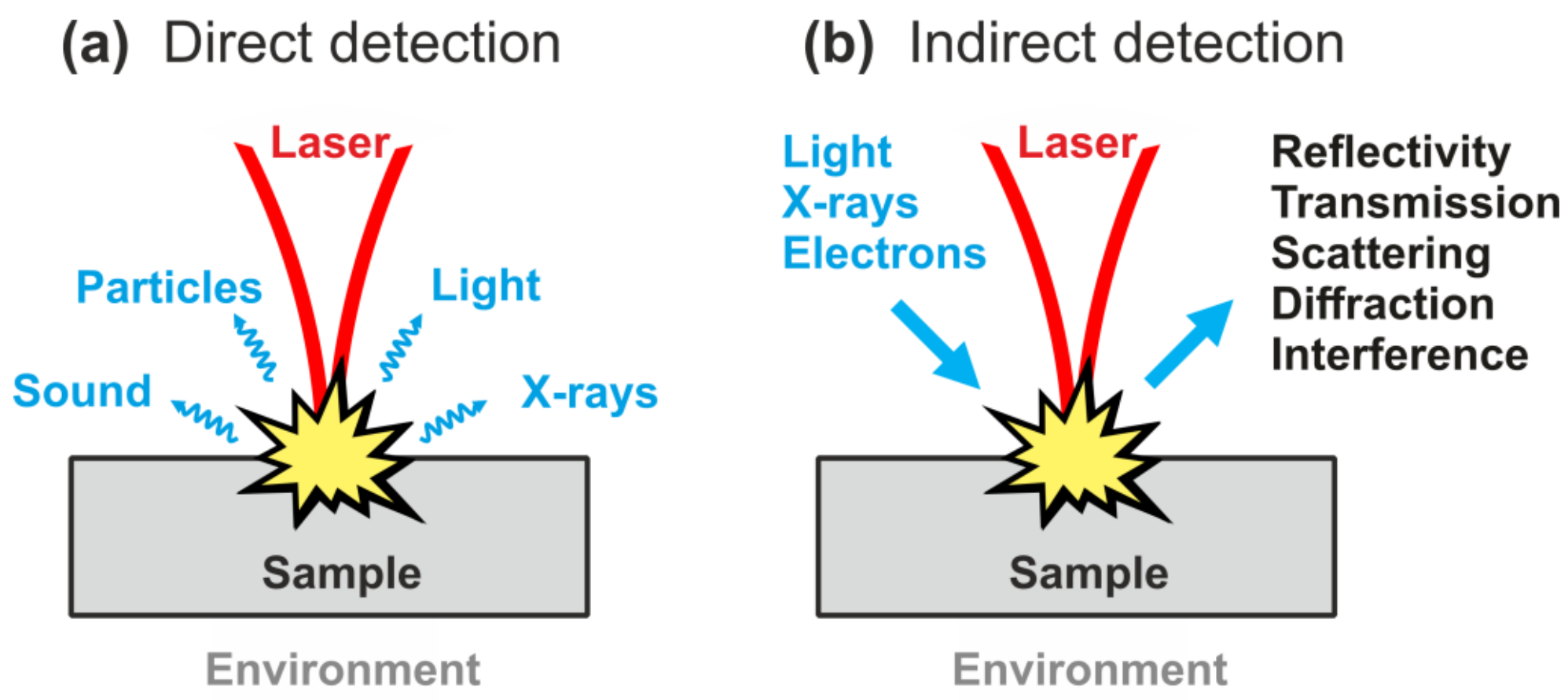

**Fig. 13.6:** Principles of observing the laser processing in an active way via direct detection of various emissions **(a)**, or in a passive way by indirect detection **(b)** probing the laser-matter interaction zone through light, X-rays or electrons via recording transient reflectivity, transmission, scattering, diffraction, or interference signals

## 1.3 Pump-Probe Technique

For the case of laser processing with pulsed radiation some remarks should be made regarding the potential necessity of temporal synchronization of the time-resolved probing and the laser system: during *in-situ* detection, the sensor signal can be acquired either in a free running way (i.e., lasers and sensors are operating independently with an own time-base) or it can be performed in a temporally synchronized way (where a signal derived from the laser system is setting the time base for the signal recording of the sensors). Particularly for ultrashort pulse laser radiation this may be relevant for a proper detection of *in-situ* signals.

### Principles of Time-Resolved Probing

The synchronized *in-situ* detection by two different methods is sketched in Figure 13.7. Here, a strong laser pulse (red peak, referred to as pump pulse) is hitting the sample surface at a time that is set here as the “zero delay” ($\Delta t = 0$), see the top panel. It triggers subsequent processes that dynamically change a specific property (blue curve) of the sample/system and should be detected *in-situ* by synchronized sensors.

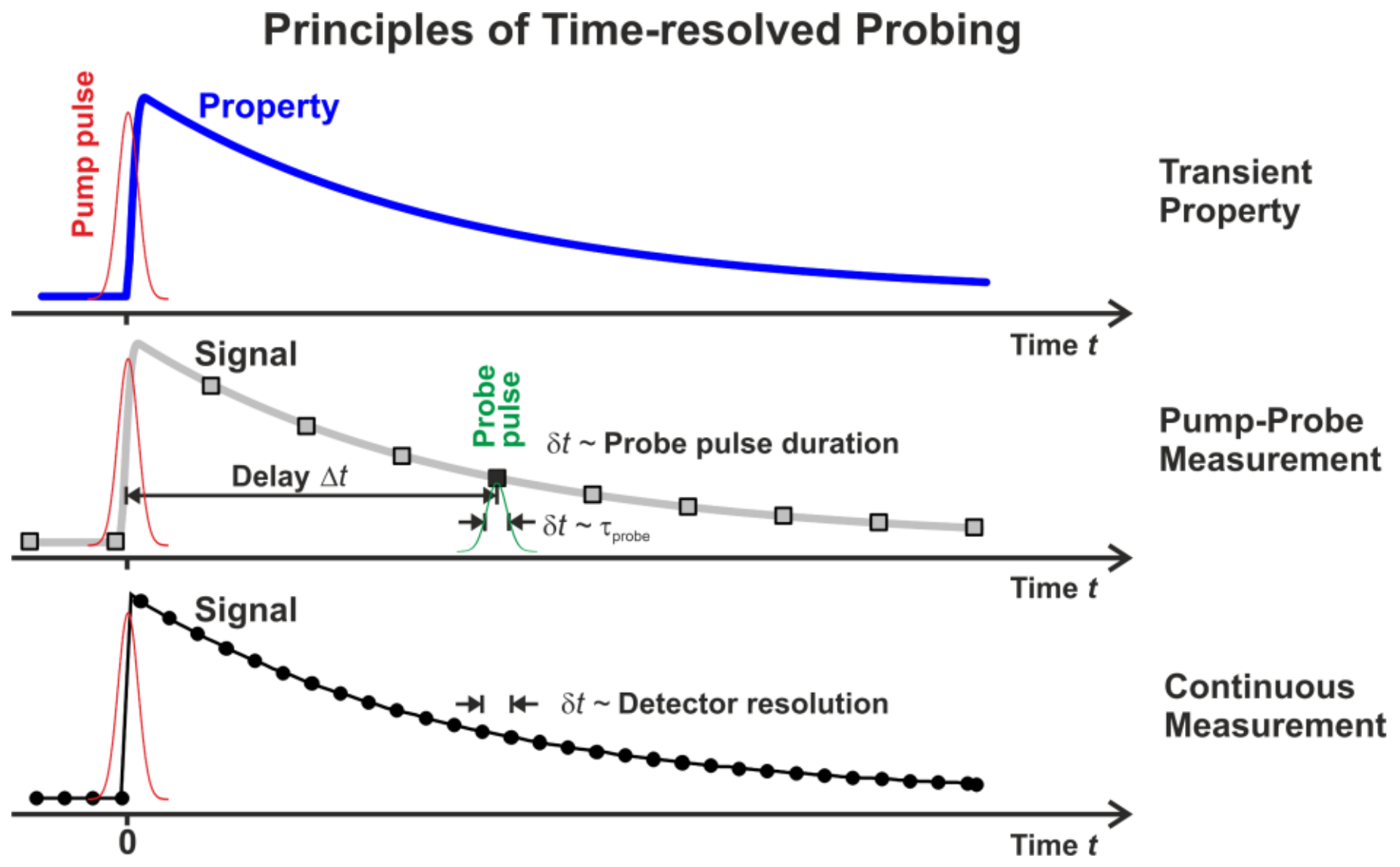

**Fig. 13.7:** Principles of time-resolved probing a specific property (blue curve) transiently changed by the laser-matter interaction, initiated and synchronized via the impact of a strong pump pulse (red peak) to the sample surface (top panel). The experimental probing can be done via pump-probe measurements (middle panel), where a weak probe pulse (green curve) interrogates the state of matter for constructing a signal curve via several measurements at different delay times ($\Delta t$). Alternatively, continuous measurements can record the sensor signal at its given temporal resolution $\delta t$ over the selected time span (bottom panel)

In *pump-probe measurements* (PP, middle panel), one (or more) delayed laser probe pulse(s) are delayed with respect to the pump pulse and directed to the region of interest, interrogating a desired property and converting it to a process-specific sensor signal at a certain delay time $\Delta t$. The signal recording of the sensors has to be activated prior to the arrival of the probe pulse(s) by a trigger signal derived electronically from the laser system that is temporally synchronized with the emission of the laser pulse(s). In most cases, a second weak probe pulse (green curve) is split-off from the primary pump pulse and delayed by adding an additional optical beam path $\Delta s = c \cdot \Delta t$ for the probe beam (compared to the pump beam path), with $c$ the speed of light, controlled via linear motorized translation stages as part of a delay-line. The probe pulse must be kept at an intensity level that does not affect the measurement of the property itself. If the strong pump pulse triggers irreversible modifications in the irradiated sample region, the sample must be moved to a pristine laser irradiation spot between successive pump-probe irradiation events. Then the measurement must be performed through a sequence of identical pump pulse irradiation events (each at a fresh spot), recording the sensor signal for selected values of $\Delta t$. If the intensity/fluence of the pump pulse stays below the permanent modification threshold, the same sample spot may be probed repetitively, and *lock-in* techniques can be used to significantly improve the detection sensitivity of the signal. If properly controlled, the temporal resolution $\delta t$ of pump-probe experiments is given by the duration of the probe pulse and of the order of its duration ($\tau_{probe}$). Pump-probe measurements are typically complex but necessary to obtain ultimate time-resolution that cannot be reached by conventional electronics. Extremely fast processes with durations in the attosecond range were successfully resolved through this technique, e.g. atomic transitions of electrons. In the context of laser material processing temporal resolutions $\delta t$ of a few tens up to several hundreds of femtoseconds, i.e., below the electron-phonon relaxation time $\tau_{e\text{-}ph}$ in solids (see Chap. 1, Nolte et al.), are usually sufficient.

For *continuous measurements* (bottom panel), the sensors are triggered by an electronic signal derived from the laser system too. The sensor signal is then continuously recorded with its intrinsic temporal resolution $\delta t$ over a selected time span. In such an approach, $\delta t$ is usually limited by the electronic detection system (sensors and data acquisition) in the sub-ns range or above. In most cases, continuous probing methods are suitable and applied for *in-situ* measuring and monitoring the laser processing – regardless, whether it is performed with single or multiple laser pulses per irradiation spot at the sample.

## General Aspects

Conceptually, the pump-probe methodology is already very old [Horn Diss, 2003] and known to most of us: it has been used to tune musical instruments for hundreds of years. When tuning the instrument, the musician uses a tuning fork that has been calibrated to a specific acoustic tone that is assigned to a specific frequency. The sound wave of the tuning fork represents the probe signal. If the musician now wants to tune a string of an instrument to the unknown frequency, he makes it vibrate by plucking it – an event that can be regarded as the pump pulse. The hearing organ (ear) is the detector. It is very sensitive to frequency differences and sound amplitudes (but rarely calibrated). To tune the instrument, the musician uses the tuning fork and listens to the superposition of the two sound waves that is manifesting in an acoustic beating if the frequencies are too different. Although the ear cannot distinguish between two acoustic tones that are very similar in frequency, the musician is able to identify the difference with this "pump-probe" method by means of their beating in the *spectral domain* and, thus, can tune them to each other in a defined way.

For further internalizing the general concept of the pump-probe methodology when using light as an information carrier, it is also instructive to recall again the function of a flashlight in photography as it represents one of the fundamental ideas of *in-situ* visualization of extremely fast processes in the *temporal domain*: In this scenario, an object (event) is imaged to a photosensitive sensor (traditional film or an electronic camera chip) that is waiting for an exposure event with an opened mechanic or electronic shutter. A flashlight with a duration of typically $10^{-3}$ to $10^{-6}$ s is electronically created and illuminates the scenery (object) at an intensity much brighter than the surrounding light level. The flashlight reflected or scattered from the object is then recorded by the activated sensor and "freezes" the whole object dynamics that is occurring over the duration of the momentary optical flash. Thus, the duration of the flashlight rules the temporal resolution (and not the sensors exposure time). Extending this idea to repetitive sequence of temporally separated flashes and image recordings neatly allows to create a stroboscopic movie out of the image sequence.

Replacing the electronic flashlight by laser pulses that are synchronized with the event enables to discover, capture, and study in detail even extremely fast physical or chemical processes manifesting in the micro- to attosecond temporal range. On short timescales (fs – ns), optically synchronized laser pulses can be easily generated by splitting of a part of the main (pump) pulse and delaying this probe pulse geometrically with respect to the pump pulse (that is used for initiating and dressing the event). On larger timescales (ns – $\infty$), the synchronized laser pulses can be generated also by different pulsed laser sources that are synchronized with each other via their electronics.

Through these examples it becomes obvious that different "domains" of the pump-probe methodology can be distinguished:

- **Temporal Domain:** The probe pulses are interrogating the pump-excited system at different delay times ($\Delta t$, quantified in s). The minimum temporal resolution is limited by the probe pulse duration ($\tau_{probe}$).
- **Spectral Domain:** The probe pulses are interrogating the pump-excited system at different spectral frequencies ($\nu$, quantified in Hz).
- **Direct Space Domain:** The probe pulses are interrogating the pump-excited system at different locations and may spatially average the over different areas, depending on the spatial resolution of the probing system (quantified $m^2$).
- **Reciprocal Space Domain:** The probe pulses are interrogating the pump-excited system in reciprocal/momentum space at different spatial frequencies ($k$ or $q$, quantified in $m^{-1}$).

For investigating laser processing, pump-probe approaches are typically established in the temporal domain, sensing signals either in direct or reciprocal space, or even spectrally. This can be either realized via probing localized spots (when spatially confining the probe beam), or via imaging larger spatial domains (when illuminating the entire domain with the probe beam).

## Experimental Aspects

For realizing pump-probe experiments with high temporal and spatial resolution typically thorough preparations must be made for properly considering and controlling all practical and theoretical constraints. This includes environmental aspects, stability constraints, optical laser beam management and pulse delivery aspects, the search for the zero delay, separation strategies for the pump and probe radiation, potential limitations through (spatial and temporal) coherence, etc. In the following, these aspects are briefly recalled and commented.

- **Environment:** Most ultrashort pump-probe experiments split-off a part of the main laser pulse to be used as a delayed probe pulse. Typically, this is realized in Michelson or Mach-Zehnder interferometer like configuration, where the pump and the probe beams pass though different interferometer arms before meeting again at the sample. Both interferometer arms usually have macroscopic optical arm lengths ($s$) and are spatially separated. Thus, upon propagation of the laser pulses through the different arms, the pump- and the probe pulses may experience small but different optical path lengths changes through the local environment, for example by density fluctuations in the atmosphere, mechanical vibrations of the optical table, etc. When assuming an interferometer arm optical path length of $s = 1$ m, the time light needs to pass through it accounts to $s/c_0 \approx 3.3$ ns at least here (just when containing air only). Even if most experimental conditions are essentially static during such a short time lapse, environmental long-term drifts (e.g. through local thermal or air flows, acoustic changes, etc.) over the time required to conduct the entire pump-probe experiment (sometimes several hours) can play an important role by affecting the two interferometer arms differently.

A prominent effect to be discussed here is the temperature drift. Even if the temperature in the laboratory of the pump-probe experiment changes just by a few tenth of °C (global room temperature or even local variations), the thermal expansion of the optical table, sample holder, optics posts, etc. can change on the micrometer scale. As an example, a piece of steel (aluminum) of 1 m length thermally expands by ≈12 µm (≈24 µm) when experiencing a temperature change of $\Delta T = 1$ °C only. Thus, for successful pump-probe experiments good control and equilibration of the room temperature are essential (ideally in a non-noisy environment and built on a damped optical table).
One strategy to mitigate the effect of long-term drifts during pump-probe measurements in the temporal domain is to randomize the sequence of delay times used for data acquisition. In this way, the systematic errors arising from the long-term drifts during the measurements manifest over the entire delay range, instead of manifesting preferentially at one of its borders.

- **Optical components:** Pump-probe experiments are usually set up with high quality optical components ensuring a minimal phase front distortion of the pump- and the probe beams. Often dielectric multi-layer coatings are used for specialized components such as lenses, high reflective mirrors, beam splitters, thin-film polarizers, anti-reflection coatings, etc. [Bass, 1995]. Especially for ultrashort pulse durations in the sub-ps range care must be taken that all components exhibit a spectral width larger that the intrinsic bandwidths of the ultrashort pulses [Diels, 2006]. Otherwise, specific spectral components may be suppressed, resulting according to the theorem of Fourier in the creation of pre- or post-pulses in the temporal domain. Also, linear dispersion effects should be minimized. Special attention should be paid to the proper selection of optical polarizers to find the optimum type (polarizing vs. non-polarizing) along with compromises between extinction/reflection ratios, dispersion effects, and damage thresholds. Zero-order achromatic designs are favorable as optical quarter- or half-wave plates for ultrashort laser pulses.

- **Optical pulse management and beam delivery:** For each element placed in the optical beam path of the experimental setup, the question should be addressed whether and how much it may change the characteristics (pulse duration, spatial beam divergence, phase, etc.) of the pump and the probe beams? Similar it should be checked if the optical components all can withstand the local peak intensities, keeping them wellbelow their respective damage thresholds. Beam expansion through Galilean telescopes (avoiding intermediate beam foci) may help to fulfil that constraint. Uncontrolled back-reflections should be avoided for working safety reasons by properly dumping all residual reflections via beam blocks, housings, etc. Beam clipping at edges of optical components, post or internal apertures should be avoided as it causes diffraction effects which are detrimental particularly for beam focusing or for imaging.
Some important remarks have to be made regarding the focusing of ultrashort laser pulses by lenses, objectives, telescopes, or mirrors. According to Bor and Kempe and their co-workers, the focusing of a collimated ultrashort pulsed laser beam may lead to a temporal prolongation of the pulse duration in the focal region of such elements [Bor, 1988 / Diels, 2006]. Two different effects are contributing to that temporal broadening upon focusing by a lens: the first effect, the *group velocity dispersion* (GVD) effect is caused by the dispersion of the lens material and the radially varying amount of lens material that the ultrashort pulse is passing through. Along with a parabolic function of the input beam radius in a lens geometry factor the temporal broadening scales with $\lambda \cdot \Delta\lambda \cdot d^2n/d\lambda^2$, with λ being the center wavelength, Δλ the spectral bandwidth of the pulse, and $d^2n/d\lambda^2$ the second-order derivative of the refractive index [Bor, 1988]. As second effect, the pulse front, which coincides with the intensity peak of the laser pulse, moves with the group velocity $\upsilon_g := d\omega(k)/dk$ and is, thus, delayed with respect to the phase front that propagates at the speed of light. This propagation-induced effect is referred to as *propagation time difference* (PTD) and depends also on the beam radial coordinate. It scales linearly with λ and the dispersion $dn/d\lambda$ of the lens material [Bor, 1988]. For most lens materials, the axial part of the beam is then delayed with respect to the marginal parts of the beam. As a general rule, such ultrafast pulse broadening effects upon focusing may increase for very short laser pulses, at short center wavelengths and for focusing with high (numerical) aperture elements.
Another important aspect is related to nonlinear effects, such as the Kerr effect, that may manifest during the propagation of pump and probe beams through the experimental setup. For example, if the strong pump beam travels through a large amount of optical glass, its large intensities $I(r, t)$ may cause non-negligible self-focusing through a transient lensing effect, when the nonlinear refractive index ($n_{nl}$) becomes radially effective $n_{nl}(r, t) = n_2 \cdot I(r, t)$. As an implication, the beam divergence may change and transiently focus or defocus the laser beam during its further propagation. In turn, this may result in altered excitation conditions in the probed region. Alternatively, the pump beam may cause locally laser-modified zones that are transiently diffracting the probe beam in collinear pump-probe experiments, eventually causing ring-pattern-shaped imaging artifacts. Such intensity-dependent self-focusing effects may be identified and quantified by measuring the pulse energy dependence of a beam-clipping aperture placed in a suitable location.
For extremely short (few cycle) optical pulses special care must be taken due to their extremely broad spectral bandwidth Δλ that cannot be handled by dielectric multi-layer coatings or normal dispersive optical materials anymore. Then, purely reflective optical elements, such as plane, spherical, or (off-axis) parabolic mirrors, having metal coatings made of gold, aluminum, or silver, offer an alternative for the laser beam management and its delivery to the pump-probe interaction region (at the cost of a reduced reflectivity of the elements).

- **Pump & probe beam discrimination:** An essential aspect in pump-probe experiments is the proper separation of the probe beam radiation from other parasitic light, for example emerging from scattering of the pump beam radiation, nonlinear interactions with the sample material, radiative recombination effects, fluorescence emitted from the ablation plasma, ambient light in the laboratory, etc. Three different concepts of pump-probe beam separation are common, which are often combined: (i) *spatial filtering*, (ii) *polarization filtering*, (iii) and *spectral filtering*.
  *Spatial filtering* takes benefit that the pump and the probe beam have different propagation characteristics (direction, divergence, etc.). Hence, the probe beam detection may be performed at a location not reached by the pump beam radiation.
  For *polarization filteri*ng, one may use different linear polarization states for the pump- and the probe beams by employing "polarizing beam splitters" for initially separating them and then keeping this characteristic. In that case, an additional linear polarizer with a high extinction ratio may be placed in front of the probe beam detector, allowing only the probe beam polarization to pass.

In most cases it is key to use additionally *spectral filtering*. Here, a spectrally narrow optical bandpass filter for to the probe beam radiation wavelength is placed in front of the probe beam detector, ideally suppressing all spectral components that are not matching its transmission characteristics. This approach efficiently removes secondary light emissions generated during the laser processing through thermal emissions, plasma fluorescence, second harmonic generation, etc. Often it is beneficial to convert the probe beam wavelength through a coherent nonlinear process (e.g. second harmonic or continuum generation) into another spectral range (while mostly keeping the probe pulse duration), thus allowing an efficient suppression of scattered pump beam radiation via spectral filtering,

- **Zero delay:** The optical path difference of the two interferometer arms ($\Delta s$) has to be controlled and adjusted in with a very high accuracy that must be smaller than the spatial extent of the laser pulses in propagation direction, i.e., typically $\Delta s < c_0 \cdot \tau_p$). As an example, a laser pulse of $\tau_p = 100$ fs duration exhibits a spatial extent of ~30 μm along its propagation direction.
The optical path length difference between the pump and the probe beams is typically controlled by a delay line (DL). When using an optical retroreflector mounted on a mechanical linear translation stage (ideally motorized) in either the pump or the probe beam path, its movement of the distance $\Delta x$ can be used to control the delay time, which then accounts $\Delta t = 2 \cdot \Delta x / c_0$. Such linear translation stages are available with a sub-micrometer precision, high linearity of the motion, and good repeatability. They can be equipped with stepper or linear motors or even combined with additional piezo translator elements for enabling nanometer precision.
For the experimental determination of the zero delay, typically a multi-step iteration is conducted. i.e., first using a ruler allowing $\Delta s$ to be reduced to the millimeter range as a coarse adjustment. Alternatively, a fast photodiode-based detection system may be used for recording simultaneously light scattered from both beams in the pump-probe overlap region, thus allowing to adjust the optical path difference $\Delta s$ with centimeter accuracy. If possible, in a second step the direct interference between the pump and the probe beams may be visualized when spatially superimposing the pump and the probe beams close to the sample position. This interference becomes only visible if the two pulses do also temporally overlap, i.e., when $\Delta s < c_0 \cdot \tau_p$. This, however, requires the potential ability of both beams to interfere, i.e., the same polarization state and wavelength of the radiation, and the capability to resolve and detect the interference pattern. If this is not possible, in a final (fine) adjustment step, typically a nonlinear ultrafast interaction process between the pump and the probe beam radiation at the sample site is chosen for detecting the spatiotemporal overlap of the pump and the probe pulses. This can be a material-imposed second harmonic generation or a two-photon absorption (effects that are both most efficient when two equally weak pump and probe pulses temporally overlap their intensity peaks). Or it can be the generation of electrons in the conduction band of a semiconducting or dielectric sample used as test material upon exposition to a very strong pump beam that is strongly changing the local reflectivity or transmission, when probed in the pump-beam excited region.

- **Coherent artifact:** In pump-probe experiments often a "*coherent artifact*" is observed for pump-probe delay times $\Delta t < \tau_p$. It manifests as non-meaningful (often oscillatory) probe signal variations when the pump and the probe beams are able to interfere coherently. It may have different interference-triggered microscopic physical origins, such as probe beam diffraction at a transient grating generated by the pump-probe interference pattern in the probed material, modulated absorption, cross-phase modulation, stimulated Raman scattering, etc. [Vardeny, 1981 / Lebedev, 2005 / Bresci, 2021]. The coherent artifact is strongly reduced, when the pump and the probe beams are not able to interfere coherently, for example when crossed linear polarization states or different wavelengths are used for the two beams [Vardeny, 1981]. On the other hand, its occurrence may be used to precisely determine the zero delay. For pump-probe delay time $\Delta t > \tau_p$ the pump and probe pulses are temporally separated, avoiding implications of their direct interference at the sample.

- **Beam coherence:** The lateral coherence of laser radiation used as probe beam radiation may cause the formation of undesired *speckle patterns*, particularly when large magnification microscopy modes are used. For reducing such speckle effects, several probe beam images may be accumulated, while slightly influencing the probe beam phase front, e.g., via rotating scattering plates. Alternatively, the coherence may be reduced by employing so-called "random lasers" as probe beam radiation allowing for an almost speckle-free imaging [Mermillod, 2013].

- **Stability:** As a general constraint, pump-probe experiments should be placed as close as possible to the relevant laser sources, to minimize the effects of laser beam-pointing instabilities. Particularly for pump-probe microscopy approaches, already very small differences in the beam direction, propagated along large distances to the experiment may cause problems in data normalization, when several individual image registrations are required for it (e.g., when computing quantitative reflectivity images from reflected intensities/signals). Image-shifting post-processing prior to normalization may help to compensate the beam-pointing instabilities.
  Moreover, pump-probe experiments at fluences exceeding the material modification threshold excite for each delay point a previously unexposed sample region and rely on a precise event repetition. Thus, the experiments require small pulse-to-pulse energy fluctuations and a spatially homogeneous sample to ensure that almost identical ablation events are probed.
  Apart from such optical instabilities, also electronic effects can generate a jitter of electric signal and, thus, affect the temporal resolution. As an example, jitters caused by laser electronics may affect the exact trigger moment of other electronic devices, such as oscilloscopes, that are finally used to record photodetector signals.

- **Resolutions (temporal / spatial / spectral / phase):** In most cases, the temporal and spatial resolutions of a pump-probe setup are not simply given by the focus spot size, the pixel-resolution of the camera, or the pulse duration of the probe beam. Some of the optical elements may change the beam characteristics or limits to the corresponding entity, e.g., through Abbe's diffraction limit or by changing the duration of the optical pulses. The temporal, spatial, and spectral/phase resolution of the pump-probe setup should be tested with suitable reference samples. For optical microscopy certified standards for the (lateral or axial) optical resolution are available, such as the classical "Siemens star", the "1951 USAF" resolution test chart, or other test patterns. Ultrashort laser pulse durations cannot be measured anymore by slow optoelectronic devices such as photodiodes, but the duration can be measured by advanced optical devices, such as second-order autocorrelators, SPIDERs (*Spectral Phase Interferometry for Direct*

*Electric-field Reconstruction*), FROGs (*Frequency-resolved Optical Gating*), etc. [Trebino, 2021]. However, the true temporal resolution of the pump-probe experiment is often longer that the probe pulse duration itself and can be tested with processes of known duration or with extremely fast durations, such as optical absorption triggered by strong electronic excitation, providing an information on the upper limit. Optical phase changes can be quantified and tested against objects of known refractive index stratigraphy, such as optical fibers, where the material and the fiber core diameters are exactly specified by the manufacturer. Alternatively, optical wave plates may be used. Spectral resolution can be estimated with different samples of precisely known wavelength-characteristics, e.g. ruled by specific atomic transition in gases, narrow bandwidth laser lines, etc.

## Echelon Technique

An important extension the stroboscopic interrogation approach of the pump-probe method is the "Echelon technique". The central idea of this technique is to transform a single ultrashort probe pulse into a train of temporally separated sub-pulses ("echoes") that sequentially interrogate the transient object with a defined optical pulse burst. This can be straightforwardly realized through a so-called *Echelon grating* (EG). That represents a transparent glass plate that exhibits at the top-surface a staircase-like, typically equally stepped thickness variation (see Figure 13.8a). When the EG is permeated by a laterally extended laser beam that illuminates many of these individual steps, the optical path, i.e. product of refractive index and geometrical path, varies stepwise across the spatial beam profile. Hence, an ultrashort probe beam can be laterally split into sub-pulses that, segment-wise, experiences individually different stepped optical path differences (OPDs). These differences of the product of the geometric path length with the local refractive index along the individual beam path segments are then translating directly into a probe pulse sequence with stepped delays $\Delta t$ (see Figure 13.8a).

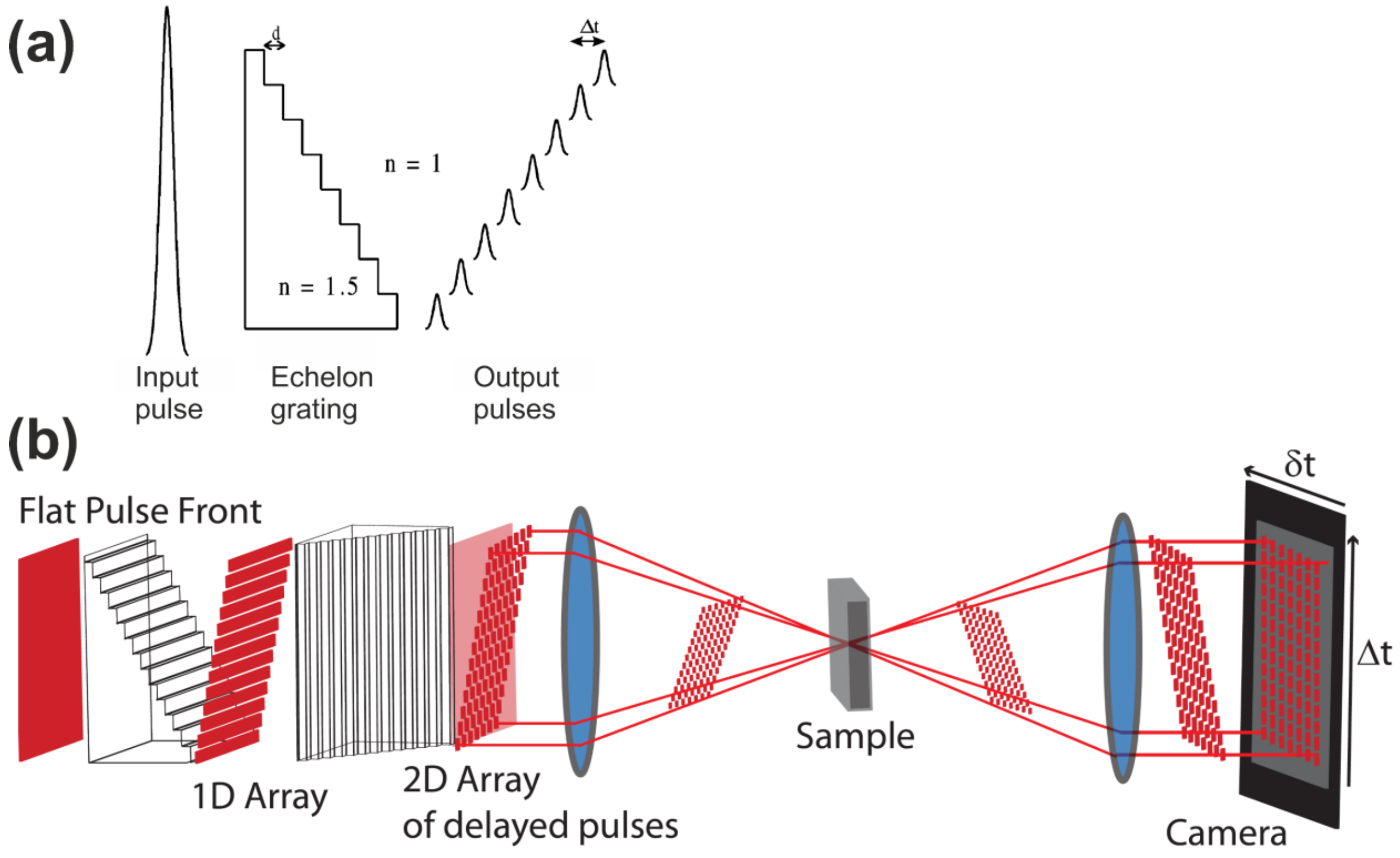

**Fig. 13.8:** Schemes of the echelon optics. (**a**) Scheme of a single transmissive echelon grating, generating multiple, stepwise spatially and temporally separated pulses from a single incident laser pulse. (Reprinted from [Wakeham, 2002], Thermochim. Acta, **384**, G.P. Wakeham et al., Femtosecond time-resolved spectroscopy of energetic materials, 7 – 21, Copyright (2002), with permission from Elsevier). (**b**) Scheme of crossed dual Echelon transmission optical setup. (Reprinted from [Teitelbaum, 2016], with permission and courtesy of Samuel W. Teitelbaum)

Initially, Echelon gratings were invented by A.A. Michelson for improving the resolution power of spectrometers (awarded the Nobel Prize in 1907) [Michelson, 1898 / Michelson, 1907]. During the early 1970s, Topp et al. recognized the potential of EGs in time-resolved ultrafast spectroscopy implementing a stepwise delay line with picosecond resolution [Topp, 1971]. By cascading two orthogonally arranged EGs, Keith Nelson's research group at the *Massachusetts Institute of Technology* (USA) extended the approach as "dual Echelon" technique (see Figure 13.8a and Figure 13.8b) towards 2D photodetection capabilities, while implementing spot probing ultrafast pump-probe experiments for studying nonlinear interaction and laser damage in fused silica samples upon irradiation with Ti:sapphire fs-laser pulses, and in other materials [Wakeham, 2000 / Wakeham, 2002 / Poulin, 2006]. Recently, the Echelon technique gained additional relevance in the field of ultrafast imaging (see Sect. 6).

# 2 Spot Probing

Often it is sufficient, of special interest, or simply ruled by the limited available laser intensity or available sample area to spatially confine the laser beam radiation for analyzing a single spot within the laser-excited region. This analysis mode is referred to as *spot probing*. The pump-probe technique was first implemented and realized in such spot probing approaches. In the context of laser processing, it turned to be key for studying laser-induced phase transitions such as melting, evaporation (ablation), which can manifest already on ultrashort time scales. This topic was particularly boosted in 1983 by the observation of the phenomenon of *nonthermal melting* of semiconductors [Shank PRL-1, 1983 / Shank PRL-2, 1983], where via strong electronic excitation the solid can turn into a structurally disordered (liquid) state on times faster than the *electron-phonon coupling time* $\tau_{e\text{-}ph}$. (see Chap. 1, Nolte et al.). Excellent review articles and book chapters on the subject of spot probing optical pump-probe ultrafast spectroscopy were already available during the 1980's, see [Shank Science, 1983 / Lowndes, 1984 / von der Linde, 1988, Mazur, 1997].

In the following section, the most common spot-probing optical pump-probe approaches are reviewed.

## 2.1 Reflectivity

Figure 13.9 presents an example of spot probing in reflection at a non-zero angle of incidence. The probed spot is placed within the center of a larger surface area that is excited at normal incidence, here. The scheme in Figure 13.9a can be considered as a prototype of a non-collinear spot probing pump-probe experiment at a single wavelength and makes use of the typical optical, optoelectronic, and mechanical elements.

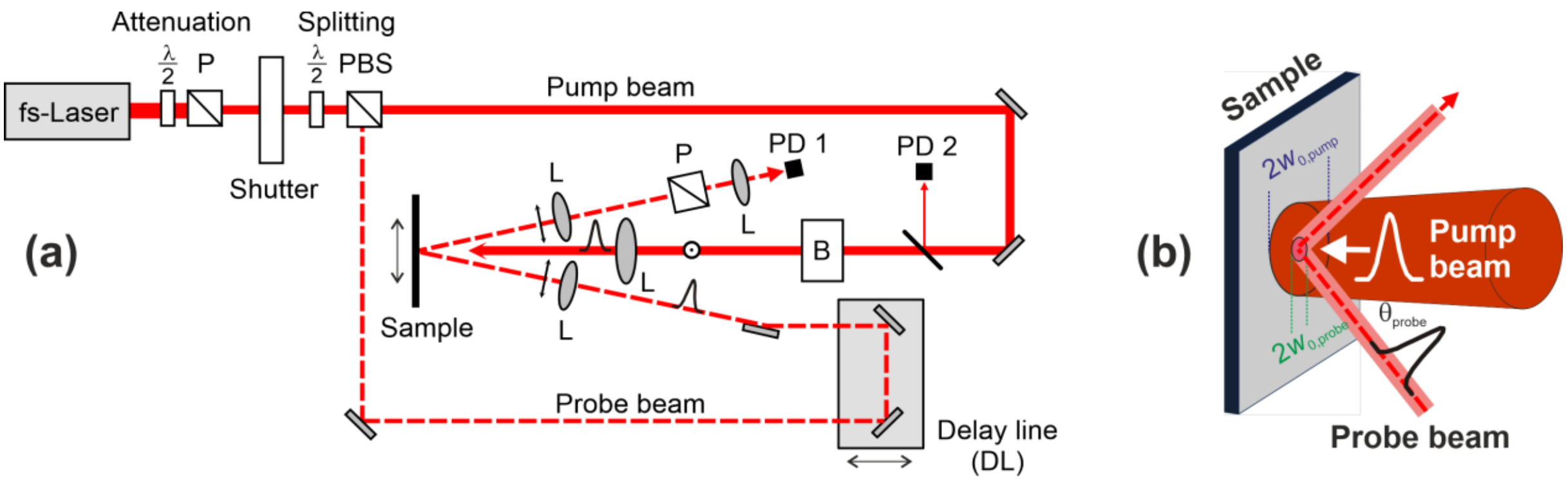

**Fig. 13.9:** (**a**) Scheme of a non-collinear spot probing pump-probe setup for time-resolved measurements of the surface reflectivity. Abbreviations: B – beam block; DL – delay line; λ/2 – half-wave plate; L – lens; P: polarizer; PBS: polarizing beam splitter; PD – photodiode. (Reprinted from [Bonse, 2005], J. Bonse et al., Dynamics of phase transitions induced by femtosecond laser pulse irradiation of indium phosphide, Appl. Phys. A 80, 243 – 248, 2005, Springer Nature). (**b**) Detailed magnification of the overlap region of the pump beam (diameter $2w_{0,pump}$) and the centered probe beam (diameter $2w_{0,probe}$) that is incident at an angle $\theta_{probe}$ to the surface normal

In this case, a pulsed (fs-)laser system generates short pulses that will be used to irradiate the surface of a sample for probing the surface reflectivity in the laser-excited spot. A combination of a half-wave plate (λ/2) and a linear polarizer (P) is used for global adjustment (attenuation) of the optical pulse energy emitted from the (fs-)laser system. An electro-mechanical shutter that is synchronized with the laser system allows the selection of a single laser pulse from a continuous pulse-train (typically operated at repetition frequencies < 1 kHz to allow proper laser pulse selection). Behind the shutter, another half-wave plate allows the rotation of the direction of the linear polarization state that determines the energy splitting ratio imposed by the following fixed polarizing beam splitter (PBS). The latter separates a weak probe beam from the strong pump beam by reflecting/transmitting distinct linear polarization components. The approach uses orthogonal polarization directions for the pump and probe pulses for reducing their coherent interaction near the zero delay and to enable a discrimination of scattered pump beam radiation with a linear polarizer placed in front of the photodiode (PD 1) that is recording the probe beam signal. The pump pulse is then focused by a lens (L) at near normal incidence directly to the sample surface. After passing through an optical delay line (DL), the p-polarized probe pulse is focused under an angle of incidence $\theta_{\text{probe}} \neq 0$ into the center of the sample pump pulse irradiated region. Subsequently, the reflected probe beam is recollimated by another lens before being recorded by a photodiode-oscilloscope detection system. Pump-to-probe-spot-size ratios $w_{0,\text{pump}} / w_{0,\text{probe}} > 5$ are recommended to obtain good quality data that are not limited by the spatially Gaussian beam shapes typically used. A second photodiode (PD 2) monitors a fraction of the pump pulse energy and the pulse-to-pulse energy stability. Motorized linear translation stages (indicated as double arrows) are used for moving the sample, and also to add the necessary optical beam path for obtaining the desired delay $\Delta t$. Each surface region should be irradiated only once if the pump can laser generate any permanent modification in the irradiated sample spot.

Material systems that naturally call for spot-probing pump-probe experiments of the surface reflectivity are *optical phase-change media*. In such materials, at a localized position (pit), a binary information (representing the states 1 or 0) is encoded via the material´s state that exhibits a different surface reflectivity in the amorphous (A) or crystalline (C) state [van Houten, 2000]. Typically, the reflectivity of the amorphous state is reduced by several up to a few tens of percent. These two states (A or C) can be ‘switched’ upon irradiation with single laser pulses, when selecting suitable irradiation parameters (laser wavelength $\lambda$, pulse duration $\tau$, irradiation geometry, etc.) and controlling the laser fluence ($\phi_0$). In case of GeSb thin films deposited on a glass substrate, the question was raised how fast the complete phase transition cycles (A $\rightarrow$ C or C $\rightarrow$ A) can be rendered and whether, in general, ultrashort pulsed lasers can help to speed-up the information encoding in such phase change systems?

Figure 13.10 exemplifies some results obtained through spot-probing pump-probe experiments on single-pulse laser irradiated 25 nm thick $Ge_{0.07}Sb_{0.93}$ thin films that were previously deposited by magnetron sputtering on 1 mm thick glass substrates (Figure 13.10a) [Wiggins, 2005]. Laser pulses at a wavelength of 600 nm and with a duration of 20 ps were used for both, pumping and probing. The surface reflectivity $R$ of p-polarized probe pulses was measured at an angle of incidence $\theta_{\text{probe}} = 12°$ by a fast photodiode and normalized by the value of the crystalline GeSb film materials ($R_C$). At the desired spots, the amorphous deposited films were locally crystallized by multiple pump laser pulses at a low laser fluence (15 mJ/cm$^2$) prior to the single-pulse pump-probe experiments performed at a pump fluence level of 50 mJ/cm$^2$. At each delay time $\Delta t$, several measurements were performed to improve the signal-to-noise ratio.

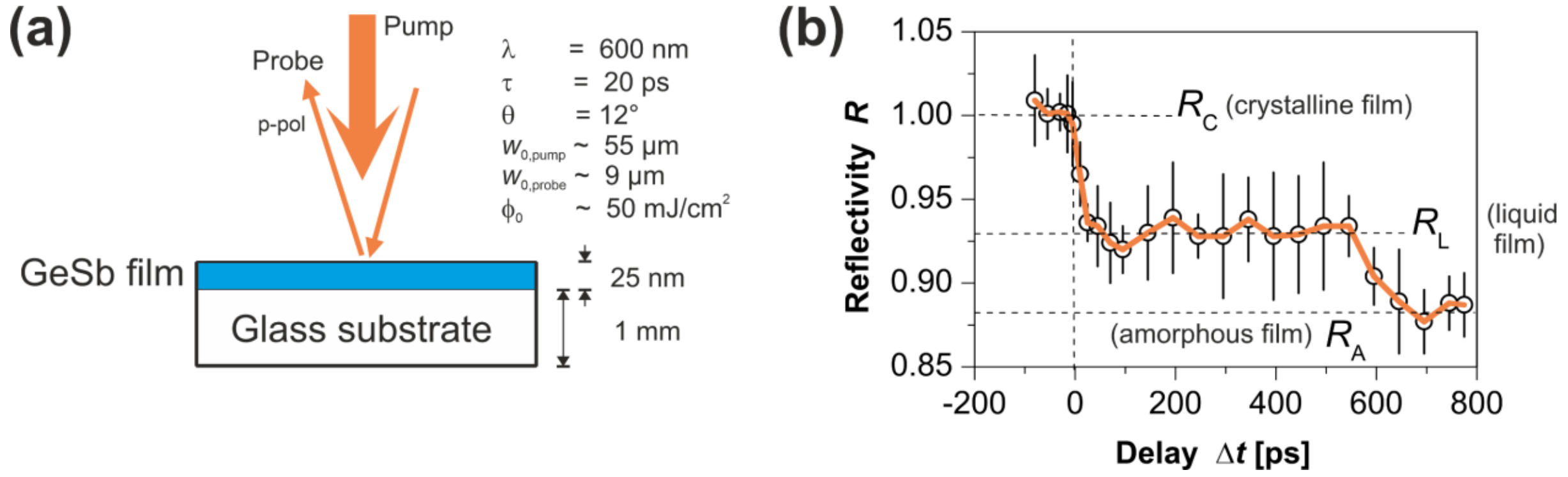

**Fig. 13.10:** (**a**) Scheme of the GeSb film sample and the non-collinear pump-probe experimental geometry along with relevant experimental parameters. (**b**) Normalized reflectivity change $R$ on crystalline GeSb films as a function of probe delay $\Delta t$ for irradiation by a single 20 ps, 600 nm pump pulse ($\theta_{pump} = 0°$). The vertical dashed line indicates the arrival of the pump pulse to the surface. The horizontal dashed lines mark the normalized reflectivity levels of the crystalline (C), liquid (L), and amorphous (A) state of a 25 nm thick GeSb film probed at 600 nm (20 ps) and an angle of incidence $\theta_{probe} = 12°$ (p-pol.). (Reprinted from [Wiggins, 2005], S.M. Wiggins et al., The influence of wavelength on phase transformations induced by picosecond and femtosecond laser pulses in GeSb thin films, J. Appl. Phys. **98**, 113518 (2005), with the permission of AIP Publishing)

Figure 13.10b presents the results of the normalized surface reflectivity $R$ measured for probe pulse delays $\Delta t$ ranging between -100 ps and +800 ps. The overall optical contrast between the amorphous and crystalline phases is ~12%. Upon irradiation by a single pump laser pulse, the surface reflectivity decreases within some tens of picoseconds (note that the temporal resolution is given here by probe pulse duration of 20 ps) from the normalized value of the crystalline film material ($R_C = 1$) to the widely constant normalized reflectivity level $R_L \sim 0.93$, indicating the formation of an optically thick liquid GeSb layer that is lasting for delays up to $\Delta t$ ~550 ps. Subsequently, the normalized reflectivity decreases further, reaching within additional ~100 ps the corresponding value of the amorphous phase $R_A \sim 0.88$. Thus, the overall C → A transition time accounts to only 650 ps here, a value that is remarkably lower than for other pump laser wavelengths and even shorter pulse durations, where several ns were reported for the complete phase cycling [Wiggins, 2005].

## 2.2 Transmission

For suitable (sufficiently transparent) samples, the pump-probe setup previously shown in Figure 13.9a can be straightforwardly adapted for transmission measurements by placing the detector in the direction of the transmitted probe beam. Moreover, it may be useful to operate the experiment in a dual mode where both the reflectivity $R(\Delta t)$ and the transmission $T(\Delta t)$ are simultaneously recorded at the same sample spot.

During the second half of the 1980's the question was explored whether the ablation of polymers with the available ns-pulsed ultraviolet lasers is of photothermal or photochemical nature [Bäuerle, 2011]? In that frame, ns-time-resolved spot probing pump-probe transmission experiments were performed by different groups for exploring the absorption dynamics in thin polymer films and that of the laser-induced photofragments in the ablation plasma plume [Koren, 1987 / Frisoli, 1991].

Pettit and Ediger performed spot probing ns-pump-probe transmission experiments during the ArF Excimer laser ablation ($\lambda_{pump}$ = 193 nm, $\tau_{pump}$ = 10 ns) of corneal tissue in a delay time regime between 10 ns and 1 ms [Pettit, 1993]. Their measurements at the probe wavelength of $\lambda_{probe}$ = 355 nm ($\tau_{probe}$ = 7 ns) revealed significant scattering at nanosecond delay times, suggesting that the onset of the ablation process occurs during the ArF Excimer laser pulse.

The group of W. Rudolph explored the pre-ablation regime of ultrashort pulse laser irradiated optical coatings (dielectric $Ta_2O_5$, $TiO_2$, and $HfO_2$ sub-µm thin films) on fused silica substrates via spot probing fs-pump-probe measurements of the reflectivity and the transmission [Jasapara, 2002 / Mero, 2005]. Their measurements ($\lambda_{pump} = \lambda_{probe}$ = 800 nm, $\tau_{pump} = \tau_{probe}$ = 25 fs, $\theta_{pump}$ = 0°, $\theta_{probe}$ = 45°, $w_{0,pump}/w_{0,probe}$ = 3, $\phi_{0,pump} = 0.3 - 0.6\cdot\phi_{th}$) allowed to retrieve the time-dependent dielectric function, while taking into account optical standing-wave effects of both the pump and the probe beam. The authors identified a very fast (sub-100 fs) transient that is followed by a slower (1 ps) transient during which a sign reversal from negative to positive was observed in the real part of the fs-laser-induced change in the dielectric permittivity. These findings were attributed to the formation of deep defect states in the band gap, possibly self-trapped excitons [Mero, 2005]. Such mid-gap defect states are typical precursors of permanent point defects and can promote material damage upon irradiation with many successive laser pulses, an effect termed "*incubation*", see Chap. 6 (Lenzner and Bonse).

The fs-laser ablation dynamics of bulk wide bandgap dielectrics (fused silica, sapphire, calcium fluoride) was studied by Chowdhury et al. through collinear fs-pump-probe experiments [Chowdhury, 2005 / Chowdhury, 2006]. The simultaneous reflectivity and transmission measurements revealed the rapid formation of dense free-electron plasma at the surface due to nonlinear absorption of a single pump laser pulse ($\lambda_{pump}$ = 800 nm, $\tau_{pump}$ = 90 fs, $\theta_{pump}$ = 0°), followed by structural damage on the timescale of a few picoseconds ($\lambda_{probe}$ = 400 nm, $\tau_{probe}$ ~ 130 fs, $\theta_{probe}$ = 0°) [Chowdhury, 2005]. In a following study, the authors extended the capabilities of their pump-probe setup for studying bi-color ablation on fused silica upon spot-processing and line scanning with delayed IR/UV double-pulses [Chowdhury, 2006]. Their results indicate that using a delay-controlled combination of the fundamental and the second harmonic beams can result in enhanced optical absorption effects, eventually manifesting in higher ablation rates.

De Haan et al. performed fs-time-resolved pump-probe transmission and reflection measurements on percolating gold films (6 nm thickness equivalent) on glass in the ablation regime [DeHaan, 2020]. The experiments showed a strong reflectivity increase accompanied by a decrease in transmission immediately after the optical excitation ($\lambda_{pump} = \lambda_{probe}$ = 1300 nm, $\tau_{pump} = \tau_{probe}$ = 56 ± 5 fs, $\theta_{pump}$ = 0°, $\theta_{probe}$ = 15°, $2w_{0,pump}$ ~ 100 µm, $2w_{0,probe}$ ~ 50 µm, $\phi_{0,pump}$ = 0.013 – 1.27·J/cm$^2$). This behavior was attributed to the increasing electron temperature and the redistribution of the electron population. The redistribution of the electrons enables excitations from the d-band to the hybridized s/p-band, increasing the optical absorption. On longer time scales and at the highest pump laser intensities, the transmission and reflection both approached the values of the bare substrate on a timescale of ~150 ps, indicating that the gold layer became optically transparent upon complete vaporization and disintegration of the film material [DeHaan, 2020].

## 2.3 Second Harmonic Generation

A disadvantage of surface optical reflectivity measurements is that sometimes no direct information about the materials structural state, i.e., changes in the atomic microstructure, can be obtained – for example when the ensemble of laser generated carriers in the conduction band of the solid exhibits a very similar reflectivity signature as the molten material, or when the reflectivity changes between the solid and the molten state are very small. This drawback may be overcome if either higher order optical processes are used, e.g., Raman scattering or nonlinear optical effects for monitoring changes in the microscopic symmetry, or by exploiting shorter radiation wavelength in the X-ray range [Siders, 1999 / Rousse, 2001 / Lindenberg, 2005] for interacting with inner shell electrons instead of valence or conduction band electrons of the solid.

In the optical spectral range, the nonlinear optical effect of *second harmonic generation* (SHG) was exploited to provide information on ultrafast structural phase transitions, particularly on the above mentioned nonthermal melting. Shank et al. used fs-time-resolved *surface SHG* for observing the structural dynamics during laser-induced melting of silicon [Shank PRL-2, 1983]. Since silicon represents a centrosymmetric crystal, second harmonic contributions from the bulk are not generated and solely a near surface layer strongly confined to the silicon wafer interface contributes to the SHG. Thus, the method is very surface sensitive in this case.

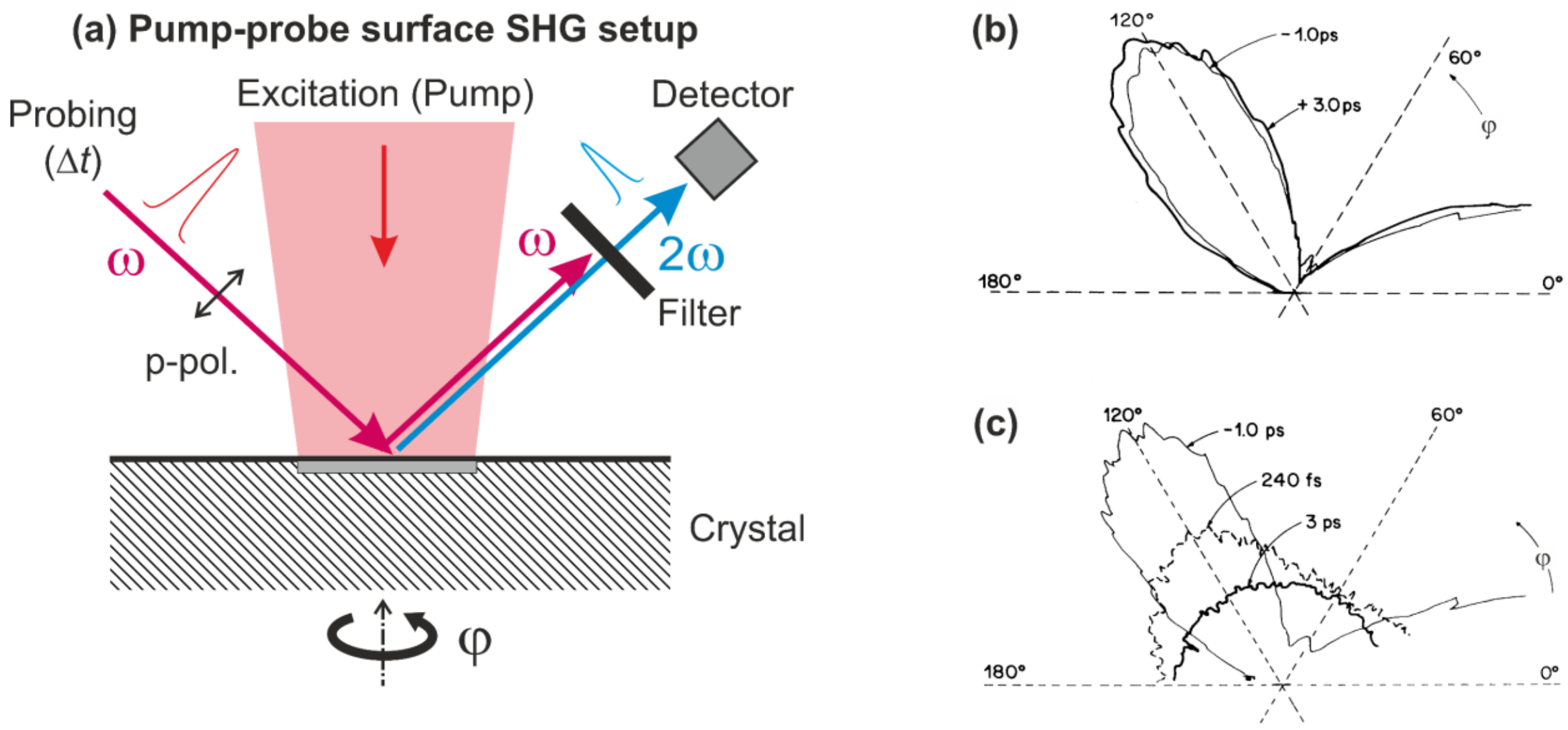

**Fig. 13.11:** (**a**) Scheme of pump-probe setup for time-resolved detection of surface-generated second harmonic generation (SHG) [$\lambda_{pump}$ = 620 nm, $\tau_{pump}$= 90 fs, $2w_{0,pump}$ ~ 100 µm, $\theta_{probe}$ = 45°]. $\varphi$ denotes the polar angle for sample rotation around the surface normal of the probed (111) silicon crystal. (**b**) Polar plot of SHG intensity as a function of polar angle $\varphi$ for two different pump-probe delay times of $\Delta t$ = –10 ps and +3.0 ps for low pump fluences of $\phi_{0,pump} = 0.5\times\phi_{th,am}$. (**c**) Polar plot of SHG intensity as a function of polar angle $\varphi$ for three different delays of $\Delta t$ = –10 ps, +0.24 ps, and +3.0 ps for high pump fluences of $\phi_{0,pump} = 2.0\times\phi_{th,am}$. (Figures (b,c) reprinted with permission from [Shank PRL-2, 1983], C.V. Shank et al., Phys Rev. Lett. **51**, 900 – 902, 1983. Copyright (1983) by the American Physical Society)

Shank and co-workers used a (111) oriented polished surface of a single-crystalline silicon wafer that was irradiated under normal incidence ($\theta_{pump}$ = 0°) by single fs-laser pulses ($\lambda_{pump}$ = 620 nm, $\tau_{pump}$= 90 fs, $2w_{0,pump}$ ~ 100 µm). The structural state of the surface was then probed in the center of the laser-excited area by measuring the surface reflected SHG of a weak delayed p-polarized probe pulse incident at an angle of incidence of $\theta_{probe}$ = 45°, see Figure 13.11a. The reflected light of the pump pulse was removed by an optical filter.

The results of the surface-generated SHG were measured at different delay times $\Delta t$ as a function of the polar angle $\varphi$ and are visualized through polar plots provided in Figure 13.11b for a pump beam fluence of $\phi_{0,pump} = 0.5\times\phi_{th,am}$ (below the threshold of surface amorphization $\phi_{th,am}$), and in Figure 13.11c for a four times higher pump-beam fluence of $\phi_{0,pump} = 2\times\phi_{th,am}$. As it can be seen in Figure 13.11b, the measured SHG signal of non-excited or weakly-excited silicon exhibits characteristic lobes of threefold rotational symmetry in $\varphi$, being characteristic for the (111) surface of cubic crystalline silicon. In contrast, for the strongly excited material (Figure 13.11c), distinct changes of the SHG signal develop after the exciting pump pulse. Already at a delay time of $\Delta t$ = 240 fs after the arrival of the pump pulse, the width of the lobe at $\varphi$ = 120° broadens and its minimum becomes less distinct. After $\Delta t$ = 3 ps the SHG distribution is completely isotropic.

These observations unquestionably proved the onset of structural transformation of a crystalline silicon surface into an isotropic liquid layer, in excellent agreements with spot-probing fs-time-resolved surface reflectivity measurements of the same group of authors [Shank PRL-1, 1983]. Later SHG was used also to study ultrafast phase transformations manifesting in other (non-centrosymmetric) semiconductor crystals, such as GaAs [Sokolowski-Tinten, 1991 / Govorkov, 1992 / Mazur, 1997] and InSb [Shumay, 1996].

## 2.4 Diffraction

Apart from the specular surface reflectivity or SHG, also transiently diffracted signals can be analyzed in time-resolved pump-probe schemes – provided there is a diffracting structure formed at the surface or in the laser processed volume. At the surface, such dynamically diffracting structures can be intentionally generated via laser-generated *interference gratings*, or they may form in a self-ordered way in the form of grating-like permanent *Laser-induced Periodic Surface Structures* (LIPSS) during the laser processing (see Chap. 9, Ancona et al.) and may be even used for tailored surface functionalization (see Chap. 22, Brock et al. and Chap. 27, Zielinski et al.), data storage, and security applications (see Chap. 35, Killaire et al.).

Historically, transient optical gratings were first explored during the early 1930’s in the form of ultrasonic waves — as known from the historic ultrasound-induced (incoherent) light diffraction experiments of Debye and Sears [Debye & Sears, 1932], and Lucas and Biquard [Lucas & Biquard, 1932]. These experiments allowed to determine the velocity, dispersion, and damping properties of ultrasonic waves, providing the ground for *Acousto-Optic Modulators* (AOM’s) that are nowadays widely used in laser technology [Römer, 2014]. With the advent of the laser, its coherence and high intensities allowed via two-beam interference the efficient generation of grating-like structures in a medium, and even the confinement to micrometer-sized spots. Simultaneously, the coherence of laser radiation enabled the sensitive probing such structures via diffraction. A comprehensive review on laser-induced dynamic gratings be found in the book of Eichler et al. [Eichler Book, 1986].

The fact that light can couple to matter on extremely short (sub-fs) timescales in turn allows to probe the transient state of matter via ultrafast diffraction. Such pump-probe dynamic diffraction studies were initially performed for analyzing the dynamics of laser-generated carriers in semiconductors (generation, state-filling, intra-band relaxation, diffusion, recombination, etc.) and related effects at pump laser fluences below the melting threshold [Smirl, 1983 / Vaitkus, 1986 / Eichler, 1987 / Sjodin, 1998], but were later also applied to explore ultrafast phase transitions, such as thermal and nonthermal melting initiated at pump laser fluences exceeding the melting threshold [Govorkov, 1992].

In more detail, Govorkov et al. combined the approach of surface SHG with that of interference-based transient diffraction for studying in a spot probing pump-probe experiment the ultrafast phase transitions in the semiconductor GaAs upon irradiation with intense femtosecond laser pulses [Govorkov, 1992]. Two pump pulses ($\lambda_{pump}$ = 620 nm, $\tau_{pump}$ = 100 fs) of equal energy were used to create a microscopic transient interference grating that was then interrogated by a weak, delayed probe pulse ($\lambda_{probe}$ = 620 nm, $\tau_{probe}$ = 100 fs). Photodetectors were placed in the directions of reflections of the fundamental (620 nm) and the SHG (310 nm) as well as in the 1$^{st}$ diffraction-orders for both wavelengths, recording the linear and non-linear material response. The authors showed that a ~13 nm thick surface layer of GaAs transforms rapidly into a centrosymmetric semiconductor-like phase that exists over ~ 300 fs after the impact of the strong fs-laser pump pulse excitation. This early excitation stage is followed by a state of metallic linear optical response, which is typical for a molten material. These measurements further confirmed that nonthermal melting manifest also in III-V semiconductors.

Other intrinsically light diffracting structures are LIPSS. They are a universal phenomenon [van Driel, 1982] manifesting often during processing with coherent linearly polarized laser radiation. Under study for more than five decades after their first experimental observation in 1965 by Birnbaum [Birnbaum, 1965], LIPSS can certainly be considered as a scientific evergreen [Bonse IEEE, 2017]. LIPSS usually manifest as quasi-periodic 1D-gratings imprinted either in the surface topography or in other surface properties of the irradiated material. Upon irradiation with ultrashort laser pulses, they feature either spatial periods ($\Lambda$) of the order of the laser irradiation wavelength – so-called *low spatial frequency LIPSS* (LSFL, $\Lambda \sim \lambda$) – or, as *high spatial frequency LIPSS* (HSFL), they are significantly smaller ($\Lambda << \lambda$) and can have periods as small as a few tens of nm only. For more details on the electromagnetic origin of these prominent surface structures, the reader is referred to the specialized literature [Sipe, 1983 / Siegman, 1986 / Bonse & Gräf, 2020 / Höhm, 2013 / Rudenko, 2017 / Rudenko, 2020].

Probing the dynamics of LIPSS formation on non-transparent metals or semiconductors with optical radiation in a specific diffraction order in a reflective geometry is very challenging since the ablation of absorbing material sets in after some tens of ps, subsequently screening the surface of interest for several hundreds of ps up to several ns via the effect of *"plasma-shielding"*, see Chap. 1 (Nolte et al.). Moreover, the sub-µm spatial periods of LIPSS are close to the *optical diffraction limit* that sets a lower limit for far-field detection methods at feature sizes of ~λ/2. Thus, HSFL are usually not accessible via far-field optical probing methods.

Despite these two challenges, the formation of the near-wavelength-sized LSFL on transparent materials (dielectrics) was successfully studied with optical radiation and high temporal resolution in a transillumination diffraction geometry [Höhm, 2013]. The corresponding fs-pump-probe diffraction setup is sketched in Figure 13.12.

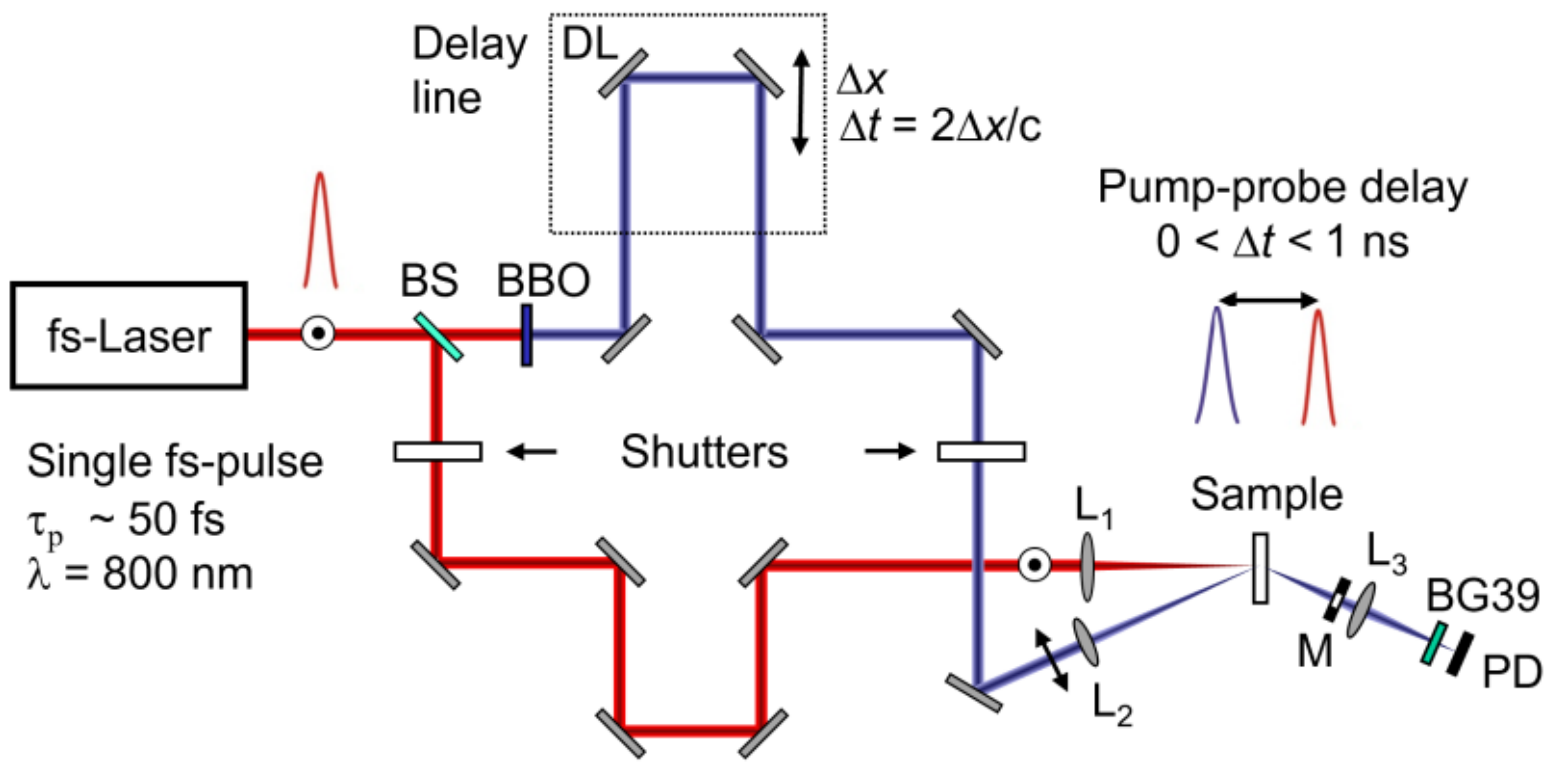

**Fig. 13.12:** Scheme of pump-probe setup for fs-time-resolved diffraction experiments in transillumination geometry. Abbreviations: BS – beam splitter; BBO – barium borate crystal for frequency doubling; DL – delay line; $\Delta t$: delay of the probe pulse; $L_{1,2,3}$ – lenses with focal lengths $f_1$ = 200 mm, $f_2$ = 60 mm, $f_3$ = 25 mm; M – mask; BG39 – optical bandpass filter; PD – photodiode. The samples are 10×10×1 mm$^3$ double-side polished fused silica plates. (Reprinted from [Höhm, 2013], S. Höhm et al., Femtosecond diffraction dynamics of laser-induced periodic surface structures on fused silica, Appl. Phys. Lett. **102**, 054102 (2013), with the permission of AIP Publishing)

Linearly polarized laser pulses with a duration of $\tau_p$ = 50 fs were provided by a Ti:sapphire fs-laser system that was operated at a pulse repetition frequency of 250 Hz. The pump beam at the fundamental wavelength ($\lambda_{pump}$ = 800 nm) was focused directly onto the front surface of a double-side polished fused silica plate (10×10×1 mm$^3$, Suprasil) using a 200 mm focal length lens ($L_1$) under normal incidence to generate regular low spatial frequency LIPSS (LSFL) with spatial periods $\Lambda_{LSFL}$ between 550 and 700 nm and lines parallel to the pump beam polarization over the entire central irradiated area. That periods are close to the value $\Lambda \sim \lambda/n$ (with $n$ the refractive index of the material), as expected for transparent dielectrics [Höhm, 2012]. The Gaussian beam diameter ($1/e^2$-decay) of the pump beam was approximately $2w_{0,pump} \approx 92$ µm. A fraction of the original laser beam was separated by a beam splitter (BS), subsequently frequency-doubled in a barium borate (BBO) crystal ($\lambda_{probe}$ = 400 nm), and then focused by an achromatic lens ($L_2$, 60 mm focal length) under an angle of incidence of $\theta \approx 15°$ into the center of the pump beam excited spot ($2w_{0,probe} \approx 20$ µm). The far-field propagated intensity pattern resulting from first-order diffraction of the p-polarized probe beam on the grating-like LSFL was collected in transmission geometry with a 25 mm focal distance lens ($L_3$) and focused onto a fast photodiode (PD). A tailored mask (M) helped to spatially suppress scattered stray light. An infrared-cutting bandpass filter (BG 39) placed in front of the photodiode removed residual pump beam radiation. Two electromechanical shutters allowed to select independently the desired number of pump or probe pulses [Höhm, 2013].

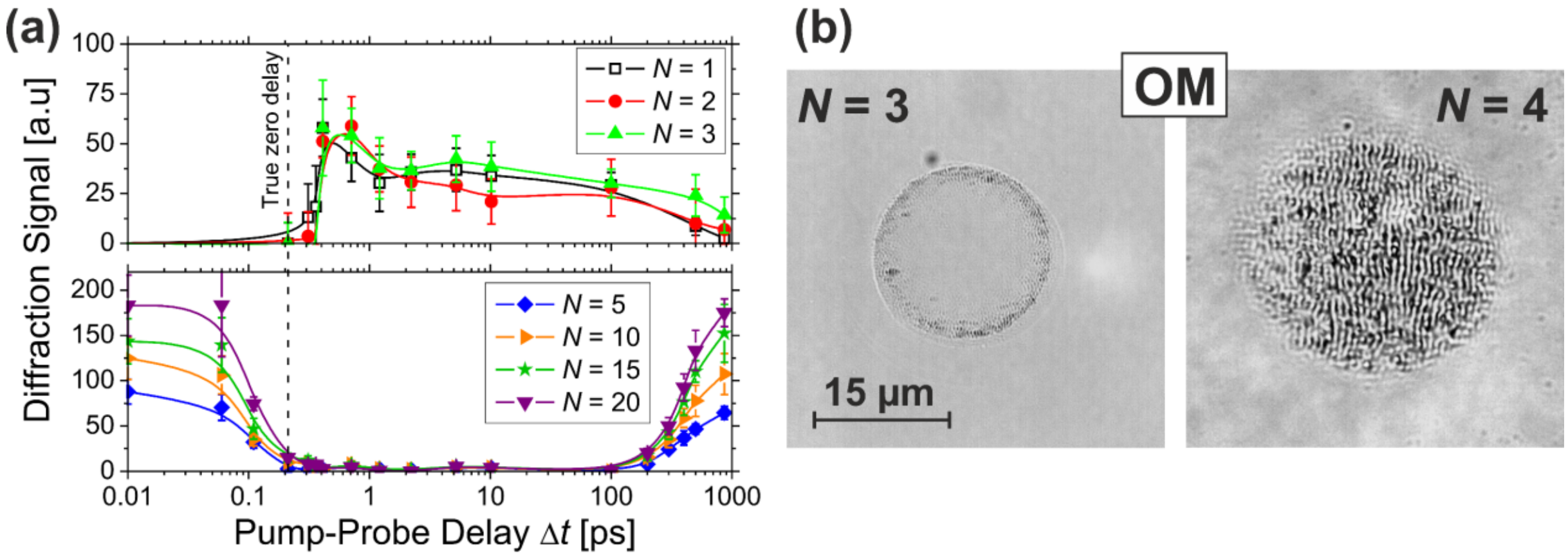

**Fig. 13.13:** (**a**) Pump-probe diffraction signal as a function of the delay time $\Delta t$ upon irradiation at a peak fluence of $\phi_0 = 3.9$ J/cm$^2$ by different numbers of pump pulses ($N = 1$ to 20, $\lambda_{pump} = 800$ nm, $\lambda_{probe} = 400$ nm, $\tau = 50$ fs) [Höhm, 2013]. For using a logarithmic scaling of the ordinate, all delays are shifted by +0.21 ps. The vertical dashed line indicates the zero delay. (**b**) Optical micrographs representing the two different pump pulse number regimes [$N \leq 3$ (left image) without LSFL and $N \geq 4$ (right image), where pronounced LSFL are visible] (Reprinted from [Höhm, 2013], S. Höhm et al., Femtosecond diffraction dynamics of laser-induced periodic surface structures on fused silica, Appl. Phys. Lett. **102**, 054102 (2013), with the permission of AIP Publishing)

Figure 13.13 presents results of the first-order diffraction signal recorded in transmission geometry at 400 nm probe wavelength as a function of the probe pulse delay. The pump-probe measurement was performed after $N$ pre-irradiating pump pulses (50 fs duration, 800 nm wavelength) had “dressed” the surface at a suitable peak laser fluence ($\phi_0 = 3.9$ J/cm$^2$) below the single-pulse damage threshold [Höhm, 2013]. At a low number of surface dressing pump pulses ($N = 1$ to 3) and even before a permanent LSFL-related surface relief was observed, an ultrafast transient diffraction at the LSFL spatial frequencies was evidenced in the transparency regime of the sample (Figure 13.13a, top panel). After the 4$^{th}$ pump pulse, a permanent surface relief of LSFL was formed in the probed spot (see the optical micrograph displayed in Figure 13.13b), accompanied with the excitation of an optically thick laser-induced free-electron-plasma at the sample surface that was dynamically shielding the probe beam (Figure 13.13a, bottom panel) for $N > 5$. These effects were attributed to a complex interplay of a transient refractive index grating formed within less than 300 fs by *self-trapped excitons* (STE’s) leading to incubation, local heating, electron plasma relaxation, and ablation [Höhm, 2013].

The measurements demonstrate the capabilities of spot-probing ultrafast pump-probe diffraction experiments, particularly in the case that the signal is not screened by other (slower) effects, such as plasma-shielding. However, care must be taken to ensure that the probe beam spot illuminates a sufficiently large number of diffracting surface features (here the LSFL ridges) for establishing a macroscopic far-field diffracted signal. Thus, a compromise between the pump and the probe spot beams must be found ($w_{0,pump} > w_{0,probe} >> \Lambda_{LSFL}$). Moreover, the sensitivity to optical surface scattering must be checked to ensure the necessary specificity to the diffracted signal.

## 2.5 Scattering

Another possibility of studying the laser-matter interaction is based on scattering of the probing radiation. In this scenario, local variations of the refractive index of the laser-irradiated material at the wavelength are causing elastic scattering (Huygens' principle) that can be used in background-free detection schemes.

### Surface Scattering

Figure 13.14 provides an example of a spot probing pump-probe experiment for detecting laser-induced particles ablated from the surface of fs-laser-irradiated dielectrics (glass, crystals), while being placed in a vacuum chamber operated at a pressure of $10^{-5}$ mbar [Rosenfeld, 1998]. The polished surfaces of various oxides (fused silica, quartz, sapphire, and magnesia crystals) were irradiated (pumped) by single focused Ti:sapphire fs-laser pulses ($\lambda_{pump}$ = 800 nm, $\tau_{pump}$ = 120 fs) and probed collinearly at the same irradiation wavelength and pulse duration (Figure 13.14a). Both beams (pump and probe) were focused directly to the sample surface by the same 75 mm focal length lens. The linear polarization of the probe pulse was rotated with a half-wave plate placed in the probe beam path to be perpendicular to that of the pump pulses. A fraction of the scattered probe beam radiation was collected by a long working distance microscope that was imaging the laser-excited spot at an angle of $\theta_{probe}$ = 30° to the sample surface normal. A part of the probe beam was separated by a beam splitter and detected by a photomultiplier. A linear polarizer was placed in front of the microscope objective and aligned to the direction of the probe beam to suppress noise contributions caused by reflections/scattering of the pump beam. For a given delay time $\Delta t$, the scattered signal was averaged for 10 different single-pulse pump-probe irradiation events, each performed on a new, previously unexposed, sample site. The contribution of the pump pulse to the scattered signal was determined independently by blocking the probe beam and then used to correct the probe pulse scattering signal. The energy of the pump and the probe pulse was chosen at 50 to 54 µJ and 9 µJ, respectively, resulting in a pump beam fluence of ~10 J/cm$^2$, i.e., more than twice the single-pulse surface damage threshold fluence.

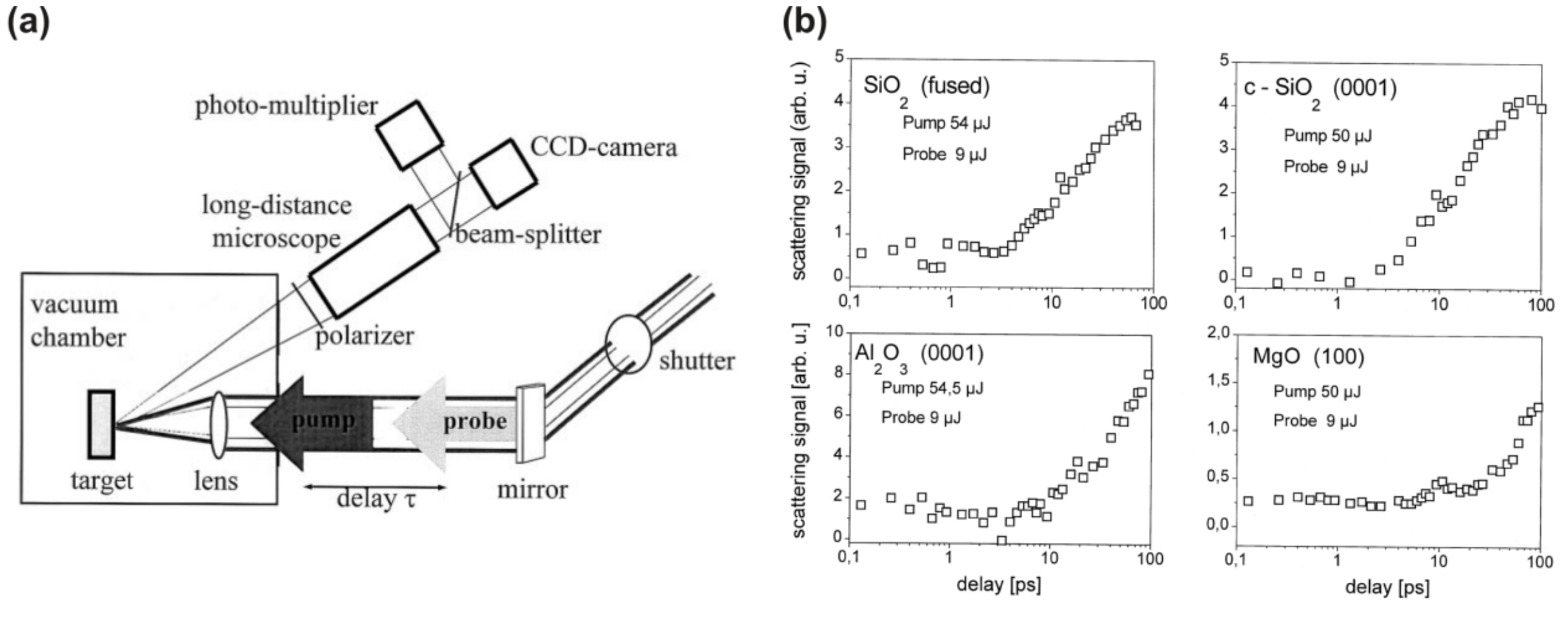

**Fig. 13.14:** (**a**) Scheme of the setup of the spot probing surface-scattered light pump-probe experiment ($\lambda_{pump}$ = 800 nm, $\tau_{pump}$ = 120 fs, $\theta_{pump}$ = 0°, $E_{pump}$ = 54 µJ, $\lambda_{probe}$ = 400 nm, $\tau_{probe}$ = 100 fs, $\theta_{probe}$ = 30°, $E_{probe}$ = 9 µJ) [Rosenfeld, 1998]. (**b**) Scattered light signal of the probe pulses as a function of the pump-probe delay time for four different dielectrics (fused $SiO_2$, crystalline c-$SiO_2$, c-$Al_2O_3$, and c-MgO). Note the logarithmic scaling of the abscissa. The data points were averaged over ten single-pulse irradiation events at the same delay. (Reprinted from [Rosenfeld, 1998], Appl. Surf. Sci., **127–129**, A. Rosenfeld et al., Time resolved detection of particle removal from dielectrics on femtosecond laser ablation, 76 – 80, Copyright (1998), with permission from Elsevier)

The results of these optical pump-probe measurements are summarized in Figure 13.14b, covering pump-probe delays between 0.1 and 100 ps for all four materials. Depending on the material, several ps after the impact of the pump pulse a significant increase of the signal scattered at the probed surface spot can be detected, which was attributed to the onset of ablation. These delays account to ~3 ps for fused or crystalline silica (a-$SiO_2$, c-$SiO_2$), whereas for crystalline sapphire (c-$Al_2O_3$) and c-MgO, the material removal starts after ~12 and ~20 ps, respectively [Rosenfeld, 1998]. The variations among the materials were attributed to the differences in the electron-phonon coupling strength, and additionally related to the distinct surface morphologies of the irradiated spots being indicative for melting and vaporization on sapphire, amorphous and crystalline quartz, while the surface of MgO showed also cracking.

## Volume Scattering

Schaffer et al. studied the optical breakdown induced by tightly focused fs-laser pulses in liquid water via two complementary time-resolved collinear pump-probe experiments, i.e., through brightfield transmission microscopy (see Sect. 3.1 below), and also via spot-probing scattering detection [Schaffer PhD, 2001 / Schaffer, 2002]. The setup for the pump-probe scattering and deflection experiment is sketched in Figure 13.15a. A single Ti:sapphire pump laser pulse ($\lambda_{pump}$ = 800 nm, $\tau_{pump}$ = 100 fs, $E_p$ = 1 µJ) is tightly focused by a microscope objective ($NA$ = 0.65) into an optical cell filled with water. A frequency-doubled delayed single probe pulse ($\lambda_{probe}$ = 400 nm, $\tau_{probe}$ = 100 fs) is collinearly coupled into the pump beam path in front of the focusing objective. The probe beam interrogates an area of about 30 µm in diameter around the center of the pumped region. The divergent probe beam is then collected by another microscope objective and convergently directed to a photodiode-based detection system. The directly transmitted probe beam radiation is stopped by a beam block, so that the photodiode registers only the light that is scattered out of the undisturbed probe beam path and propagates around the beam stop.

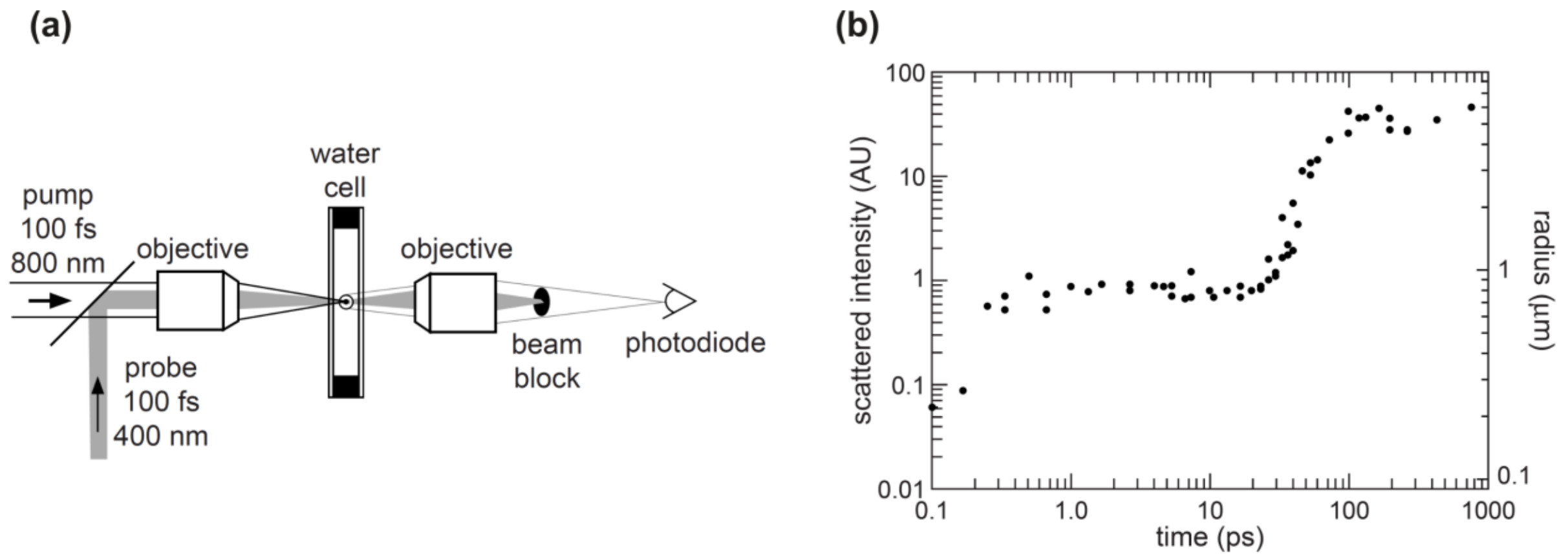

**Fig. 13.15:** (**a**) Scheme of the setup of the spot integrating volume-scattered/deflected light pump-probe experiment ($\lambda_{pump}$ = 800 nm, $\tau_{pump}$ = 100 fs, $\theta_{pump}$ = 0°, $E_p$ = 1 µJ, $\lambda_{probe}$ = 400 nm, $\tau_{probe}$ = 100 fs, $\theta_{probe}$ = 180°). The directly transmitted probe beam is blocked so that only the scattered/deflected probe radiation reaches the photodetector. (**b**) Time-resolved scattered signal (left ordinate) of single probe pulses as a function of the pump-probe delay time for the single laser pulse induced optical breakdown in water. The right ordinate corresponds to the estimated radius of the scattering super-critical (metal-like) focal plasma region. Note the double-logarithmic scaling. (Figures (a,b) reprinted from [Schaffer PhD, 2001], with permission and courtesy of Christopher B. Schaffer)

Figure 13.15b provides corresponding time-resolved measurements of the scattering signal from the pump laser pulse generated plasma in the delay range between 100 fs to 1 ns. Within a delay of ~200 fs the scattering signal sharply increases by approximately one order of magnitude to a plateau value that persists for additional $\Delta t$ ~ 30 ps before the scattering signal strongly rises again by one to two orders of magnitude towards another saturation level. The early increase of the scattering signal indicates that a free-electron plasma is formed within ~200 fs due to the strong excitation of the focal volume by the pump pulse. This rise time can be considered as the temporal resolution of the setup. When assuming that the laser-induced electron plasma density always exceeds the critical density for the probe beam radiation ($\approx 10^{22}$ $cm^{-3}$), the pump-beam generated focal plasma acts like a metallic ball [Schaffer, 2002]. The scattered amplitude is then proportional to the cross-sectional area of the plasma. In combination with the co-axial time-resolved imaging (not shown here) the authors calibrated the scattering signal against the radius of the metal-like focal plasma (see the right ordinate in Figure 13.14) [Schaffer PhD, 2001]. The radius of ~0.8 µm associated with the plateau region that is observed for delay times between 0.2 and 30 ps then approximately reflects the lateral dimension of the early focal metal-like electron plasma. This plasma is very hot compared to the surrounding water. For delays exceeding $\Delta t$ ~30 ps it expands forming a microscopic bubble, doing work on the surrounding material until the kinetic energy available in the breakdown-plasma is balanced out. When the breakdown-plasma expansion stops around $\Delta t$ ~800 ps, a pressure wave is launched and continues to propagate outward [Schaffer, 2002 / Juhasz, 1996 / Noack, 1998].

Time-resolved experiments like these were important to clarify the mechanisms and dynamics of the laser-induced optical breakdown in water, subsequent bubble-formation, cavitation effects, and shockwave emission during the laser-processing and nanosurgery of soft biological tissues (that mainly consist of water). For more details, the reader is referred here to review articles and book chapters available in the pertinent literature [Vogel, 2003 / Vogel, 2005 / Linz, 2023].

Conceptually close to these transmission scattering/deflection experiments is the *Transient Lens* (TrL) method. This pump-probe method was initially developed to detect the heat evolution in aqueous solutions in the picosecond time range, manifesting through the temperature dependence of the refractive index [Terazima, 1996]. The TrL method detects a refractive index lens that is created locally in a transparent sample upon irradiation with a spatially inhomogeneous (e.g. Gaussian-shaped) pump laser beam. Sakakura, and co-workers implemented the method for studying fs-laser-induced heating, Kerr-effect, and pressure waves in various dielectrics (glasses) [Sakakura, 2003 / Sakakura, 2005 / Sakakura OPEX-1, 2007 / Sakakura OPEX-2, 2007 / Sakakura, 2011].

## Coherent XUV- and X-ray Scattering

Taking benefit of the spatial and temporal coherence of the probe beam radiation with even shorter wavelengths enables to study laser-induced matter-reorganization with unprecedented spatio-temporal resolution. Currently, a spatial resolution in the sub-µm range and a temporal resolution in the sub-ps range can be achieved by using 3$^{rd}$- and 4$^{th}$-generation light sources, such as *Synchrotrons*, *Linear Accelerators*, or *Free-Electron Lasers* (FELs), which are capable of generating XUV and X-ray pulses with high peak brightness, short wavelengths and femtosecond pulse durations (see Sect. 7.2). Moreover, their photon energy can be controlled over a wide range allowing to be suitably selected and adjusted to the absorption properties of the probed sample material.

In order to demonstrate the capabilities of FELs, Figures 13.16 and 13.17 show a proof-of-principle example for the investigation of the dynamics of the formation of nanometric LIPSS by coherent scattering of ultrashort pulsed (fs) XUV radiation. The text is an excerpt of Refs. [Sokolowski-Tinten, 2010 / Sokolowski-Tinten, 2023 / Bonse & Sokolowski-Tinten, 2024].

Figure 13.16 provides a scheme of an optical pump / XUV probe scattering experiment performed at the FLASH-facility at DESY (Hamburg, Germany). The beamline was operated in single bunch, ultrashort-pulse mode, delivering pulses with 13.5 nm wavelength, 10 – 20 fs pulse duration, and about 10 – 20 µJ mean pulse energy [Sokolowski-Tinten, 2010 / Bonse & Sokolowski-Tinten, 2024]. These probe pulses were focused onto the sample surface to a full-width-at-half-maximum (FWHM) spot size of about 20 µm at normal incidence ($\theta_{probe} = 0°$). The scattered radiation was detected in transmission geometry on a suitable CCD camera. All irradiation experiments were performed in a high vacuum environment.

The samples were 100 nm thick polycrystalline silicon (pc-Si) films deposited on large arrays of 100 × 100 µm$^2$ sized 20 nm thin $Si_3N_4$ membrane windows, supported by a silicon wafer frame. This target design allowed a flexible replacement of the sample window between consecutive irradiation events. The movement of the sample to a fresh surface spot was necessary due to the damage induced upon single-pulse irradiation with the laser or the XFEL radiation.

The silicon films were irradiated at an angle of incidence of $\theta_{pump} = 47°$ by single pump laser pulses ($\lambda_{pump}$ = 523 nm, $\tau_{pump}$ = 12 ps), focused to an elliptical beam spot with FWHM-dimensions of approximately 40 × 30 µm$^2$. The laser peak fluence was set at $\phi_0 \approx 1.7$ J/cm$^2$ sufficient to completely ablate the silicon film (ablation threshold fluence $\phi_{abl} \approx 0.5$ J/cm$^2$).

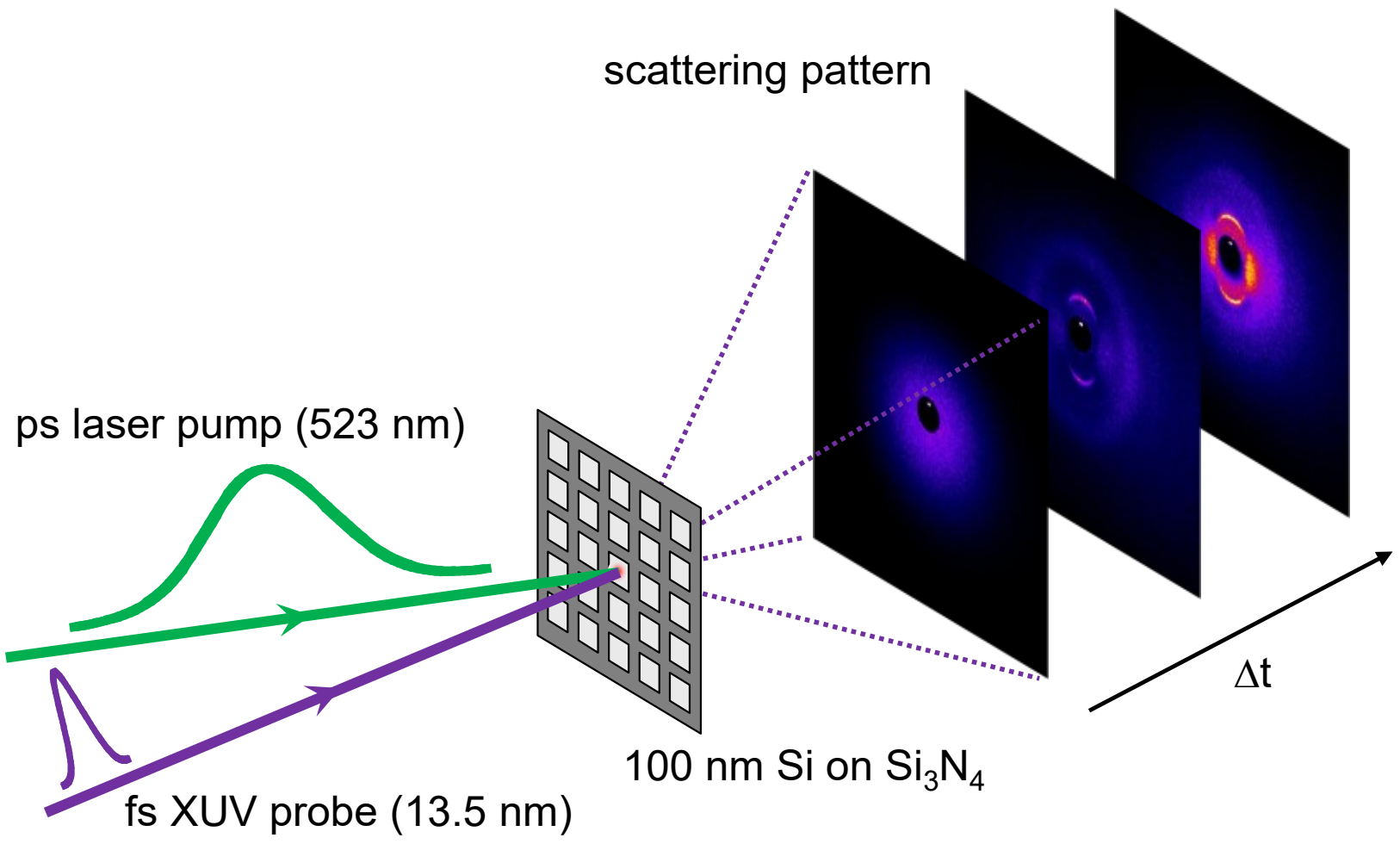

**Fig. 13.16:** Scheme of the non-collinear fs-time-resolved single-pulse coherent scattering experiments performed at the XUV-FEL (FLASH, DESY, Hamburg, Germany) [Sokolowski-Tinten, 2010]. Single focused ps-laser pump pulses ($\lambda_{pump}$ = 523 nm, $\tau_{pump}$ = 12 ps) are incident at an angle of ($\theta_{pump}$ = 47°). Delayed fs-XUV pulses ($\lambda_{probe}$ = 13.5 nm, $\tau_{probe}$ = 10 – 20 fs) are probing the laser excited region at 100 nm thick polycrystalline Si film on a 20 nm thick $Si_3N_4$ membrane under normal incidence ($\theta_{probe}$ = 0°). The scattering patterns are recorded by a CCD camera that is synchronized with the fs-XUV probe pulses. The black disks in the scattering patterns represents a beam stop that is blocking the directly transmitted XUV radiation (zero order). Scattered optical laser radiation is blocked by a filter (not drawn). (Reprinted from [Bonse & Sokolowski-Tinten, 2024], J. Bonse et al., Probing laser-driven structure formation at extreme scales in space and time, Laser & Photon. Rev. **18**, 2300912 (2024), Copyright 2024 under Creative Commons BY 4.0 license. Retrieved from https://doi.org/10.1002/lpor.202300912)

Figure 13.17a compiles a set of six transient scattering patterns (frames) that were recorded at selected pump-probe delay times $\Delta t$ ranging between −10 ps and +4.5 ns. The vertical white bar in the frame in the upper left corner ($\Delta t$ = −10 ps) indicates the projection of the p-polarized pump laser beam polarization direction to the sample surface. The intensity scale of the false-color encoding is identical for all scattering patterns. The scattering pattern acquired at the negative delay time (see $\Delta t$ = −10 ps) reveals the signature the non-irradiated polycrystalline silicon film. At positive delays the scattering patterns rapidly develop into an intricate structure. Three characteristic features can be identified (labelled f1 – f3 in the frame recorded at $\Delta t$ = 840 ps) and evolve with different dynamics. Already during the laser pulse irradiation the most prominent feature (f1) becomes visible, having a characteristic double-sickle shape (see $\Delta t$ = 15 ps). The intensity of the f1-feature increases with the delay time and reaches a maximum around ≈1 ns. It persists until the sample gets destroyed completely after a few ns (see $\Delta t$ = 4.5 ns).

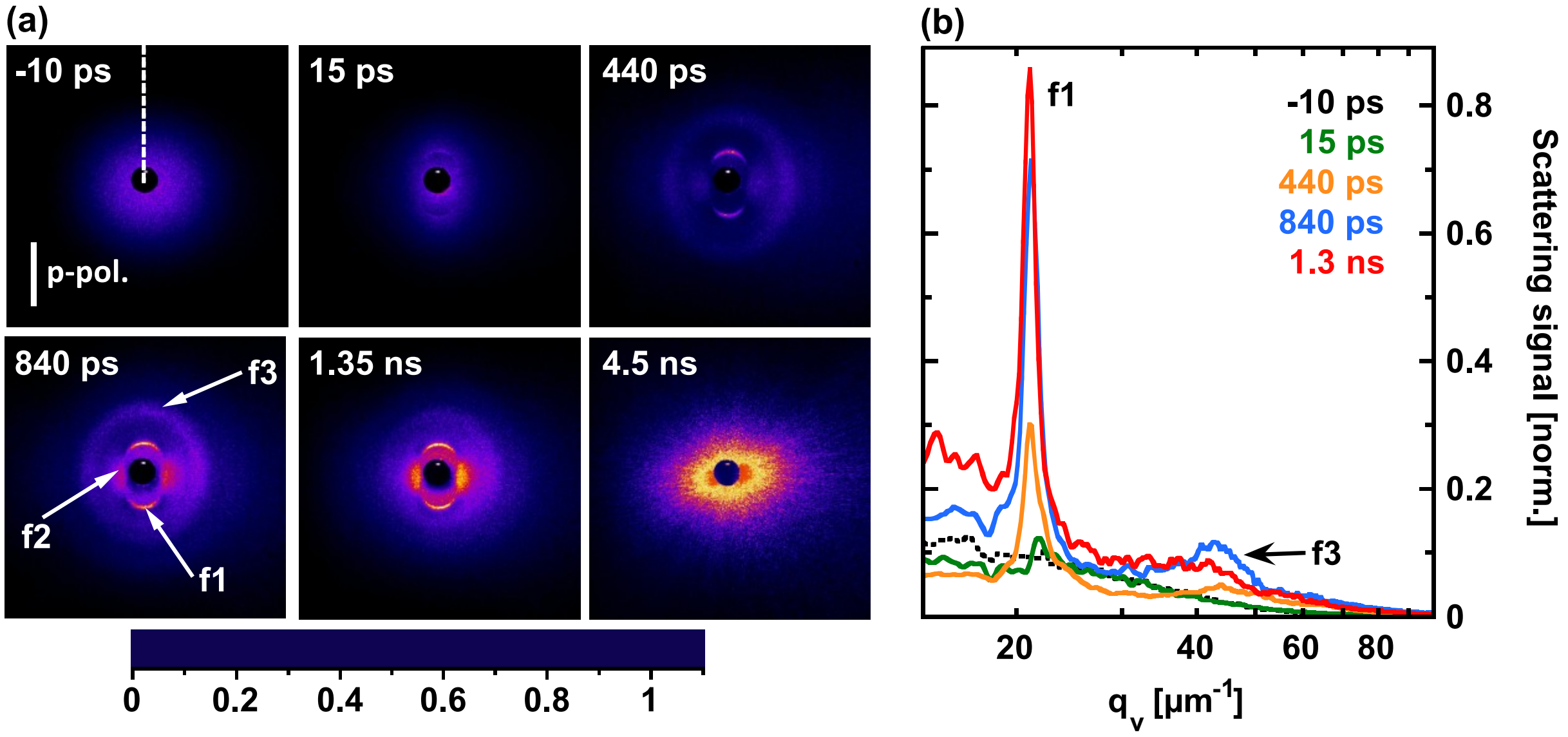

**Fig. 13.17:** (**a**) Transient XUV scattering patterns (false-color representation) of a 100 nm thick polycrystalline silicon film as a function of the pump-probe delay time $\Delta t$ between the optical pump pulse (peak fluence $\phi_0 \approx 1.7$ J/cm$^2$, pulse duration 12 ps, wavelength 523 nm, $\theta_{\text{pump}} = 47°$), and the XUV probe pulse (pulse duration 10-20 fs, wavelength 13.5 nm $\theta_{\text{probe}} = 0°$). The displayed range of spatial frequencies $q_{h/v}$ (in horizontal and vertical directions) is between −92 µm$^{-1}$ and +92 µm$^{-1}$. The false-color intensity scale is normalized and the same for all frames. The solid white bar in the delay frame for $\Delta t = -10$ ps indicates the projection of the laser beam polarization on to the surface (p-pol.) [Bonse & Sokolowski-Tinten, 2024]. (**b**) Vertical cross sections along the white dashed line in the image for $\Delta t = -10$ ps in (a) for different pump-probe time delays. (Reprinted from [Bonse & Sokolowski-Tinten, 2024], J. Bonse et al., Probing laser-driven structure formation at extreme scales in space and time, Laser & Photon. Rev. **18**, 2300912 (2024), Copyright 2024 under Creative Commons BY 4.0 license. Retrieved from https://doi.org/10.1002/lpor.202300912)

This dynamics can also be seen in the vertical cross-sections provided in Figure 13.17b. Here, the feature f1 corresponds to a sharp peak with the maximum at the wave number $q_v \approx 21$ µm$^{-1}$. It is indicative of the formation of a well-oriented structure perpendicular to the laser beam polarization that has a spatial period of $\Lambda_{f1} \approx 300$ nm in real space. The feature f1 can be associated with type-s LSFL-I [Bonse & Gräf, 2020]. It is split here into two branches due to the inclined angle of incidence (the second branch is hidden/blocked here by the beam stop). Almost quantitative agreement the measured diffraction pattern and the calculated efficacy factor η based on Sipe's theory of the electromagnetic absorption of a microscopically rough surface [Sipe, 1983] was obtained when optical constants of liquid Si (which behaves metallic) were used as input for the calculations [Sokolowski-Tinten, 2023]. In view of the high (pump) laser fluence well above the melting and ablation thresholds, this assumption is reasonable here since melting will occur already during the 12 ps duration of the optical laser pulse.

Feature f2 exhibits two broad maxima at $q_h$ between 15 and 25 µm$^{-1}$ scattered in the direction perpendicular to the laser beam polarization. It starts to develop significantly later than f1 and f3, i.e., after ≈ 300 ps and increases in intensity until the irreversible disintegration of the sample after a few ns.

Feature f3 emerges almost as fast as feature f1, represented by a significantly weaker broad and slightly elliptical diffraction ring. The elliptical shape of f3 indicates the formation of randomly oriented periodic structures with spatial periods ranging from about 170 to 180 nm in the horizontal direction and from 140 to 150 nm in the vertical direction. The integrated scattered signal amplitude of the f3-feature increases until $\Delta t \sim 0.5$ ns before falling off and vanishing after about 1 ns (see Figure 13.17a and also Figures 6.2 and 6.3 in Ref. [Sokolowski-Tinten, 2023]). However, the size of the f3-feature exhibits a small but continuous decrease in momentum space (see Figure 13.17b). In view of the spatial periods and the temporal dynamics of the f3-feature it was associated to *capillary waves* excited through the ablation at the surface of a previously molten silicon surface [Sokolowski-Tinten, 2023]. These time-resolved ps-optical pump / fs-XUV probe scattering experiments provided the first unambiguous proof of Sipe's first principles theory as origin of LIPSS (the f1-feature) under true single-pulse laser irradiation conditions. They opened-up some new perspectives for probing the dynamics of laser-matter interaction at extreme temporal and spatial scales [Bonse & Sokolowski-Tinten, 2024].

Conventional light detectors can only measure intensity variations of the incoming light wave. Phase information that encode for example optical delays or directions are lost during the data acquisition process. This loss of the phase information is known as 'the phase problem' [Hendrickson, 2023]. It was first noted in the field of crystallography, where the real-space crystal structure can be recovered if the phase of the diffraction pattern can be determined in reciprocal space. In turn, if the phase of the diffraction pattern in reciprocal space is not known, the real-space diffracting structure usually cannot be retrieved. Thus, it not possible to directly reconstruct the surface topography causing the different features (f1 – f3) just from the recorded coherent scattering patterns presented in Figure 13.17). However, with some additional efforts, an image of the diffracting object may be obtained, see Sect. 6.4 on *Coherent Diffraction Imaging* (CDI).

In a very similar experimental ultrafast pump-probe coherent scattering scheme and with probe photon energies in the X-ray range Sun et al. explored the dynamics of nanoscale phase decomposition of thin gold films (100 nm and 250 nm thickness) upon fs-laser ablation [$\lambda_{pump} = 800$ nm, $\tau_{pump} = 50$ fs, $2w_{0,pump} \sim 28$ µm] through fs-time-resolved *Small-Angle X-ray Scattering* (SAXS) combined with simultaneously recorded *Wide-Angle X-ray Scattering* (WAXS) [$\lambda_{probe} = 0.13$ nm, $\tau_{probe} = 30$ to 50 fs, $2w_{0,probe} \sim 1.2$ µm] [Sun, 2025]. While SAXS enabled probing spatial meso-scales, WAXS allowed to analyze the Bragg diffraction peaks reflecting the material structure on atomic scales. The measurements revealed that the gold first undergoes a phase transition to the liquid state, where the loss of long-range order manifested through the decay of the Bragg peaks, before a dramatic increase of scattered intensities in intricate SAXS patterns occurred. This joint WAXS/SAXS experimental approach was complemented with atomistic (*Molecular Dynamics*, MD) modeling, allowing to draw a comprehensive picture of the early energy deposition and subsequent redistribution, the kinetics of involved structural phase transformations, as well as the thermodynamic driving forces behind the spatially heterogeneous and highly non-equilibrium ablation process.

So far, scattering experiments in a rather simple normal-incidence transmission geometry have been discussed here. This requires samples that are transmissive for the probe pulse at the corresponding wavelength. Therefore, thin films must be used, with a thickness adjusted to the absorption length. In transmission geometry, the recorded scattering patterns then probe and "average" the transient scenario along the probe beam path across the entire film thickness, i.e. depth-resolved information are not directly accessible. For overcoming this limitation, *Grazing-Incidence Small-Angle X-ray Scattering* (GISAXS) can be performed in reflection geometry. Recently, Randolph et al. have demonstrated this approach for studying fs-laser-induced plasma formation and ablation of Ta/$Cu_3N$ thin film multilayers [Randolph, 2022]. On costs of the higher complexity of such experiments (e.g. for keeping the spatial overlap between the pump- and the probe-beams upon translation the sample, or when considering the high sensitivity to beam pointing instabilities, etc.) ultrafast GISAXS is straightforward applicable to study LIPSS-formation, particularly for sub-wavelength-sized HSFL featuring spatial periods of the order of 100 nm only [Sokolowski-Tinten, 2024 / Nakatsutsumi, 2025]. Another striking capability of the grazing incidence geometry is the possibility to tune the information depth of the probing beam through a variation of its angle of incidence for obtaining depth-resolved information.

# 3 Microscopic Imaging

Following the well-established concept of "seeing is believing", the most powerful approach of probing the laser-matter interaction is based on direct imaging methods. Scientific imaging can be realized by serial spot probing at different locations ($x$, $y$, $z$) and then constructing a 2D- or 3D-image from a measured dataset signal, $S_{2D/3D}(x, y, z)$. Such imaging approaches are referred to as *Scanning Probe Microscopy* (SPM). *Atomic Force Microscopy* (AFM) and *Confocal Laser Scanning Microscopy* (CLSM) are prominent examples for measuring the 3D topography $S_{3D}(x, y, z) = z(x, y)$ of a surface, while *Scanning Electron Microscopy* (SEM) is widely used to provide 2D surface morphologies $S_{2D}(x, y, z_{focus,SEM})$. However, such serial scanning (SPM) approaches are time consuming and typically assume that the object (sample) and the measured signal do not change with time. However, the latter assumption is often not fulfilled, extending the parameter space of interest by one additional dimension, i.e., the time ($t$).

Thus, for transient signals as typically involved in laser processing, a simultaneous data acquisition is preferred as a parallel method, where the entire object is captured at a selected moment $t_0$ and recorded at once as a full 3D-dataset $S_{3D}(x, y, z, t_0)$. For simplification, in most cases then an optical *projection* of the 3D object into a 2D plane is performed, where either a photosensitive film or an optoelectronic detector (consisting of a 2D-array of photosensitive pixel elements) is placed. This projection then represents a two-dimensional *"image"* $S_{2D}(x, y, t_0)$ of the three-dimensional object, referred to as a *"frame"* that is taken at the time stamp $t_0$. If the projected object exhibits circular symmetry along a rotation axis being perpendicular to the line of sight, a reconstruction of the object may be performed from its 2D projection via integral transforms, such as the *Abel inversion* [Smith, 1988].

Photography or optical microscopy (OM) [Davidson, 2002 / Murphy, 2013] are among the most widely used projection techniques that endow these direct imaging capabilities – provided that the signal levels and the detection technology are well controlled and suitably adapted to the object. "*Movies*" can be combined then from a sequence of image frames that were recorded at different time stamps $t_0$, $t_1$, $t_2$, etc. The specific optical microscopy methods of choice depend on the subject and field of technology and may be significantly different or optimized for specific environments, e.g. in materials science or in biotechnology.

In the following, we will present a survey of different techniques developed in ultrafast optical microscopy and photography that were successfully applied as pump-probe variants for visualizing the rapid dynamics of various transient processes that are manifesting during laser processing.

## Different Schemes of Ultrafast Microscopic Imaging

**Table 13.1** compiles a list of different ultrafast microscopic imaging schemes that were realized in pump-probe approaches for studying transient laser-matter interaction processes. It is indicated whether the specific imaging modes were applied for analyzing specific physical phenomena at the surface (S) or in the volume (V) of the laser processed materials (see also Chap. 1, Nolte et al.). Experimentally realized time ranges are included along with some links to exemplifying figures following in this section.

| **Ultrafast Microscopy Mode (pump-probe)** | **Surface (S) / Volume (V)** | **Phenomena** | **Time range** | **Examples** |
|---|---|---|---|---|
| Brightfield (BF) Microscopy (Reflectivity / Transmission) | BF-R: S<br><br>BF-T: S or V | Kerr effect, Absorption, Melting, Ablation, Newton fringes, Breakdown | 0.1 ps – 10 ns | Fig.13.18a to Fig. 13.23<br><br>Fig. 13.23, Fig. 13.24 |
| Darkfield (DF) Microscopy | DF: S | Scattering, Particles | 0.3 ns – 408 ns | Fig. 13.18b, Fig. 13.25 |
| Shadowgraphy | S | Schlieren, Optical phase changes, Pressure waves | 0.2 ns – 23 ns | Fig. 13.26 to Fig. 13.28 |
| Interference Microscopy (IFM) | IF-R: S<br><br>IF-T: V | Optical phase changes, Topography<br><br>Laser-induced breakdown in air | 0.1 ps – 5 ns<br><br>20 ps – 1.2 ns | Fig. 13.29, Fig. 13.30<br><br>Fig. 13.31, Fig. 13.32 |
| Phase-Contrast Microscopy (PCM) | V | Refractive index changes, Pressure waves, Heat flows | 0.5 ps – 100 µs | Fig. 13.33 to Fig. 13.35 |
| Differential Interference Contrast (DIC) Microscopy (Nomarski Microscopy) | V | Optical phase variations, Pressure waves, Crack formation | 30 ps – 66 ns | – |
| Polarization Microscopy (PM) | V | Stress, Pressure waves, Melting | 0.3 ps – 70 ns | Fig. 13.36 to Fig. 13.37 |
| Imaging Reflectometry/ Ellipsometry | S | Optical constants | 0.1 ps – 3 ns | Fig. 13.38 to Fig. 13.39 |

**Table. 13.1:** Ultrafast pump-probe microscopy approaches used for probing various physical phenomena during laser processing at the surface (S) or in the volume (V)

Note that there are many more advanced variants of pump-probe microscopy used in optical spectroscopy, biomedical imaging, material science, etc., for example relying on other fundamental atomic, molecular, or solid-state phenomena, such as two-photon absorption/fluorescence, excited state absorption, ground state depletion, stimulated emission, or harmonic generation. For details on such aspects being not in the main focus of laser processing the reader is referred to the review articles [Fischer, 2016 / Dong, 2017 / Gross, 2023] here.

## 3.1 Brightfield and Darkfield Microscopy

The most frequently used microscopy modes are *brightfield* (BF) microscopy and *darkfield* (DF) microscopy. In classical BF microscopy, usually the light reflected from the surface of interest is imaged via a microscope objective (MO) and a tube lens (TL) either to the eye or to a camera (BF-R mode). For transparent samples, an alternative back-illumination enables imaging with light transmitted through the sample (BF-T mode). A sample with perfectly smooth interfaces (and sufficient material transparency) generates a bright field of view, thus ruling the name “brightfield microscopy” for both, the BF-R and the BF-T modes. In DF microscopy the illumination path is adapted in a way that the directly reflected (or directly transmitted) light is not entering the imaging beam path. Therefore, the image background appears dark in the field of view. However, light of the illumination source being redirected, e.g., through omnidirectional optical scattering by the object, is captured then and can be imaged to the eye/camera. Thus, light scattering objects appear bright on a dark background in DF microscopy – again for both, reflection and transmission illumination.

Figure 13.18 presents corresponding general pump-probe microscopy schemes in BF and DF mode, respectively [Bonse LSA, 2017]. A strong pump pulse (red lines) excites a solid sample surface, which is then illuminated by a weak, delayed probe pulse (blue lines). In brightfield imaging (a, BF-R) the probe pulse is focused to the back-focal plane of the microscope objective, thus illuminating its entire field of view at the sample surface. The sample surface is imaged using the specular reflection of the probe pulse being transmitted through a beam splitter towards the tube lens and camera. Darkfield imaging (b, DF) relies on optical scattering at the sample’s surface topography and by ablated particles (black dots). If illuminated through the probe pulse, scattered waves are emitted (blue circles) and the exposed features become visible through the imaging system.

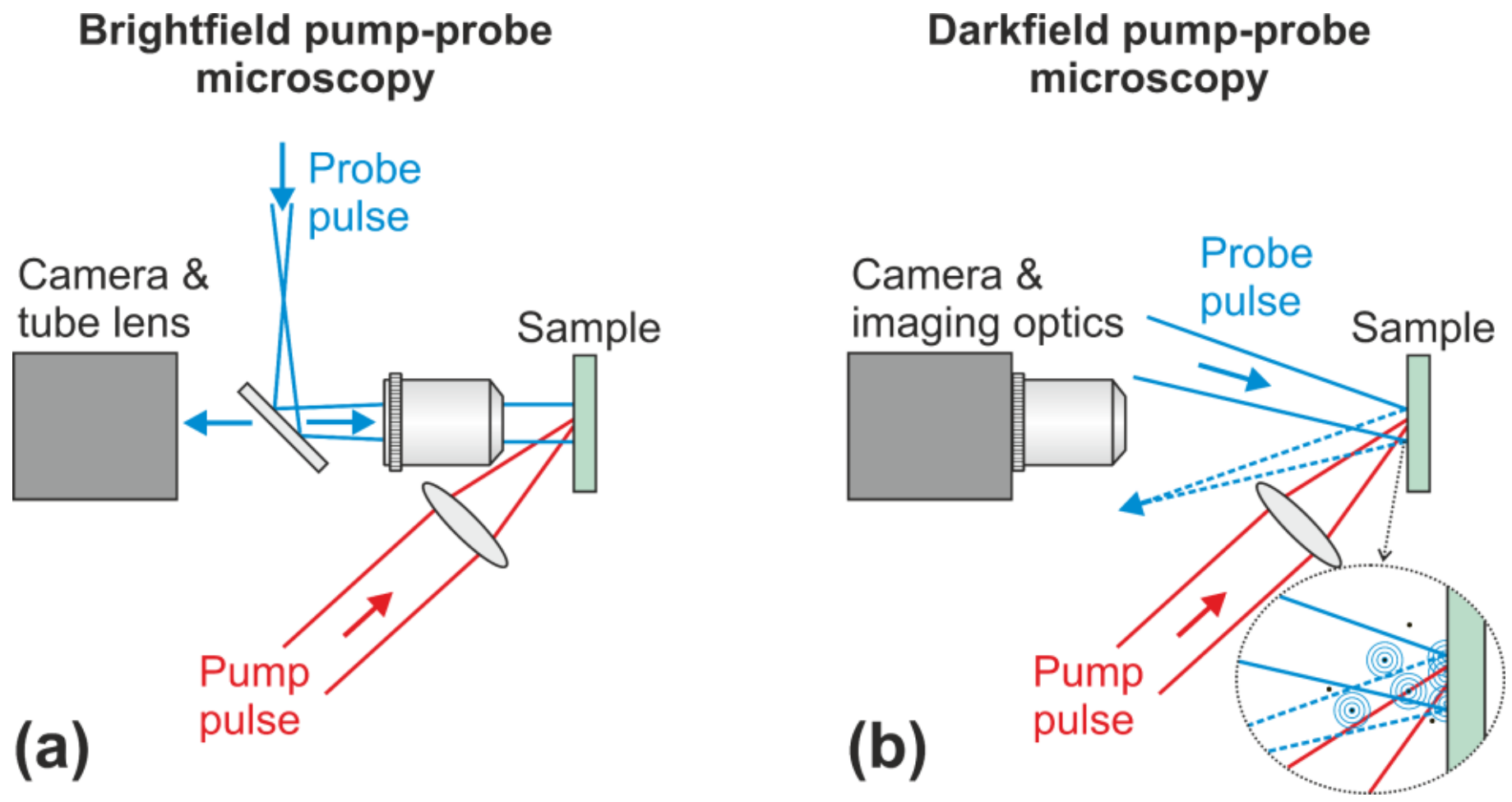

**Fig. 13.18:** Schemes of two variants of time-resolved pump-probe microscopy for studying ultrafast laser-induced processes [Bonse LSA, 2017]. (**a**) Brightfield imaging. (**b**) Darkfield imaging. (Reprinted from [Bonse LSA, 2017], J. Bonse, Scattering on scattering, Light: Sci. Appl. **6**, e17088 (2017), Copyright 2017 under Creative Commons BY 4.0 license. Retrieved from https://doi.org/10.1038/lsa.2017.88)

## Brightfield Reflection Mode

The first fs-pump-probe imaging of ultrafast laser-induced processes at the surface of silicon was realized by Downer et al. and published in 1985 [Downer, 1985]. The authors provided the first images of *nonthermal melting* (NTM) of semiconductors manifesting on sub-ps time scales shorter than the *electron-phonon coupling time* ($\tau_{e\text{-}ph}$) upon strong excitation at laser fluences several times the (thermal) melting threshold (see also Chap. 1, Nolte et al.). About ten years later, the imaging scheme was greatly improved by Sokolowski-Tinten and von der Linde, leading to a series of pioneering publications on the mechanisms of fs-laser ablation of solids [von der Linde, 1997 / Sokolowski-Tinten, 1998 / von der Linde, 2000]. Apart from fs-pump-probe microscopy of the surface reflectivity in BF-R mode, the authors also developed a polarization-sensitive brightfield imaging setup that allows to monitor the temporal evolution of optical anisotropy (birefringence) of the irradiated material [Ashitkov, 2002]. They applied the method to probe the loss of structural order in femtosecond laser-excited mono-crystalline tellurium upon ultrafast melting on the 0.5 to 3 ps timescale.

Figure 13.19 presents a scheme of a non-collinear fs-pump-probe optical brightfield microscopy (BF-R) setup that is capable of visualizing phenomena, such as thermal and non-thermal melting, ablation, and rapid solidification phenomena for solid samples upon irradiation with single ultrashort laser pulses in a temporal range between hundreds of fs up to several ns [Bonse, 2006].

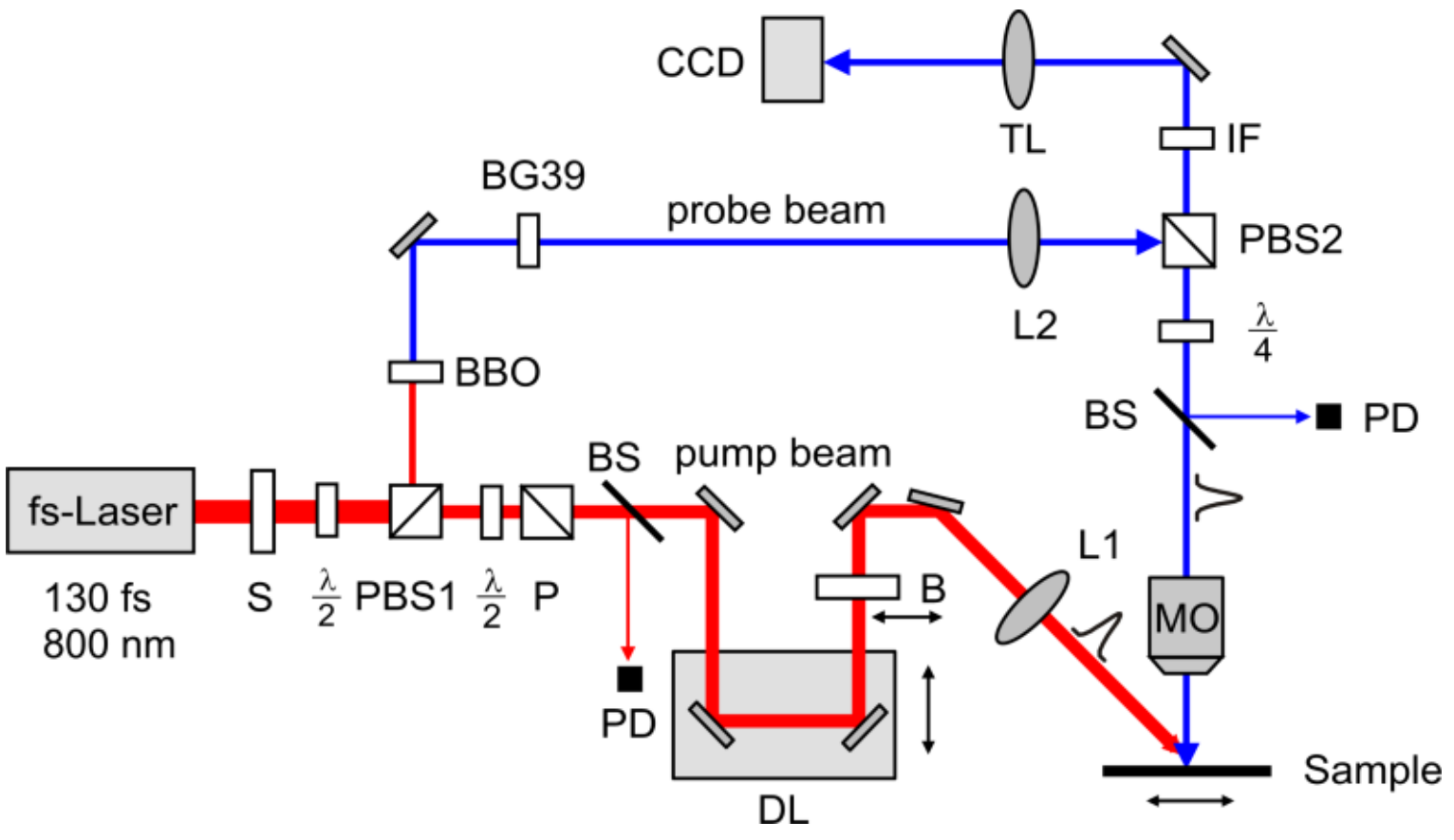

**Fig. 13.19:** Scheme of a non-collinear fs-time resolved microscopy setup in brightfield reflectivity (BF-R) mode [Bonse, 2006]. Abbreviations: B – beam stop; BBO – barium borate nonlinear crystal; BG39 – optical bandpass filter; BS – beam splitter; CCD – camera; DL – delay line; IF – interference filter; λ/2 – half-wave plate; λ/4 – quarter-wave plate; L – lens; MO – microscope objective; P – Glan laser polarizer; PBS – polarizing beam splitter; PD – photodiode; S – electromechanical shutter; TL – tube lens. (Reprinted figure with permission from [Bonse, 2006], J. Bonse et al., Phys. Rev. B **74**, 134106, 2006. Copyright (2006) by the American Physical Society)

The fs-time-resolved microscopy experiment was set up in a non-collinear geometry using different wavelengths for the pump ($\lambda_{pump}$ = 800 nm) and probe ($\lambda_{probe}$ = 400 nm) pulses. An electromechanical shutter (S) allowed to select a single laser pulse from a continuous pulse-train at 100 Hz pulse repetition frequency that was provided by a Ti:sapphire laser amplifier system. The selected laser pulse of 130 fs in duration was divided into pump and probe pulses using a combination of a half-wave plate (λ/2) and a polarizing beam splitter (PBS1) for adjusting the energy ratio between both pulses. Another combination of a half-wave plate and a Glan laser polarizer (P) was used in the pump beam path for variable energy attenuation. After passing through an optical delay line (DL), the pump pulse was focused by a lens L1 at an angle of incidence of $\theta_{pump}$ = 54° onto the sample surface. The pump beam was s-polarized with respect to the plane of incidence. At the sample surface, the elliptical focused laser beam had a Gaussian $1/e^2$-spot diameter of 100 µm and 60 µm along its long and short axis, respectively.

The low intensity probe beam previously split off from the pump beam was frequency doubled by means of a barium borate nonlinear crystal (BBO). Residual 800 nm radiation was removed by an infrared-cutting bandpass filter (BG39). The probe beam was then focused by another lens (L2) to the back-focal plane of a long-working distance microscope objective (MO, magnification 20×, numerical aperture NA = 0.4) after reflection in a polarizing beam splitter (PBS2) cube. A quarter-wave plate (λ/4) was placed in the beam path to convert its linear polarization state into a circular one. The MO finally recollimated this probe beam that then illuminated the entire sample area excited by the 800 nm pump beam.

After being reflected at the sample surface the probe beam radiation was then reconverted to a linear polarization state of orthogonal orientation by passing a second time through the same quarte-wave plate. Thus, now capable to pass through PBS2, the probe beam was used to image the laser-excited surface region by means of the MO and an additional tube lens (TL, $f$ = 200 mm) directly onto a CCD-camera that was synchronized with the laser system. For separating the 400 nm imaging radiation from scattered light of the 800 nm pump beam or from light originating from optical plasma emissions during the ablation process, a narrow band pass interference filter (IF) was additionally placed in the imaging path.

The pump-probe reflectivity measurements were performed with a pump beam fluence that was sufficient to induce phase transitions such as melting or ablation, whereas the fluence of the probe beam was kept well below the melting threshold.

At each sample location, two measurements were used to determine the laser-induced reflectivity change at the chosen delay time. The first measurement (with the pump pulse blocked by the beam stop (B) was used to acquire a reflectivity image $R_0(x, y)$ of the sample surface at the selected measurement position before the exposure to the pump pulse. The second measurement recorded the transient surface reflectivity image $R(x, y, \Delta t)$ upon pump pulse excitation at the selected delay time $\Delta t$. The normalized reflectivity change was then been calculated by subtracting both images and by normalizing with the reflectivity of the undisturbed surface, i.e. $\Delta R/R(x, y, \Delta t) = \{R(x, y, \Delta t) - R_0(x, y)\}/R_0(x, y)$. This image normalization procedure allows one to consider the spatial profile of the illuminating probe beam as well as variations in the surface reflectivity due to surface imperfections, contaminants, etc.

Figure 13.20 compiles a sequence of pump-probe brightfield microscopy images of the surface reflectivity during the irradiation of single-crystalline germanium (c-Ge) by a single fs-laser pulse (800 nm, 130 fs) at a peak fluence in the ablative regime ($\phi_0$ = 3.1 J/cm$^2$) for seven different delay times ranging between 400 fs (**a**) up to 10 ns (**f**), i.e. covering five orders of magnitude after the impact of the pump pulse. Additionally, a micrography of the final surface morphology is shown (**g**) for comparison [Bonse, 2006]. The small dark feature marked by “A” in some frames arises from an imaging artifact caused by an optical reflection in the setup in conjunction with the image normalization procedure described above. It is not related to any transient physical process at the sample surface.

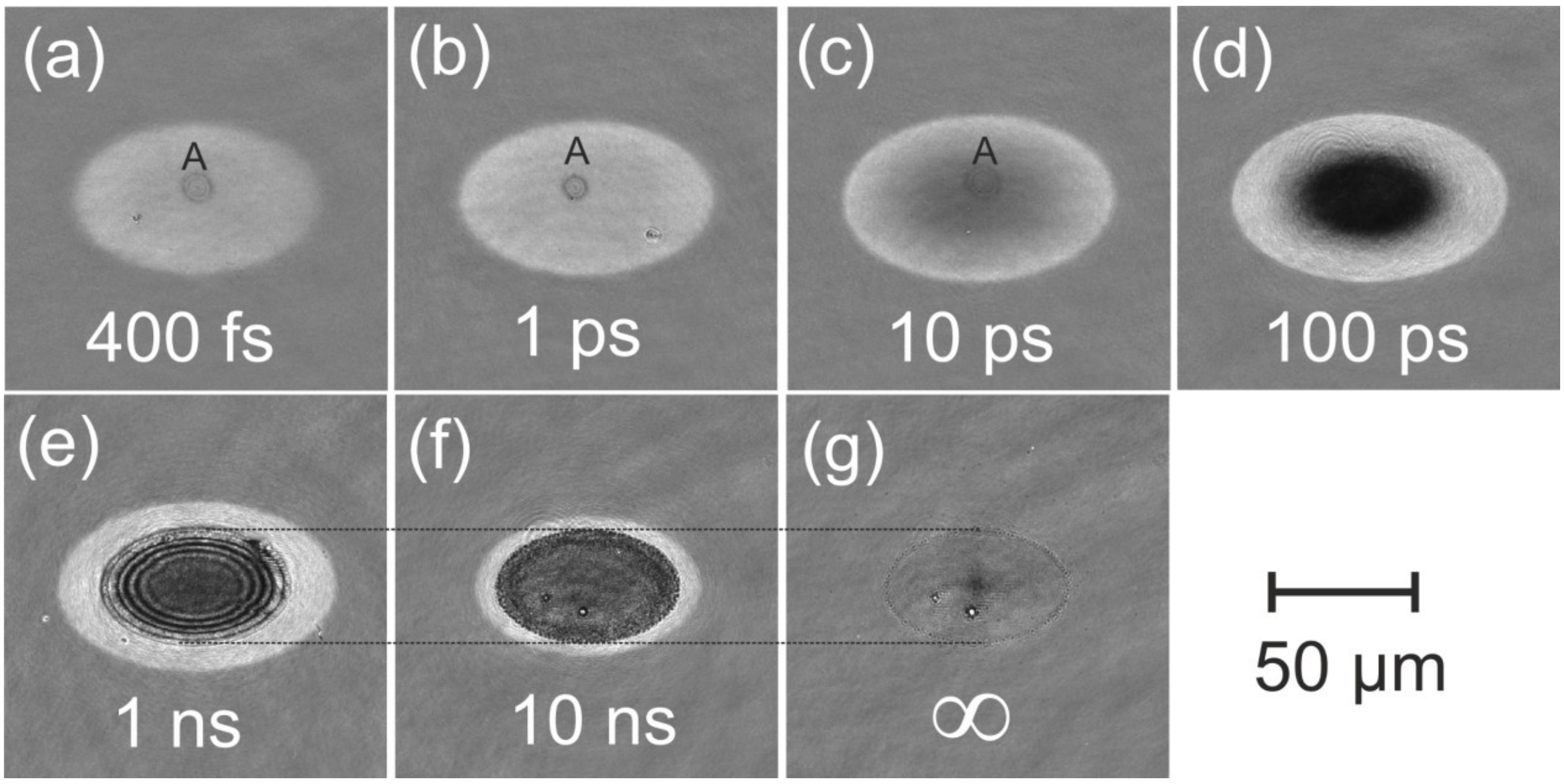

**Fig. 13.20:** Pump-probe fs-time-resolved top-view brightfield microscopy images of the surface reflectivity (BF-R mode, $\lambda_{probe}$ = 400 nm) of a germanium wafer surface at different delay times $\Delta t$ after the exposure to a single pump pulse ($\phi_{0,pump}$ = 3.1 J/cm$^2$) in the ablative regime ($\lambda_{pump}$ = 800 nm, $\tau_{pump}$ = 130 fs, $\theta_{pump}$ = 54°) [Bonse, 2006]. The image sequence is encoded in a linear gray scale with an optimized contrast. “A” denotes an imaging artifact which is not related to a modification of the sample surface. (Reprinted figure with permission from [Bonse, 2006], J. Bonse et al., Phys. Rev. B **74**, 134106, 2006. Copyright (2006) by the American Physical Society)

Immediately after the arrival of the intense pump pulse to the surface an increase of the surface reflectivity in the center of the irradiated spot can be seen (Figure 13.20a, $\Delta t$ = 400 fs). This arises from to the strong electronic excitation resulting in the formation of an electron-hole plasma in the conduction band (CB) of the solid, which can destabilize the lattice structure on a sub-ps time scale when a critical electron density in the CB is exceeded. It indicates an ultrafast phase transition that is referred to as *nonthermal melting* and is typically observed in semiconductors (see Chap. 1, Nolte et al.). After 1 ps, the reflectivity of the entire laser-irradiated region has increased to almost the same reflectivity value, forming a bright elliptical area with sharp defined edges (Figure 13.20b, $\Delta t$ = 1 ps). It can be associated with the formation of an optically thick layer of liquid germanium (ℓ-Ge) on the surface that behaves metallic. On a timescale of tens of picoseconds, the reflectivity of almost the entire spot is lowered in its center, which may arise from the optical properties of a superheated liquid (Figure 13.20c, $\Delta t$ = 10 ps). After some tens to hundreds of ps an even darker zone of strongly decreased reflectivity can be seen in the central region of the irradiated spot (Figure 13.20d, $\Delta t$ = 100 ps). It is indicative of the onset of ablation [Sokolowski-Tinten, 1998 / von der Linde, 2000]. From this central dark feature, a characteristic pattern of rings develops at later delay times (Figure 13.20e, $\Delta t$ = 1 ns) that can be associated with transient Newton interference fringes [Sokolowski-Tinten, 1998 / von der Linde, 2000]. For increasing delays that characteristic rings are moving from the center towards the outer regions of the spot, while increasing their number and decreasing their radial spacings. After $\Delta t$ = 10 ns delay, the characteristic fringes are not observed anymore in the area previously covered by them (Figure 13.20f, $\Delta t$ = 10 ns) that coincides with the area of an ablation crater finally formed at the surface (Figure 13.20g, $\Delta t$ = "∞"). The reduced reflectivity indicates that ablation still persists after 10 ns. This area of ablation is surrounded by an annulus of high reflectivity, indicating that the material is still molten there. However, the molten region exhibits a reduced lateral extent when compared to earlier delay times, resulting from the re-solidification, that has started from the outer and deeper lying regions of the melt pool and becomes visible at the surface after some ns.

It must be noted that the transient Newton fringes were observed predominantly upon ultrashort pulsed laser irradiation of metals and semiconductors at moderate laser fluences in the ablative regime, while initially they were not seen on dielectrics [von der Linde, 1996 / Siegel, 2007 / Puerto, 2008 / Puerto, 2010], see the sub-section on brightfield transmission microscopy below. However, later theses ablative Newton rings were observed also during the irradiation of different amorphous or crystalline dielectric materials (SF57 glass, lithium niobate, sapphire) in rather narrow fluences close to the ablation threshold [Garcia-Lechuga APL, 2014 / Garcia-Lechuga JAP, 2014]. They could be even detected for the irradiation at laser fluences below the ablation threshold, visualizing the propagation of a melting front via a Fabry-Perot interference effect on longer timescales in the ns- to µs-range [Hernandez-Rueda, 2014 / Garcia-Lechuga APL, 2016].

Another example for the impressive capabilities of fs-time resolved microscopy in BF-R mode is provided in Figure 13.21. The scenario is depicted in the left part of the figure. It is related to the clarification of the processes occurring during *laser cleaning* with fs-laser pulses, where a characteristic optical near-field related surface damage was observed under the surface of silica spheres that were removed by a single fs-laser pulse from the surface of silicon wafers [Mosbacher, 2001]. In Figure 13.21 the exposure of a single silica sphere with a diameter of 7.9 µm on a silicon wafer by a single fs-laser-pulse ($\lambda_{pump}$ = 800 nm, $\tau_{pump}$ = 120 fs, $\theta_{pump}$ = 54°) is studied [Kühler, 2013].

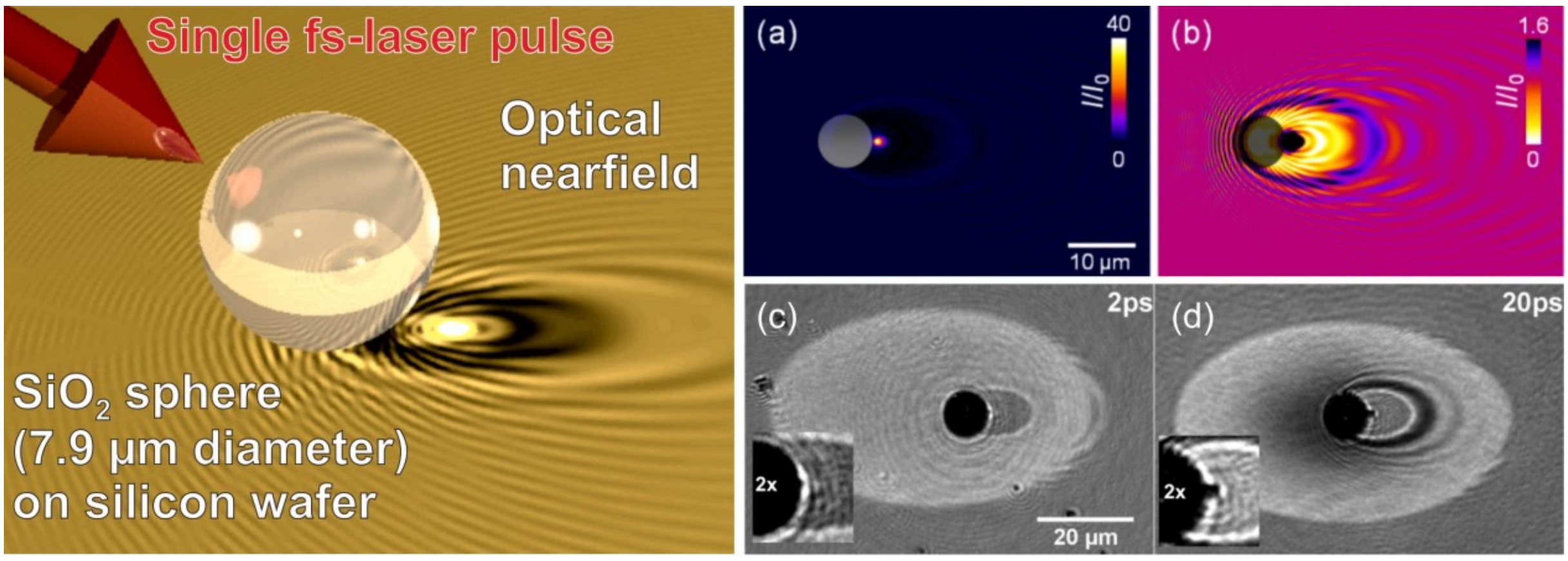

**Fig. 13.21:** Left: Scheme visualizing the optical near-field underneath of a 7.9 µm diameter silica sphere on a silicon wafer surface upon optical irradiation. [Kühler, 2013]. Right: (**a**) Calculated near field distribution at the surface of a Si wafer substrate for a $SiO_2$ dielectric particle of 7.9 µm diameter homogenously illuminated at a wavelength of 800 nm and at an angle of 54°. The spatial intensity distribution has been normalized to the incident beam intensity ($I_0$). The position of the particle is shown as a semitransparent disc. (**b**) Same data as in (a) but re-scaling the color scale to a maximum enhancement factor ($I/I_0$) of 1.6. (**c, d**) Pump-probe fs-time-resolved top-view brightfield microscopy images of the surface reflectivity (BF-R mode, $\lambda_{probe}$ = 400 nm) of a silicon wafer surface at two different delays $\Delta t$ = 2 ps (c) and $\Delta t$ = 20 ps (d) after the exposure to a single pump pulse ($\phi_{0,pump}$ = 0.87 J/cm$^2$) in the ablative regime ($\lambda_{pump}$ = 800 nm, $\tau_{pump}$ = 120 fs, $\theta_{pump}$ = 54°). Insets with a 2× zoom of the relevant near-field part are provided in the lower left corner. (Reprinted from [Kühler, 2013], P. Kühler et al., Femtosecond-resolved ablation dynamics of Si in the near field of a small dielectric particle, Beilstein J. Nanotechnol. **4**, 501 – 509 (2013), Copyright 2013 under Creative Commons BY 2.0 license. Retrieved from https://doi.org/10.3762/bjnano.4.59)

Figures 13.21a shows the result of a numerical calculation of the optical near-field intensity distribution, normalized with the homogeneous incident intensity $I_0$. A sub-micrometric bright spot indicating a near-field enhancement of a factor of $I/I_0 \sim 40$ can be seen close to the sphere at its right-hand side. The position of the particle is overlayed with a circular disk. Figure 13.21b presents the same calculated data with the intensity color scale limited at $I/I_0$ = 1.6. The image indicates the presence of a fine structure in the intensity pattern with characteristic interference maxima and minima. The observed fringe period in the beam propagation direction (forward) is larger than in the backscattered direction, corresponding to values of $\lambda_{pump}/\{1 \pm \sin(\theta_{pump})\}$, i.e. ≈ 4.0 µm and ≈ 0.4 µm, respectively, as expected for the interference of a spherical wave with a scattering source [Kühler, 2012].

Figures 13.21c and 13.21d show corresponding snapshots of the surface reflectivity ($\lambda_{probe}$ = 400 nm, $\theta_{probe}$ = 0°) recorded by fs-time-resolved microscopy 2 ps and 20 ps after the impact of the pump laser pulse. The laser peak fluence incident to the sample was set at $\phi_0$ = 0.87 J/cm$^2$ in the ablative regime, here [Kühler, 2013]. At $\Delta t$ = 2 ps (Figure 13.21c) a large molten spot

with an increases surface reflectivity can be seen around the particle (appearing dark). The tiny spot excited by the largest optical near-fields nearby the particle appear very dark, indicating strong locally confined ablation already at this very early delay time. The effect is even stronger after a delay of $\Delta t = 20$ ps (Figure 13.21d). Apart from this region of maximum optical field-enhancement, the effect of the more complex near-field pattern is also visible as a characteristic fringe pattern. It exhibits bright regions (where destructive interference has inhibited the onset of ablation) and dark regions (where constructive interference has driven the silicon material into the ablation regime) – consistently resembling the calculated optical near-field distribution presented in Figure 13.21b. Finally, the removal of the micrometric silica particle from the silicon surface requires more time and occurs after more than 20 ns (data not shown here) [Kühler, 2013].

Taking benefit of the directional scattering of surface electromagnetic waves, Garcia-Lechuga et al. employed their fs-pump-probe brightfield microscopy setup to study *in-situ* the birth and growth of laser-induced periodic surface structures (LIPSS). Specifically, they studied the branch of type-s LSFL-I with supra-wavelength spatial periods formed upon laser beam scanning in the sub-ablative regime of fs-laser-induced melting with subsequent amorphization into a very regular stripe pattern of ~3.5 µm spatial period [Gracia-Lechuga ACS, 2016]. With a similar pump-probe microscopy approach, very recently Chen et al. were able to visualize in BF-R mode the dynamics of the material-response originating from the electromagnetic waves scattered from the rim of fs-laser-induced ablation craters on a thin gold film [Chen, 2024]. In 2022, Xu et al. increased the lateral resolution by building a widefield *single-probe structured light microscopy* (SPSLM) setup for studying the temporal dynamics of sub-wavelength LSFL-I upon multi-pulse spot processing on silicon in the ablative regime [Xu PR-A, 2022]. Using $\lambda_{probe} = 400$ nm as probe wavelength, the authors demonstrated a spatial lateral resolution of ~480 nm, along with a spatial axial and a temporal resolution of ~22 nm and ~256 fs, respectively.

Conducting a pump-probe microscopy experiment in a non-collinear configuration may have additional implications on the temporal material excitation dynamics recorded for delay times $\Delta t$ in the sub-ps regime: if the loosely focused pump beam is arriving at non-normal incidence to the surface, its phase front is hitting the surface first at one of the beam sides, followed by an "in-plane lateral sweeping" across the surface, until the phase front has reached the surface also at the opposite side of the beam. Hence, along this sweeping direction a pulse front delay time $\Delta t'$ may be associated with each location at the surface. When the surface is imaged in such a scenario at normal incidence, this implies that within the field of view the sample excitation starts earlier at one of the sides of the laser-excited elliptical area than on the opposite side (along the major axis). Thus, the laser excited region may be seen first via a local reflectivity change only partly in an asymmetrical way at small $\Delta t$. For an example of the manifestation of this effect during the fs-laser excitation of silicon when studied by pump-probe optical microscopy in BF-R mode, the reader is referred here to Figure 6 in [von der Linde, 1997]. On the other hand, observing such a transient asymmetry in the surface reflectivity pattern across the laser excited region can be even used to estimate or confirm the minimum temporal resolution of a pump-probe microscopy setup just on basis of a few geometrical parameters, such as the angle of incidence and the major axis diameter of the laser-excited surface spot.

Sokolowski-Tinten et al. recognized the scientific potential of this "optical streaking" of the pump beam, i.e., the assignment of a spatial coordinate on the sample plane to a local delay time and developed it into a modification of time-resolved brightfield microscopy, that allows time- and fluence-resolved BF-R measurements at fs-laser excited surfaces in a true single-pulse configuration [Sokolowski-Tinten, 1999]. Figure 13.22a presents a scheme of the experimental setup. In its right part, an expanded view of the laser-excited surface area is provided. The pump pulse strikes the sample surface at an oblique angle of incidence (here 45°). The pump beam is focused by a cylindrical lens for generating an elongated line focus in the plane of incidence at the sample. Therefore, the intensity distribution at the surface is associated with an elliptical Gaussian shape having a large eccentricity (marked as dark grey shaded region in the right part of Figure 13.22a). In a good approximation, the local fluence in the central part of this pump beam distribution can be considered as nearly constant in the major-axis direction of the ellipse (parallel to the plane of incidence, here denoted as *x*-direction), but of Gaussian shape in the direction perpendicular to it (along the minor axis of the ellipse, here denoted as *y*-direction). The time-delayed probe beam (marked as light grey disk) illuminates the central part of the pump beam excited surface area at normal incidence. The duration of the probe pulse determines the temporal resolution then. A high-resolution microscope objective is used for imaging the laser excited surface area onto a CCD camera. A single recorded image of the excited surface contains information on the time- and the fluence-dependence of the surface reflectivity. The recorded spatial reflectivity distribution $R(x, y)$ directly represents $R(\Delta t'(x), \phi(y))$.

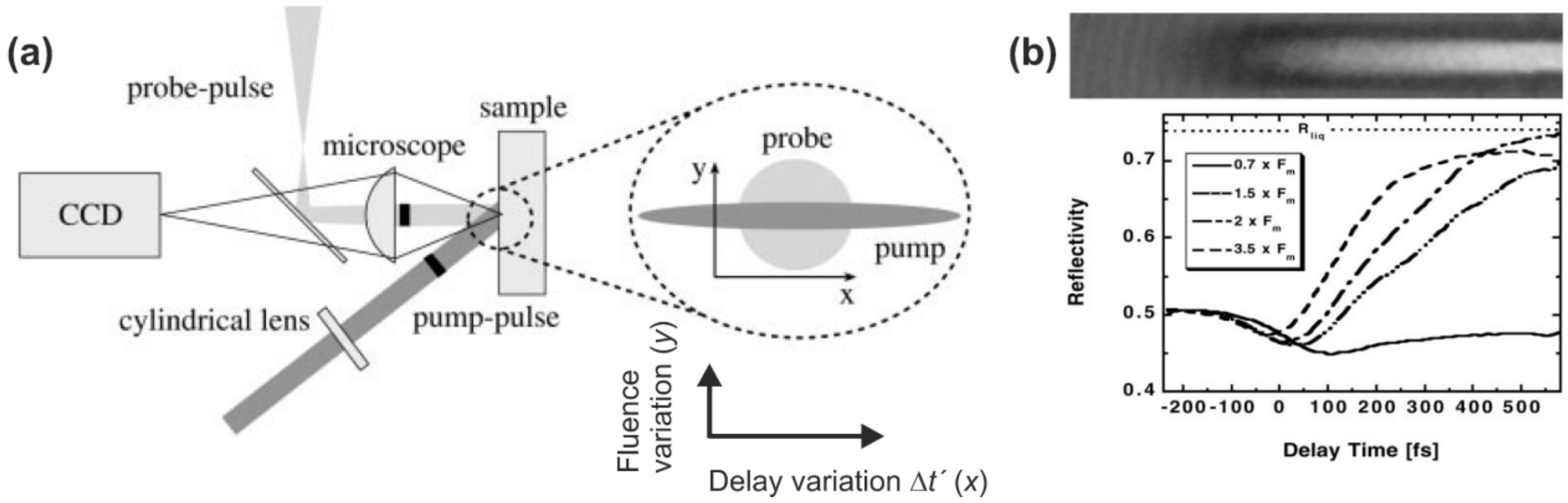

**Fig. 13.22:** (**a**) Scheme of a single-pulse non-collinear fs-time resolved microscopy setup in brightfield reflectivity (BF-R) mode. The circled area provides an expanded view of the laser-excited surface. Dark and light grey shades indicate the surface areas covered by the pump and the probe spots, respectively [Sokolowski-Tinten, 1999]. (**b**) Time evolution of the reflectivity of germanium for several pump fluences, as obtained from horizontal cross sections of the picture shown in top panel [$\lambda_{pump} = \lambda_{probe}$ = 620 nm, $\tau_{pump} = \tau_{probe}$ = 120 fs, $\theta_{pump}$ = 45°, $\theta_{probe}$ = 0°]. (Reprinted from [Sokolowski-Tinten, 1999], K. Sokolowski-Tinten et al., Single-pulse time- and fluence-resolved optical measurements at femtosecond excited surfaces, Appl. Phys. A **69**, 577 – 579, 1999, Springer Nature)

Figure 13.22b provides experimental results obtained for single-pulse fs-laser excitation ($\lambda_{pump}$ = 620 nm, $\tau_{pump}$ = 120 fs, $\theta_{pump}$ = 45°) of crystalline germanium. The delay time dependencies have been extracted from horizontal cross sections of the micrograph (*x*-direction) provided in the top-panel for different local fluences (*y*-positions). At pump fluences below the melting threshold the time evolution of surface reflectivity is determined by the generation and relaxation of the laser-induced free-electron plasma (solid curve). At higher pump fluences ultrafast melting is observed within a few hundreds of femtoseconds (dashed and dotted curves) [Sokolowski-Tinten, 1999]. As expected, the behavior is qualitatively similar to that previously shown in Figure 13.20 but was obtained here just from single laser shot.

## Brightfield Transmission Mode

With only minor modifications the fs-pump-probe time-resolved brightfield microscope previously presented in Figure 13.19 can be straightforwardly adapted to a brightfield transmission (BF-T mode) microscope for acquiring images of transiently absorbing objects. Figure 13.23 exemplifies such a system that was built to study the surface processing of transparent sample materials [Puerto, 2008].

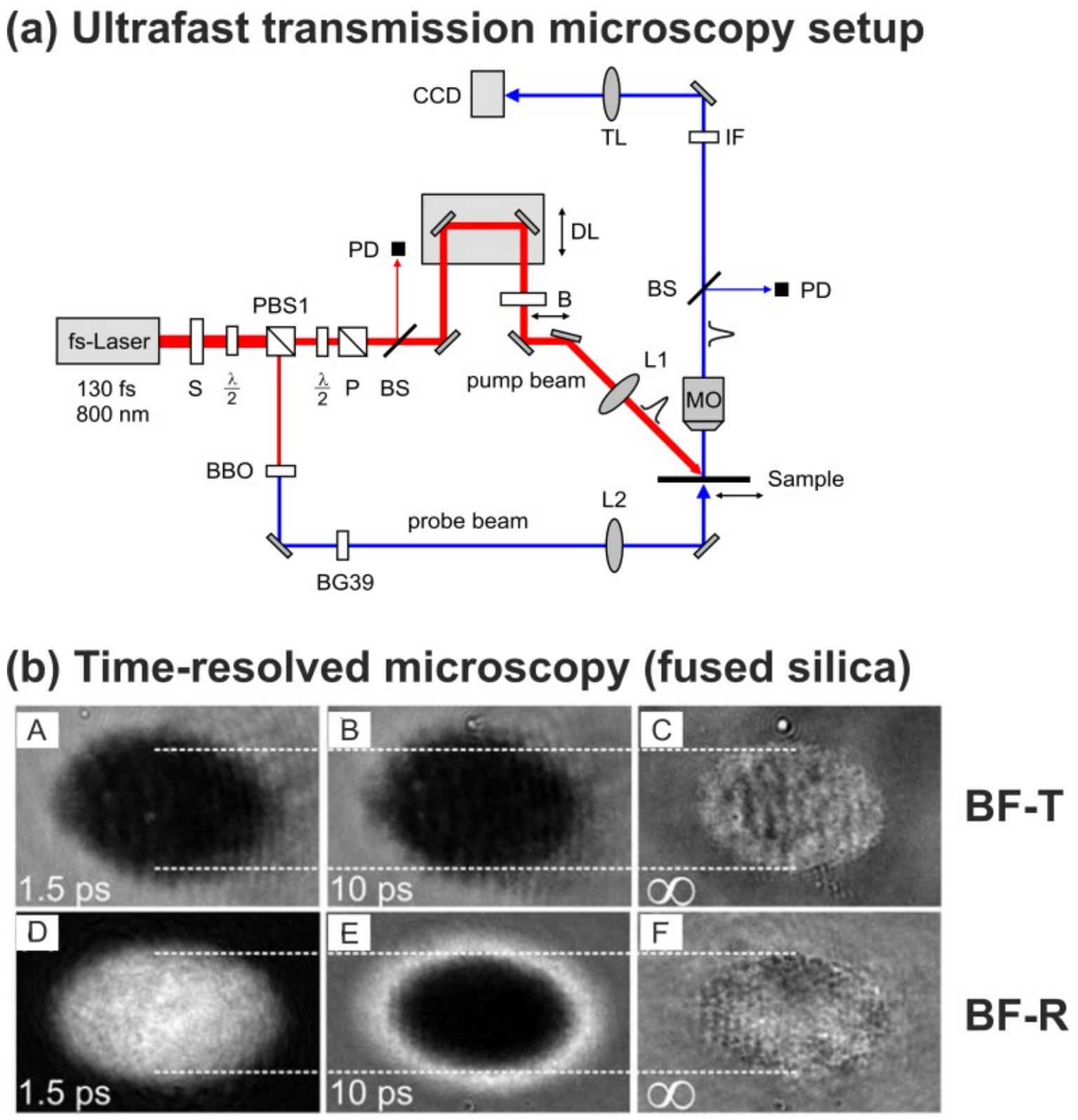

**Fig. 13.23: (a)** Scheme of a non-collinear pump-probe fs-time-resolved microscopy setup for measuring either the optical transmission through transparent samples (BF-T mode, $\lambda_{pump} = 800$ nm, $\lambda_{probe} = 400$ nm) [Bonse, 2006]. For the abbreviations see the caption of Figure 13.19. **(b)** Pump-probe fs-time-resolved top-view brightfield microscopy images ($\lambda_{probe} = 400$ nm) of the sample transmission (BF-T mode, top row) and the corresponding surface reflectivity (BF-R mode, bottom row) of a fused silica sample at different delays $\Delta t$ after the exposure to a single pump pulse ($\phi_{0,pump} = 11.9$ J/cm$^2$) in the ablative regime ($\tau_{pump} = 120$ fs, $\theta_{pump} = 54°$) [Puerto, 2008]. The dashed horizontal lines mark the diameter of the ablation crater in the vertical direction. The intensity is encoded in a linear grey scale with an optimized contrast in each case. (Reprinted (adapted) from [Puerto, 2008], D. Puerto et al., Transient reflectivity and transmission changes during plasma formation and ablation in fused silica induced by femtosecond laser pulses, Appl. Phys. A **92**, 803 – 808, 2008, Springer Nature)

Figures 13.23b.A – 13.23b.C show BF-T images of the sample transmission recorded at three different delays after irradiation with a single pump pulse with peak fluence value of $\phi_{0,\text{pump}}$ = 11.9 J/cm$^2$, i.e., well above the ablation threshold. For comparison, corresponding BF-R surface reflectivity images (at the same delays and pump fluence) are provided in Figs. 13.23b.D to 13.23b.F [Puerto, 2008]. The images at $\Delta t$ = 1.5 ps delay (A and D) indicate a strong decrease in transmission. It is accompanied by a concomitant reflectivity increase within the irradiated region, which can be attributed to the formation of a dense free-electron plasma at the surface. At $\Delta t$ = 10 ps delay, the reflectivity image (E) shows a central dark region, which is indicative of surface ablation, while in transmission (B), the dark region shows only minor changes. Figures 13.23b.C and 13.23b.F, taken several seconds after the pump pulse ($\Delta t = \infty$), show the final ablation craters at the surface. It can be seen that the region of a transient free-electron plasma clearly extends beyond the visible surface ablation crater.

Wang et al. implemented a BF-T pump-probe setup that is capable of simultaneously imaging in a cross-sectional geometry both, the crater formation at the surface and the sub-surface nonlinear self-focusing effects upon fs-laser processing of transparent materials, in the same field-of-view [Wang, 2019]. It was used for direct observation of surface-structure-assisted filament splitting upon ultrafast multiple-pulse laser ablation of fused silica, during the early stage of free-electron plasma evolution (i.e. for pump-probe delays $\Delta t$ < 300 fs, results not shown here). The experimental results revealed the influence of the pre-pulses-induced surface morphology on the surface and bulk energy distributions induced by subsequent pulses.

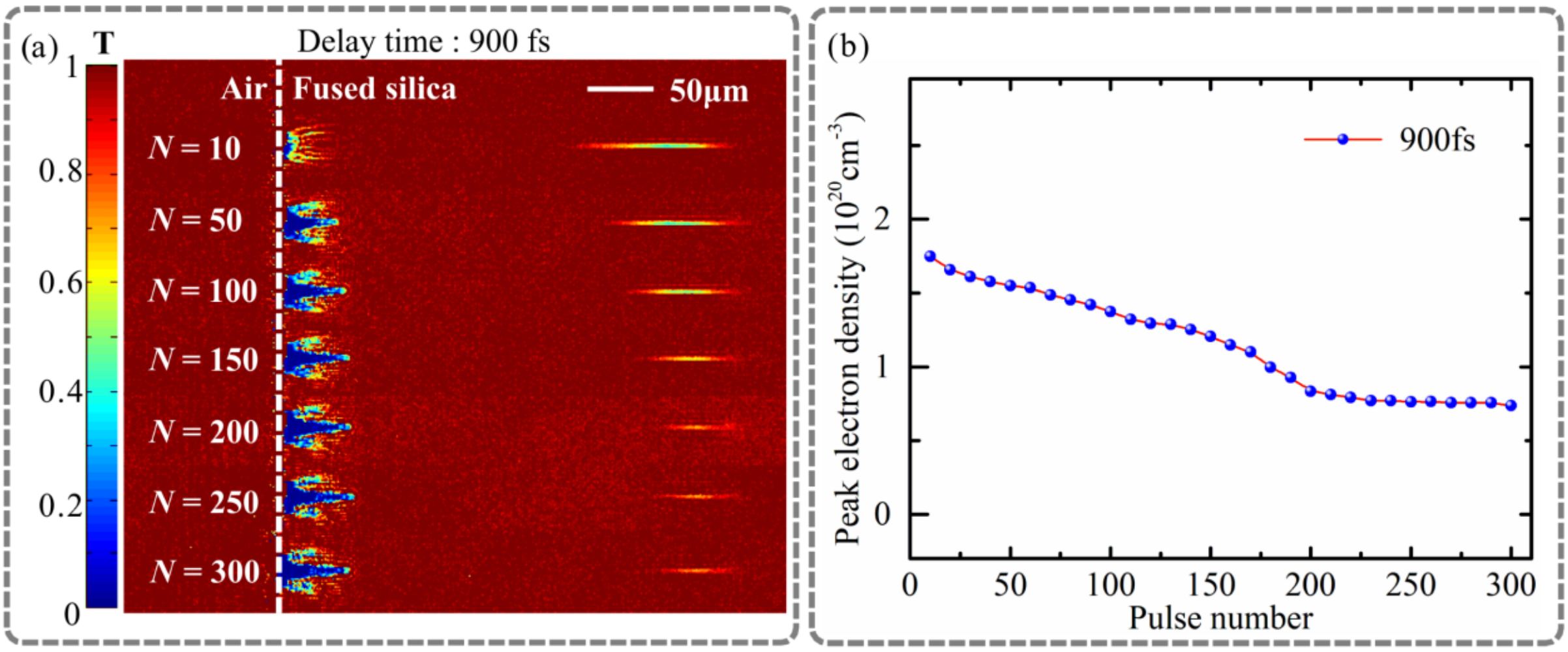

**Fig. 13.24: (a)** Side-view pump-probe fs-time-resolved microscopy of the optical transmission (BF-T mode, $\lambda_{\text{probe}}$ = 400 nm, $\tau_{\text{probe}}$ = 35 fs, $\theta_{\text{probe}}$ = 90°) through a fs-laser irradiated ($\phi_{\text{pump}}$ = 17.7 J/cm$^2$, $\lambda_{\text{pump}}$ = 800 nm, $\tau_{\text{pump}}$ = 35 fs, $\theta_{\text{pump}}$ = 0°) fused silica sample upon application to a different number of pre-irradiating pump pulses ($N$ = 10 to 300, increasing from the top to the bottom panel) prior to the single-pulse pump-probe event, all recorded at a fixed delay time $\Delta t$ = 900 fs [Wang, 2019]. **(b)** Corresponding peak electron densities in the center of the fs-laser induced sub-surface filament. The laser pulse (pump) is incident from the left. (Reprinted with permission from [Wang, 2019] © Optical Society of America)

An example for such a BF-T pump-probe microscopy ($\lambda_{\text{pump}}$ = 800 nm, $\tau_{\text{pump}}$ = 35 fs, $\theta_{\text{pump}}$ = 0°, $\lambda_{\text{probe}}$ = 400 nm, $\tau_{\text{probe}}$ = 35 fs, $\theta_{\text{probe}}$ = 90°) obtained at a fixed probe pulse delay of $\Delta t$ = 900 fs for an increasing number of pre-conditioning pump laser pulses per spot ($N$ = 10 to 300) prior to the single pump-probe event is provided in Figure 13.24 [Wang, 2019]. At the selected fluence of 17.7 J/cm$^2$, a crater was drilled at the surface, see the air/fused silica interface in the side-view micrographs presented in Figure 13.24a. For progressing $N$, the depth of the crater increased, while characteristic side branches were formed. Simultaneously, at depths > 200 µm below the sample surface another region of decreased transmission is formed. It originates from a nonlinearly excited free-electron plasma during the self-focusing of the initially divergent fs-laser beam upon propagation in the fused silica material.

Figure 13.24b presents the corresponding peak electron density in the center of the self-focused filament as a function of the number of pulses $N$ [Wang, 2019]. After an initial decrease for a low number of laser pulses, the electron density saturates at $\approx 0.75\times10^{20}$ cm$^{-3}$ for $N$ exceeding 200. Similarly, the crater shape does not change anymore significantly for $N > 200$. Thus, it was concluded that the formation and evolution of the near-surface crater morphology indirectly regulates and limits the sub-surface filament electron density [Wang, 2019]. Note that these observations were made for the moderate pump beam focusing conditions employing a 5× microscope objective with a numerical aperture of NA = 0.15. However, the balance of the surface and sub-surface effects strongly depends on the focusing condition and material and will vanish for strong focusing conditions using high NA microscope objectives.

## Darkfield Mode

A darkfield pump-probe microscope according to the principle previously outlined in Figure 13.18b was realized by Fang and co-workers [Fang, 2017]. The detailed experimental setup is presented in Figure 13.25a. A Ti:sapphire fs-laser ($\lambda_{pump}$ = 800 nm, $\tau_{pump}$ = 65 fs) provided the pump beam, which was focused to irradiate a sample at an angle of incidence of $\theta_{pump}$ = 36°. Single pump laser pulses at fluences ranging between 0.1 and 1 J/cm$^2$ were used for irradiation. The delayed probe beam was frequency doubled in a BBO crystal ($\lambda_{probe}$ = 400 nm) and focused under an angle of $\theta_{probe}$ = 18° to illuminate the pumped laser spot. Long delay times were enabled via multiple probe beam reflections in the optical delay line. A CCD camera was synchronized with the fs-laser and captured the scattered probe beam radiation from a direction normal to the surface. Optical filters were used to suppress parasitic pump and probe beam radiation.

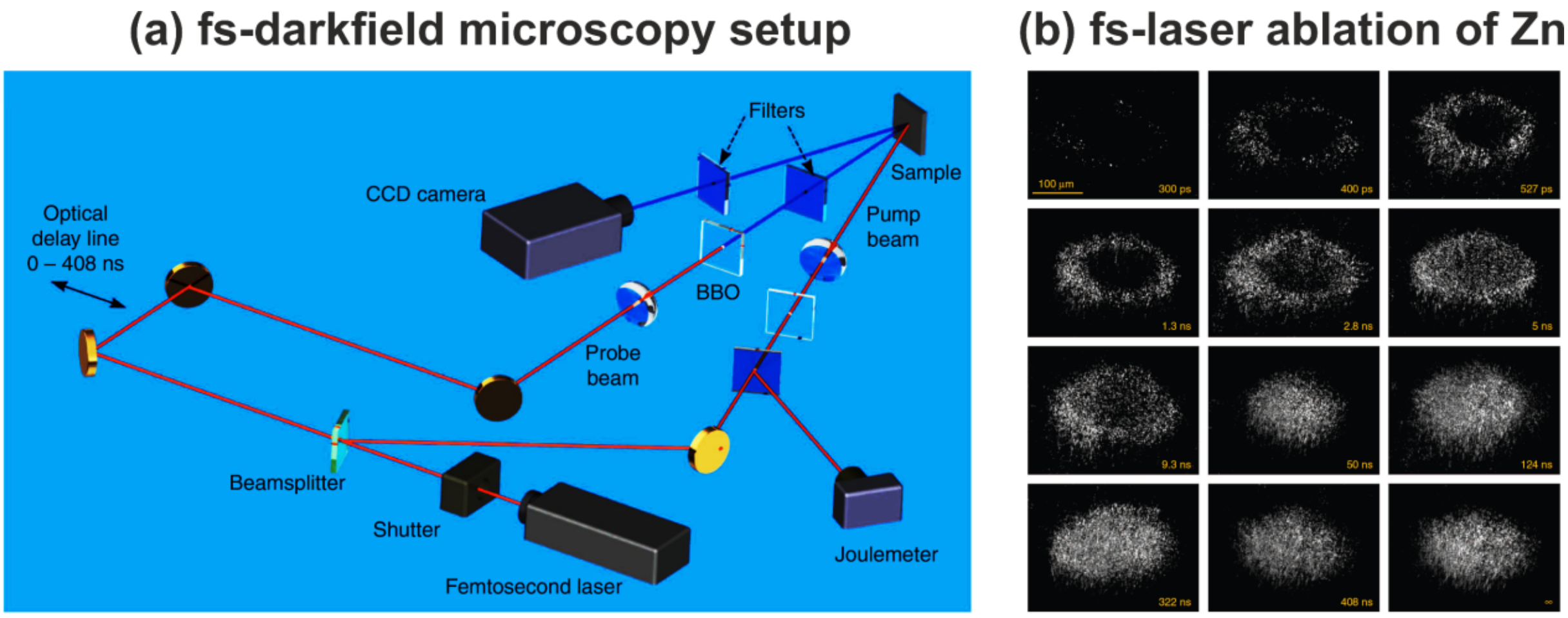

**Fig. 13.25:** **(a)** Scheme of a non-collinear fs-time resolved microscopy setup in darkfield (DF) mode. BBO – barium borate nonlinear crystal; CCD – charge-coupled device [Fang, 2017]. **(b)** Pump-probe fs-time-resolved DF microscopy images ($\lambda_{probe}$ = 400 nm) of a zinc surface at different delays $\Delta t$ after the exposure to a single pump pulse ($\phi_0$ = 1.0 J/cm$^2$) in the ablative regime ($\lambda_{pump}$ = 800 nm, $\tau_{pump}$ = 65 fs, $\theta_{pump}$ = 36°). (Reprinted from [Fang, 2017], R. Fang et al., Direct visualization of the complete evolution of femtosecond laser-induced surface structural dynamics of metals, Light: Sci. Appl. **6**, e16256, 2017, Springer Nature)

Figure 13.25b provides results of such a pump-probe experiment performed on a polished zinc sample at a pump beam fluence of 1.0 J/cm$^2$ in a delay range $\Delta t$ between 0.3 and 408 ns [Fang, 2017]. Scattering becomes visible already after a few hundred ps. In line with the brightfield microscopy results presented above, the optical pattern evolves as a ring for delays up to a few ns, before the rather homogeneous scattering over the entire irradiated spot becomes visible. The ring-shaped signature arises from ablating material (electrons, ions, atoms, etc.), which does not macroscopically scatter the probe beam radiation but that absorbs it partially. This plasma shielding of the probe beam is strongest in the center of the ablating region, where the pump beam fluence is maximal.

## 3.2 Shadowgraphy

*Shadowgraph*y and *Schlieren* techniques are powerful optical methods for visualizing phenomena in transparent media such as fluids or gaseous atmospheres that are manifesting through local changes of the refractive index [Settles, 2001]. In shadowgraphy, the probing radiation is used either in a projection or in an imaging mode for visualizing the transient ablation event on a screen or a camera. On the recorded image, the object’s shadow contrasts with a bright field formed by the light from the illumination system. Thus, conceptually, the back-illumination imaging shadowgraphy is very similar to BF-T imaging approach discussed in Sect. 3.1. Schlieren imaging requires an additional collimation of the illuminating light and its direct suppression through a knife edge placed at the focus of a field lens, resulting in an improved contrast. While Schlieren imaging is sensitive to the gradient of the local refractive index (first order derivative), shadowgraphy is sensitive to its the second order derivative [Settles, 2001]. Both methods are extremely useful to reveal atmosphere-related ablation phenomena and material propagating into the ambient air that are usually present in laser processing [Benattar, 1992 / Breitling, 1999 / Mingareev Diss, 2009 / Breitling Diss, 2010 / Gregorčič, 2011]. However, ultrafast shadowgraphy has also successfully been implemented to visualize bulk laser processing phenomena and beam propagation effects in the volume of transparent solids, e.g. via the shadow created by the cloud of laser-generated conduction band electrons in the vicinity of the focal region of a laser beam focused into a fused silica sample [Sun, 2005].

In contrast to optical emission imaging methods, in shadowgraphy the irradiated or ablating material does not have to actively emit light here. Figure 13.26a presents a typical pump-probe fs-shadowgraphy setup for recording side-view images of the laser surface ablation [Mingareev, 2008].

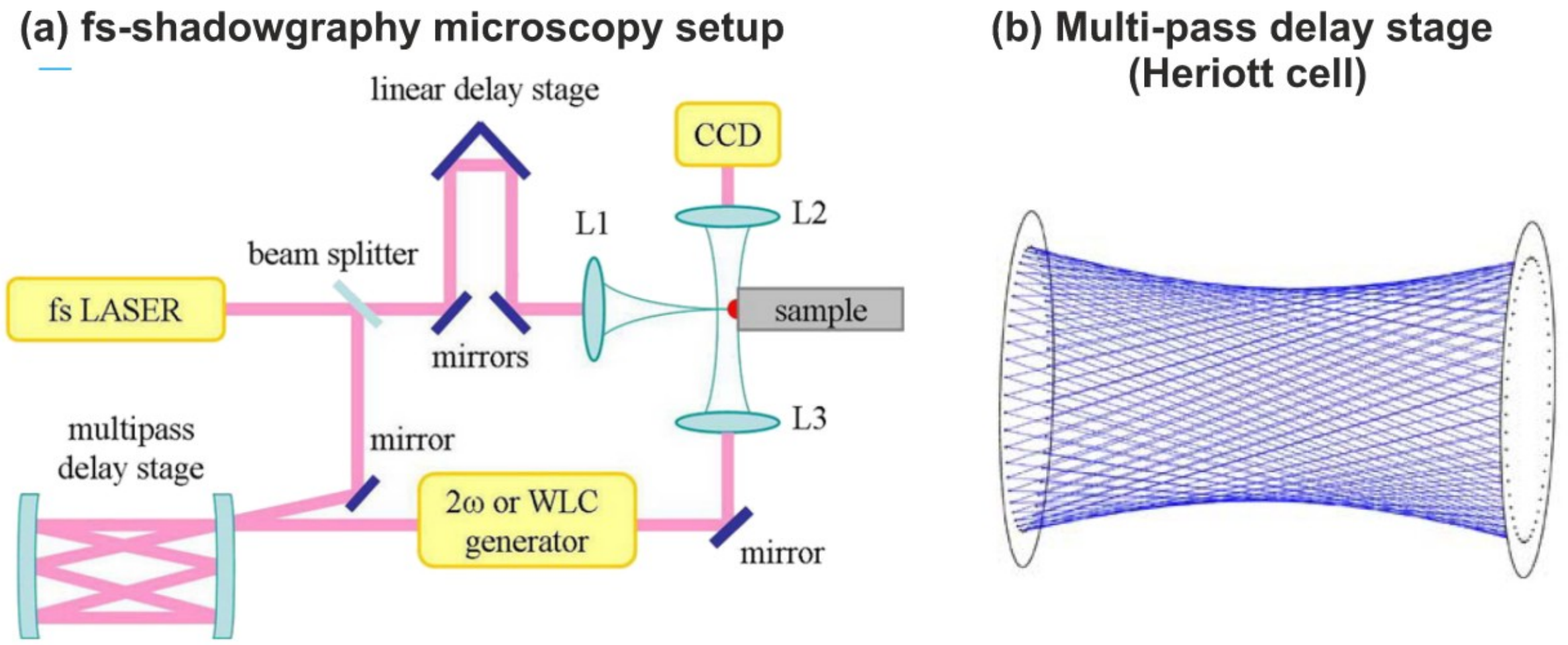

**Fig. 13.26: (a)** Scheme of pump-probe side-view microscopy setup for studying ablation processes at the surface of a sample [Mingareev, 2008]. **(b)** Optical beam pass in a Herriot cell type multi-pass delay stage for enabling delays up to ~2 µs. (Reprinted from [Fang, 2017], I. Mingareev et al., Time-resolved investigations of plasma and melt ejections in metals by pump-probe shadowgraphy, Appl. Phys. A **92**, 917 – 920, 2008, Springer Nature)

After being reflected at the beam splitter, the ultrashort probe pulses are passing through a multi-pass delay stage for generating large pump-probe delay times. A coarse adjustment can be done via selecting a desired number of round trips in the multi-pass delay stage, while the fine adjustment with high temporal accuracy is realized via a motorized linear delay-line placed in the pump beam path. An elegant implementation of such a multi-pass delay stage is a so-called *Herriot cell* [Herriott, 1965], see the beam path sketched in Figure 13.26b. Such a cell consists of a pair of facing spherical concave mirrors that are aligned with the same optical axis for multiple folding of the laser beam inside the cell. Typically, it allows to create probe pulse delay times $\Delta t$ in the ns- to few µs-range. Other variants of multi-pass delay stages with specific advantages and disadvantages are alternatively available [Robert, 2007]. After passing through the multi-pass delay stage, the probe beam may be frequency doubled (2ω) or even converted into a *white light continuum* (WLC), before illuminating the transient ablation plasma, while passing across the pump pulse excited surface region toward the camera (CCD).

An early pump-probe implementation of Schlieren imaging and shadowgraphy was developed in the context of fs-laser generation of ultrashort pulsed X-ray pulses upon irradiation of metals at laser peak intensities in the tera- to petawatt regime. Under these extreme conditions ultrashort pulsed optical pulses may generate a small amount of soft X-ray radiation [Luther-Davies, 1978 / Murnane, 1989 / Kieffer, 1993]. At such intensities the existence of a "weak" optical pre-pulses may already create a near-surface plasma that changes the interaction with the main laser pulses and, thus, affect the X-ray yield. In that context, the role of optical pre-pulses originating from the ns-background of *amplified spontaneous emission* (ASE) accompanying the 80 fs colliding pulse mode locked dye laser main pulses during the irradiation of aluminum at $10^{16}$ W/cm$^2$ was clarified by Benatar et al. in 1992 [Benattar, 1992].

Later, fs-time resolved side-view shadowgraphy was successfully used for studying the evolution of the ablation plasma plume of ultrashort single pulse laser irradiated silicon (native oxide covered single-crystalline wafers) in air environment [Russo, 1999 / Choi, 2002]. Different ablation stages of ablation were identified. Ejection of material was observed after several ps up to tens of ns after the arrival of the fs-laser pump pulse (83 fs, 800 nm, 1.5 J/cm$^2$).

Shock wave propagation indicated that the ultrafast initial ablation plasma becomes visible at delay times around 10 ps and is followed by a slower ''thermal'' contribution in the time scale of a few tens of ns [Choi, 2002].

Later, McDonald et al. explored the effect of the presence of additional thermally grown oxide layers covering the surface of single-crystalline (100)-silicon wafers upon irradiation with single fs-laser pulses at a very similar fluence level (150 fs, 780 nm, 1.3 J/cm$^2$) [McDonald, 2007]. Figure 13.27 compiles the results obtained through side-view shadowgraphy.

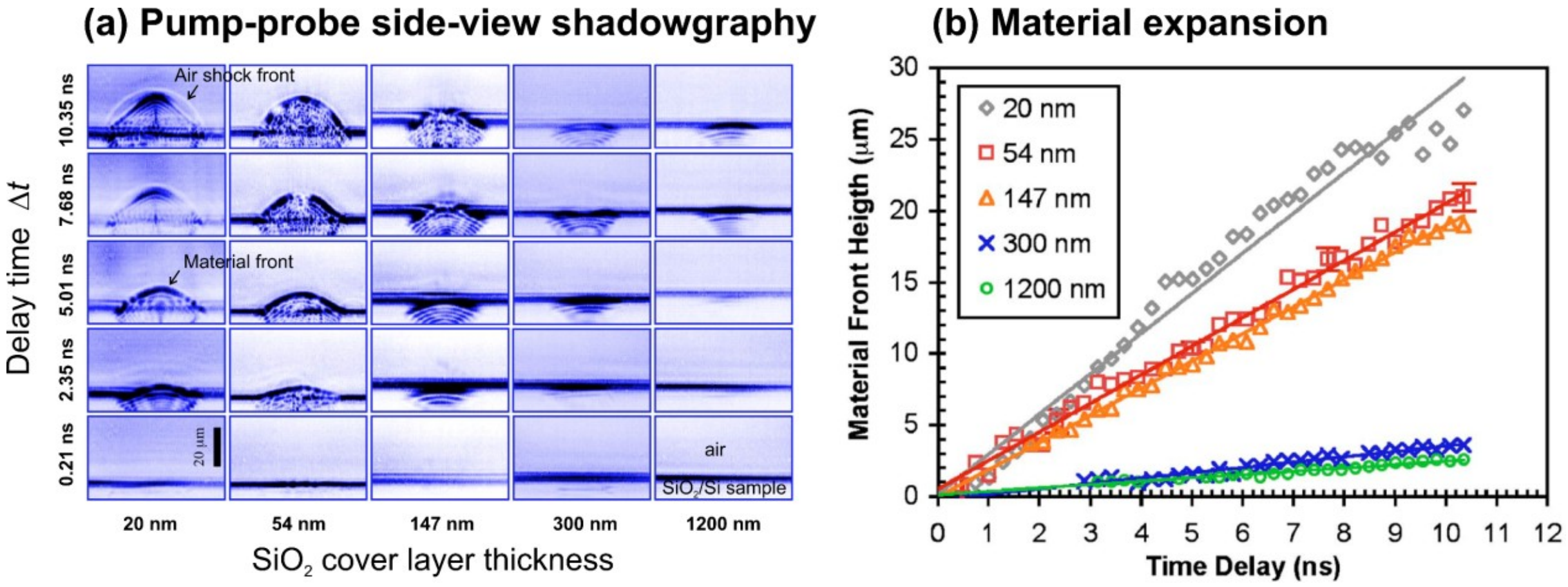

**Fig. 13.27:** Single laser pulse ablation (775 nm, 150 fs, 1.3 J/cm$^2$) of thermally grown silicon oxide ($SiO_2$) layers of various thicknesses (20 nm, 54 nm, 147 nm, 300 nm, 1200 nm; columns) from single-crystalline silicon (100) wafers [McDonald, 2007]. **(a)** Pump-probe side-view shadowgraphic images taken at five different delay times $\Delta t$ = 0.21 ns, 2.35 ns, 5.01 ns, 7.68 ns, 10.35 ns (rows). The laser pulses (pump) are incident from the top. All images have the same scale bar provided in the bottom left image. **(b)** Position (height) of the ablating material front vs the time delay $\Delta t$. The solid lines are linear least-squares-fits. (Figures a, b are reprinted from J.P. McDonald et al., Pump-probe imaging of femtosecond pulsed laser ablation of silicon with thermally grown oxide films, J. Appl. Phys. **102**, 063109 (2007), with the permission of AIP Publishing)

Figure 13.27a presents side-view shadowgraphy images for five pump-probe delays (rows) of 0.21 ns, 2.35 ns, 5.01 ns, 7.68 ns, and 10.35 ns as well as for five different $SiO_2$ cover layer thicknesses (columns) of 20 nm, 54 nm, 147 nm, 300 nm, and 1200 nm, respectively. Dark blue regions represent the sample surface, as well as ablating material. The horizontal fringes around the surface are caused by diffraction of the probe beam radiation at the sample edge. For the thinnest oxide layer thickness (20 nm), after several ns the front of a shock wave formed in air becomes visible as a bright almost semi-circular boundary. It can be clearly distinguished from the material front (dark blue) that lags behind the air shock front. For medium oxide layer thicknesses (54 nm and 147 nm), the ablated material front appears fractured at $\Delta t$ = 7.68 ns. For even larger layer thicknesses (300 nm and 1200 nm), the spatial expansion of the ablating material is strongly suppressed by the presence of the oxide.

From these pump-probe side-view shadowgraphy measurements, for each of the five oxide layer thicknesses, the position of the material front was determined. Figure 13.27b plots this material front height a function of the delay $\Delta t$. For all data sets (oxide layer thicknesses), a linear scaling with the delay time is visible during the early stage of ablation ($\Delta t$ < 11 ns). It allows to quantify the velocity of the ablating material via the slope of the least-squares-fits, which are significantly different among the oxide layer thicknesses. A supersonic value of (3010 ± 360) m/s, exceeding the speed of sound in air by a factor of approx. 10, was observed for the thinnest oxide layer (20 nm), while for the thickest oxide layer (1200 nm) subsonic ablation velocities of (200 ± 20) m/s were obtained [McDonald, 2007].

While shadowgraphic imaging of single-pulse irradiation events is very useful to visualize fundamental processes accompanying the laser ablation, practical laser processing generally relies on the application of many laser pulses. With the availability of large energy high repetition rate laser sources and pulse burst machining modes, shadowgraphic imaging gained again interest for visualizing plasma-shielding effects in a train of laser pulses that may reduce the processing efficiency and precision (see Chap. 1, Nolte et al.).

Kraft et al. implemented projection-based pump-probe shadowgraphy for visualizing high fluence, high repetition rate ultrashort pulse laser ablation ($\lambda_{pump}$ = 1030 nm, $\tau_{pump}$ = 400 fs, $f_{rep}$ = 500 kHz, $\phi_0$ = 8.6 J/cm$^2$, $\theta_{pump}$ = 0°, $\theta_{probe}$ = 90°, $\lambda_{probe}$ = 515 nm, $\tau_{probe}$ = 400 fs) of AISI316 stainless steel [Kraft AO, 2020]. For optically delaying the probe pulses in the range between $-1$ ns $\leq \Delta t \leq +50$ ns, a *Robert cell* type was used here [Robert, 2007], allowing for more flexibility and compactness. It consists of a spherical and a split mirror, whereby the number of internal reflections can be controlled by tilting one part of the split mirror. Additional even larger delays in the microsecond range could be realized electronically. Figure 13.28 assembles a sequence of pump-probe side-view shadowgraphs corresponding to a train of 10 pulses irradiating the same surface spot at 500 kHz pulse repetition rate. At that rate, the temporal separation between consecutive pump pulses accounts to 2 µs. All images are captured at the constant delay time $\Delta t$ = 23 ns after the respective pump pulse.

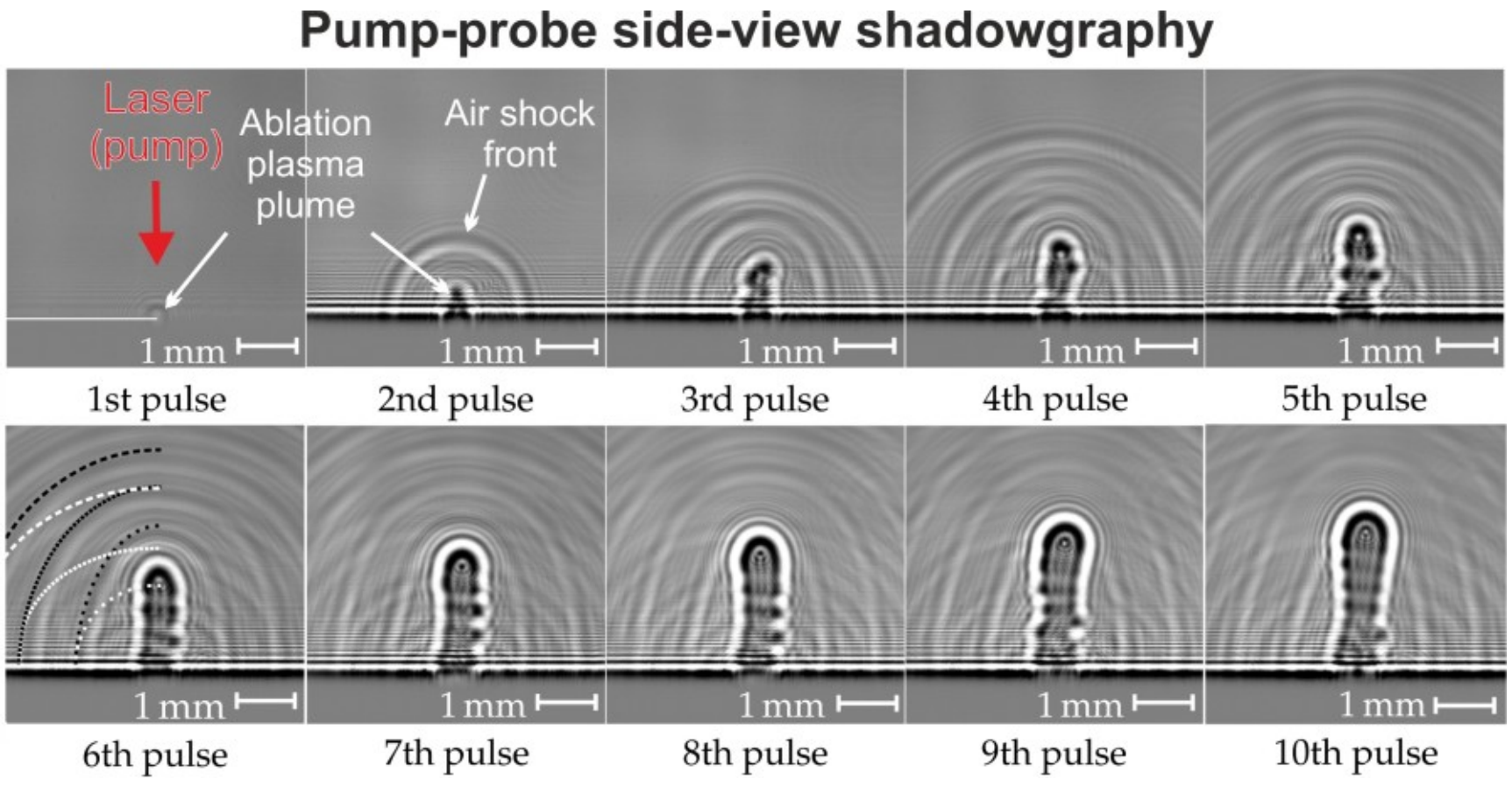

**Fig. 13.28:** Pump-probe side-view shadowgraphs recorded 23 ns after the arrival of the pump pulse ($\lambda_{pump}$ = 1030 nm, $\tau_{pump/probe}$ = 400 fs, $f_{rep}$ = 500 kHz, $\phi_0$ = 8.6 J/cm$^2$, $\lambda_{probe}$ = 515 nm) to the surface [Kraft AO, 2020]. Their corresponding number is labelled below the individual frames. The sample and its surface are visible at the bottom, the laser pulses (pump) are incident from the top. (Reprinted with permission from [Kraft AO, 2020] © Optical Society of America)

After the first laser pulse, the height of the ablation plasma plume is ~0.2 mm. The shadowgraph after the second laser pulse presents a further expanded ablation plasma plume of ~0.6 mm in height (dark region). Additionally, a semi-spherically shock wave expanding up to a height of ~1.1 mm has formed around the irradiated surface spot. Both, the air shock front and the ablation plasma plume are indicated by arrows in this frame. The shock wave was created by the first pump pulse and expanded during the 2 μs inter-pulse separation, associated with a supersonic (average) propagation velocity of ~500 m/s.

With an increasing number of pump pulses, the number shock fronts visible above the surface and their extent increments stepwise. Similarly, the height of the ablation plasma plume steadily increases. The general shape of the ablation plasma plume is like a finger, with some irregularities and stepping included. After the tenth pump pulse, the height of the ablation plasma exceeds 2 mm. In contrast to the height of the ablation plasma plume, its lateral extent increases only until the third pump pulse and then remains almost constant during irradiation with additional pump pulses.

Note that after a few pump pulses, some additional shock waves become visible in the shadowgraphs. For clarity, this is visualized in the left half of the shadowgraph taken after the $6^{th}$ pump pulse, where dashed black and dashed white lines indicate the first-order and second-order shock fronts, respectively. The second-order shock fronts are spherically shaped and less intense than the first-order ones. With an increasing number of pump pulses, the shape of the first-order shock waves rather turned into a parabolic form, while their axial expansion velocity is reduced below the speed of sound. No local deformation of the shock fronts can be identified in the shadowgraphs, indicating that the subsequent pump laser pulse does not affect the compressed air inside the shockwaves themselves.

In another publication, Kraft et al. investigated the laser processing by high repetition rate ultrashort laser pulses ($\lambda$ = 1030 nm, $\tau_{pump}$ = 600 fs, $\phi_0 \leq 3$ J/cm$^2$, $f_{rep}$ up to 48 MHz) swept by a polygon scanner across the surface of steel, copper, or aluminum as sample materials [Kraft SPIE, 2020]. A diode laser system provided the probe pulses of $\lambda_{probe}$ = 688 nm wavelength and with a duration of $\tau_{probe}$ = 13 ns was used for pump-probe side-view shadowgraphic imaging. For accessing large delay times, the pump and the probe beam could be electronically delayed within a time frame of up to 10 μs. This approach allowed to visualize the highly dynamic ablation effects occurring above the processed surface in single- and multi-pulse micro-processing by using an ultrafast moving laser beam.

In single pump pulse experiments on stainless steel, the dimensions (height and width) of the ablation plasma plume were systematically studied, revealing an increase with progressing delay times $\Delta t$ after the pump pulse irradiation, as well as the dynamics of expanding shock waves – both in good agreement with existing models. Multi-pulse experiments were performed with ultrafast laser beam deflection using a polygon scanner providing a scan speed of 250 m/s. At laser pulse repetition frequencies of 0.5 MHz or 1 MHz, corresponding to lateral inter-pulse displacements larger than the focused beam spot diameter of the pump pulses, spatially separated individual ablation plumes were observed at $\Delta t$ = 10 μs delay (data not shown here). In contrast, at higher pulse repetition frequencies of several MHz, the ablation plumes spatially overlap and merge, forming a closed cloud of ablating material. The "critical" pulse repetition frequency for merging crucially depends on the laser peak fluence and material [Kraft SPIE, 2020]. This extended cloud of merged ablated material staying for a long time above the laser processed surface causes an additional serious limitation (plasma-shielding) for the up-scaling of the laser processing rates via the pulse repetition frequency.

## 3.3 Interference Microscopy

Apart from brightness information in the micrograph that is generated by local variations of the intensity of the detected light, additional information is encoded in its optical phase. For example, the phase of the light changes in a predictable manner when it propagates along a defined beam path. Unfortunately, the phase of the light cannot be directly detected but the fundamental idea of converting specific optical path differences via interference effects into amplitude (intensity) changes has enabled fundamentally new types of optical microscopy. They are generally referred to as *interference microscopy* (IFM) but there are also some widely known specific implementations, such as *phase-contrast microscopy* (PCM) or differential interference contrast microscopy (DIC), see their detailed descriptions below. However, commercial standalone IFM devices usually do not provide the capability of performing dedicated time-resolved analyses on short temporal scales.

When equipped with the necessary time-resolution, the idea of a phase-sensitive interferometric analysis has been proven to be a very powerful tool for the quantitative analysis of laser processing. *Spectral interferometry* has been used to investigate the expansion dynamics of fs-laser-induced ablation plasma at the surface of metals [Geindre, 1994], fs-laser driven shocks in metals [Evans,1996], or even laser-induced free-electron plasmas and transient defect formation in the bulk of dielectrics [Petite, 1996 / Mao, 2004]. In brief, the approach involves the illumination of the sample surface by ultrafast pump and probe laser pulses, imaging the reflected probe pulses onto the entrance slit of an imaging spectrograph, and then analyzing the resulting spectral content. Thus, the method provides lateral spatial resolution in one dimension only.

This limitation was overcome through the development of ultrafast interference microscopy by implementing time- and space-resolved IFM in a pump-probe scheme [Gahagan, 2002 / Temnov, 2004 / Temnov Diss, 2004 / Temnov, 2006]. Temnov et al. [Temnov Diss, 2004 / Temnov, 2006] designed a Michelson-interferometry-based microscope with a Linnik imaging configuration [Linnik, 1933]. The apparatus is sketched in Figure 13.29a. It exhibits a 2D sub-micrometer (~1 µm) spatial and a ≈100 fs temporal resolution, combined with a phase-shift resolution better than $\lambda/200$ and is suitable for measuring amplitude changes of ≈1% [Temnov, 2004 / Temnov Diss, 2004 / Temnov, 2006].

A p-polarized Ti:sapphire pump laser pulse ($\lambda_{pump}$ = 800 nm, $\tau_{pump}$ = 100 fs) incident at $\theta_{pump}$ = 45° was focused on the surface with a lens of $f$ = 50 cm focal length. The pump laser excited surface region was illuminated by a weak, time-delayed probe pulse (second harmonic, $\lambda_{probe}$ = 400 nm, $\tau_{probe}$ = 100 fs) normally incident onto the sample surface through a high-resolution objective lens (20×, NA = 0.3, 17 mm working distance). For that, a strongly divergent probe beam was obtained by placing an additional lens ($f$ = 25 mm) in front of the beamsplitter, confocally arranged with the back-focal plane of the objective. This enabled the collimated illumination of a large surface area on the sample. The probe pulse reflected from the surface partly passed the beamsplitter and was imaged to the CCD camera. A second reference arm of the Michelson interferometer was realized behind the other direction of the beamsplitter with an identical objective and a reference mirror placed at the equivalent distance instead of the sample.

The formation of the interferogram is illustrated in Figure 13.29b. It is formed in the image plane by the interference of the "object beam" with the "reference beam", i.e. on the chip of the CCD camera. The object beam is the light of the probe pulse reflected at the sample surface (drawn as red arrows in the figure). The reference beam is a fraction of the incident probe pulse that passed through the reference arm of the Michelson interferometer. The orientation and the spacing of the interference fringes (a pattern of periodic bright and dark stripes) can be adjusted by tilting the reference mirror. The transient information is encoded in the phase front of the object beam. Its deviation from a plane wave leads to a deformation of the periodic fringe pattern that can be evaluated for quantifying the laser-induced phase changes. Each set of measurement consists of three recorded interferograms. (1) an interferogram of an unexcited surface prior to the pump pulse exposure, (2) a transient interferogram for a given pump-probe delay time $\Delta t$, and (3) an interferogram of the final surface state taken a few seconds after the excitation.

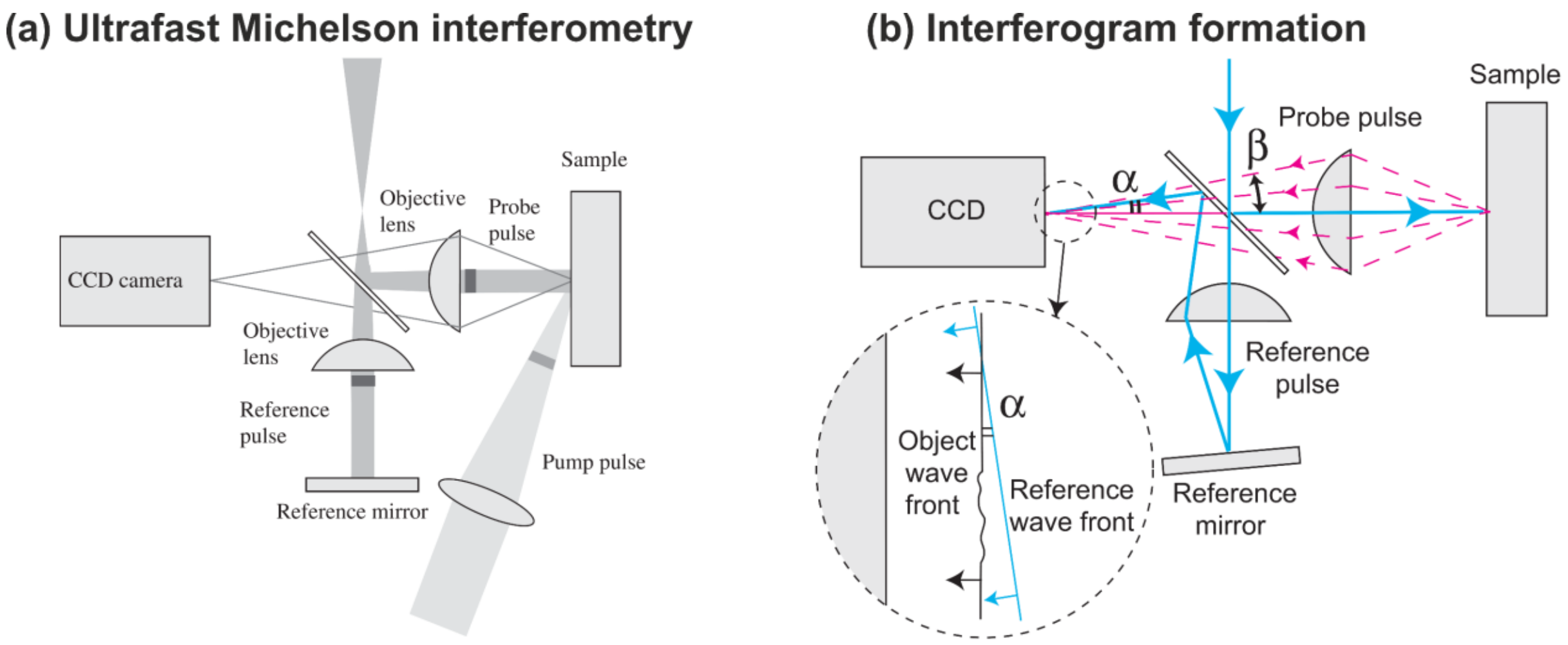

**Fig. 13.29: (a)** Scheme of pump-probe top-view Michelson interference microscopy setup for studying ablation processes at the surface of a sample. (Reprinted from [Temnov, 2004], V.V. Temnov et al., Femtosecond time-resolved interferometric microscopy, Appl. Phys. A **78**, 483 – 489, 2004, Springer Nature) (**b**) Detailed optical paths of the reference beam and object beam forming an interferogram [Temnov Diss, 2004]. (Figure reprinted from [Temnov Diss, 2004], with permission and courtesy of Vasily V. Temnov)

Figure 13.30 exemplifies two interferograms recorded for the single fs-laser pulse irradiation ($\lambda_{pump}$ = 800 nm, $\tau_{pump}$ = 100 fs, $\phi_{0,pump} = 1.4\times\phi_{abl}$ = 0.28 J/cm$^2$, $\theta_{pump}$ = 45°, $\lambda_{probe}$ = 400 nm, $\tau_{probe}$ = 100 fs, $\theta_{probe}$ = 0°) of a polished GaAs(100) wafer surface after a pump-probe delay time of $\Delta t$ = 1.4 ns (a) and of the final (permanent) surface state (d). In the transient interferogram (a) considerable deformations of the interference pattern are visible, indicating the involvement of large phase shifts. Inspection of the final interferogram (d) reveals straight interference stripes with a distinct shift along a sharp elliptical boundary (horizontally oriented), which surrounds the crater formed at the surface upon laser ablation. The analysis and processing described in [Temnov, 2004] allowed to retrieve the phase $\Psi_{ind}(x, y)$ and amplitude $r_{ind}(x, y)$ images of the laser-induced transient (b, c) and final surface states (e, f).

Note that the transient phase image (b) is scaled in rad, while the phase image of the final surface (e) is expressed in nm, allowing to display a surface topography. The latter indicated a smooth ablation crater with sharp boundaries and a rather constant depth between 40 and 50 nm [Temnov, 2004]. The analysis of phase surfaces generated at different pump-probe delay times $\Delta t$ indicated a sharp ablation front expanding at a nearly constant velocity with the maximum in the center of the ablating area accounting to ≈400 m/s [Temnov, 2004] (data not shown here). This is in good agreement with a previous study of the authors [von der Linde, 2000] and consistent with the shadowgraphy results presented in Figure 13.27 for silicon wafers irradiated at 3 - 4 times larger peak laser fluences.

The amplitude images $r_{\mathrm{ind}}(x, y)$ carry information on the attenuation (or increase) of the probe beam radiation, caused for example by (transiently) changes surface reflectivity. The amplitude distribution derived from the transient interferogram shown in Figure 13.30a reveals a series of alternating dark and bright rings (c). They represent the transient Newton fringes already discussed above in the context of Figure 13.20 and are a manifestation of a sharp ablation front. In contrast, the amplitude map reconstructed from the final interferogram shown in Figure 13.30d shows a fine dark ring at the location of the crater wall (f). Otherwise, the optical reflectivity inside and outside of the crater is constant and corresponds to the reflectivity value of solid GaAs.

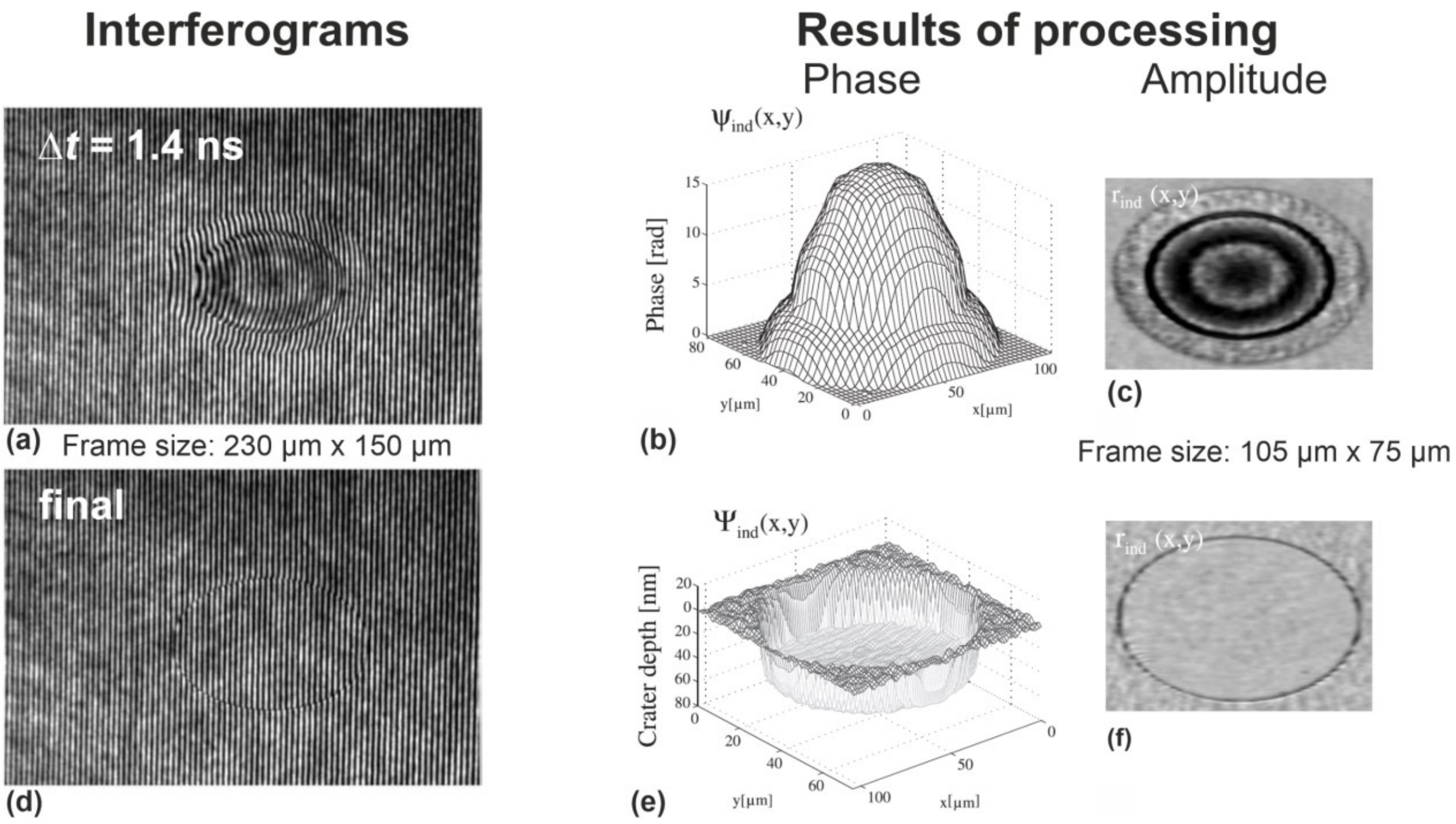

**Fig. 13.30:** Pump-probe side-view interferograms recorded 1.4 ns (**a**) and several second (**d**) after the arrival of the pump pulse ($\lambda_{\mathrm{pump}}$ = 800 nm, $\tau_{\mathrm{pump}}$ = 100 fs, $\phi_{0,\mathrm{pump}} = 1.4\times\phi_{\mathrm{abl}}$ = 0.28 J/cm$^2$, $\theta_{\mathrm{pump}}$ = 45°, $\lambda_{\mathrm{probe}}$ = 400 nm, $\tau_{\mathrm{probe}}$ = 100 fs, $\theta_{\mathrm{probe}}$ = 0°) [Temnov, 2004]. (**b**, **c**, **e**, **f**) Results of retrieval of the induced optical phase image $\Psi_{\mathrm{ind}}(x, y)$ and amplitude images $r_{\mathrm{ind}}(x, y)$ from the corresponding interferograms. (Reprinted from [Temnov, 2004], V.V. Temnov et al., Femtosecond time-resolved interferometric microscopy, Appl. Phys. A **78**, 483 – 489, 2004, Springer Nature)

The fs-IFM surface measurements of the fs-laser ablation of semiconductors presented here are fully consistent with the results obtained by time-resolved BF-R brightfield microscopy but enhanced by some additional quantitative information gained by the interferometric approach (on costs of increased experimental and data evaluation complexity). This dual mode capability was recently employed also by Wei et al. for visualizing through a similar Michelson-interferometer-type pump-probe IFM the dynamics of the formation of LIPSS (type LSFL-I) on silicon wafer surfaces upon irradiation with single and multiple fs-double pulse sequences generated by another Michelson interferometer implemented the pump beam path ($\lambda_{pump}$ = 800 nm, $\tau_{pump}$ = 120 fs, inter-pump pulse delay ~19.8 ps, $\theta_{pump}$ = 0°, $\phi_{av,pump}$ = 0.21 J/cm$^2$ per single pulse) [Wei LAM, 2025]. The authors demonstrated a lateral resolution of ~240 nm of their imaging system along with a temporal resolution of 260 fs. The accuracy of the topographical height measurements was better than 25 nm in their measurements. Note that the ultrafast pump-probe IFM method is not restricted to the analysis of surface processing and can be adapted also to study the bulk laser processing of transparent samples [Sun, 2005 / Bergner AO, 2018 / Bergner Diss, 2019].

Another prominent variant of IFM is the *white light interference microscopy* (WLIM). Straightforwardly implementing the above outlined general idea, the technique allows to acquire the surface topography by quantitatively evaluating beam propagation related phase difference along the optical beam path through the interference effects [DeGroot, 2015]. However, the conventional (commercial) WLIM does not provide the necessary time-resolution for the *in-situ* analysis of laser processing.

Thus, in a first step a suitable ultrashort supercontinuum white light source was developed [Horn, 2010], as depicted in Figure 13.31a. It is based on principle that a fs-laser beam focused into the bulk of dielectrics can generate via the effect of self-phase modulation new spectral components in the form of a so-called "supercontinuum" (SC) [Brodeur, 1999]. Fundamentally, that SC radiation arises from the Kerr nonlinearity. It exhibits a high degree of spatial coherence and can simultaneously cover a large part of the visible spectral range, i.e. representing ultrashort pulsed white light.

Technically, such a SC source was realized by directing a Ti:sapphire fs-laser beam ($\lambda$ = 800 nm, $\tau_p$ = 80 fs) onto an array of 127 hexagonally-arranged convex micro lenses (focal length $f$ = 18 mm, numerical aperture NA ≈ 0.008, lens diameter 300 µm), resulting in an array of laser foci in an sapphire rod placed right behind the micro lens array. Each of these foci leads to self-focusing and causes beam filamentation with a single filament that is capable to generate a SC. These separated broadband SC emissions are spatially and temporally overlapped at the exit of the sapphire cylinder. A highly reflective dielectric mirror separated the residual 800 nm radiation towards a beam dump. The transmitted SC radiation ($E_p$ ≈ 10 µJ) was then recollimated and expanded by a lens system. Cross-correlation measurements were used to determine the pulse duration of the SC. Depending on the length of the sapphire rod used for the SC generation, and without any chirp compensation the white light SC pulse duration has varied between be 3.5 ps ($L_{rod}$ = 6 mm) and 9.0 ps (for $L_{rod}$ = 50 mm) [Mingareev, 2010]. The corresponding spectral range of the SC covered the wavelength range between 380 and 750 nm [Horn, 2010 / Mingareev, 2010].

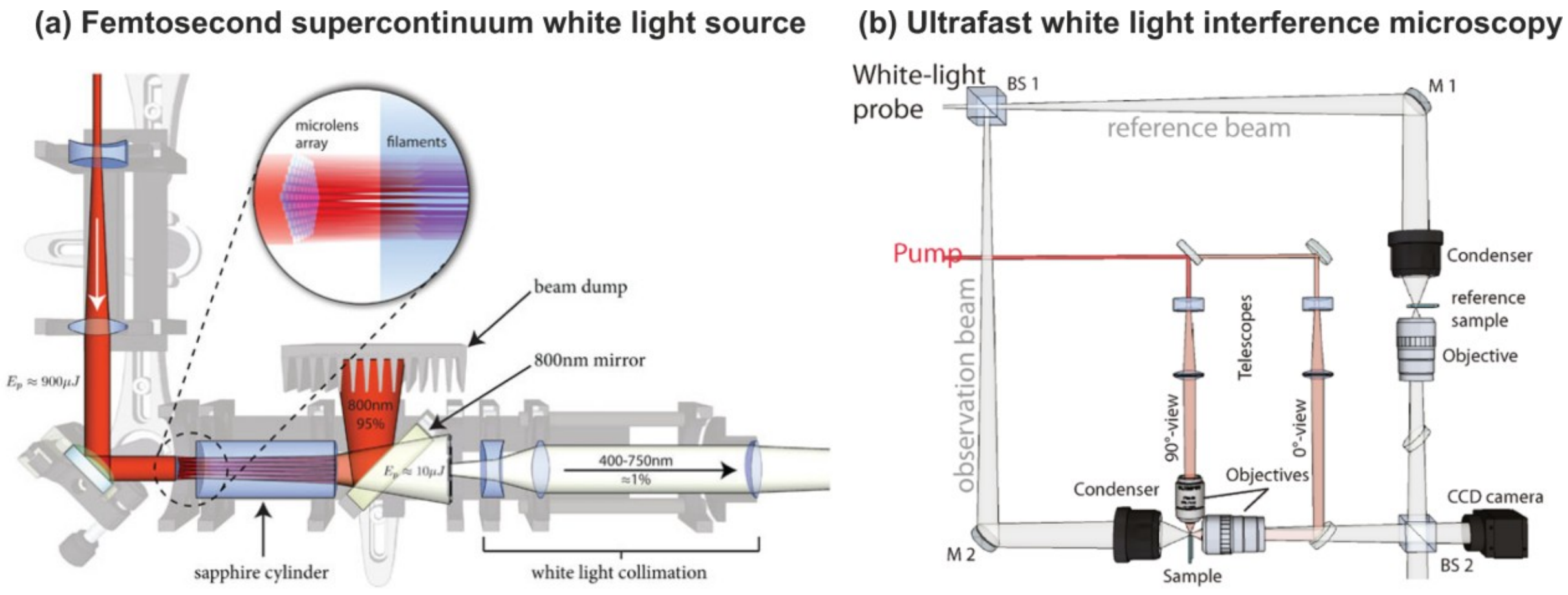

**Fig. 13.31: (a)** Scheme of a micro-lens array based femtosecond laser driven supercontinuum white light source [Horn, 2010]. **(b)** Scheme of ultrafast pump-probe white light interference microscopy setup. (Reprinted from [Horn, 2010], A. Horn et al., Development of a time-resolved white-light interference microscope for optical phase measurements during fs-laser material processing, Appl. Phys. A **101**, 231 – 235, 2010, Springer Nature)

Based on this ultrashort SC source, a pump-probe white light interference microscope was implemented, as sketched in Figure 13.31b [Horn, 2010]. It is based on a Mach-Zehnder interferometric setup, where a beamsplitter (BS1) divides the probe beam into an "observation beam" and a "reference beam". The latter one acts as an (undisturbed) reference, while the observation beam passes through the laser-processed object of interest, where its optical phase gets shifted according to the sample's diffractive and refractive properties caused by the irradiation with the pump laser beam. For visualizing the pump-laser-induced phase changes as an interferogram, the two probe beams must be precisely superimposed and realigned via a second beamsplitter (BS2) before an interferogram is recorded by a CCD camera. Mirror M2 can be used to adjust and control here the fringe pattern of the interferograms. The difference between a common Mach-Zehnder interferometer and the Mach–Zehnder interference microscope implemented here basically lies in the two sets of identical microscopes being added to the observation and the reference beams, each consisting of an objective (20×) and a condenser. The setup presented in Figure 13.31b can be operated in two different modes: when the pump beam and the probe beam are perpendicular, a $\theta_{pump} = 90°$ (side) view can be encoded through the interferograms. In contrast, an additional dielectric mirror placed in the observation beam path allows the pump beam to be focused anti-collinearly to the observation probe beam, thus, allowing a coaxial $\theta_{pump} = 0°$ detection scheme. Considering the spectral properties of the SC and the color depth and pixel size of the CCD camera, the resolution of the optical path difference (OPD) was estimated as ~1 nm [Mingareev, 2010].

The ultrafast pump-probe white light SC interference microscope described above was successfully used to characterize the dynamics of fs-laser-induced melting tracks in the bulk of BK7 glass during scan processing. [Horn, 2010]. As another application, the optical breakdown generated by a single fs-laser pulse being focused in the ambient air atmosphere was studied [Mingareev, 2010]. Figure 13.32 summarizes results of these measurements, where the delay $\Delta t$ was varied between 0 and 1.5 ns. At larger delays, emerging optical plasma emission disturbed the fringe analysis of the interferograms and, thus, prevented a proper data analysis.

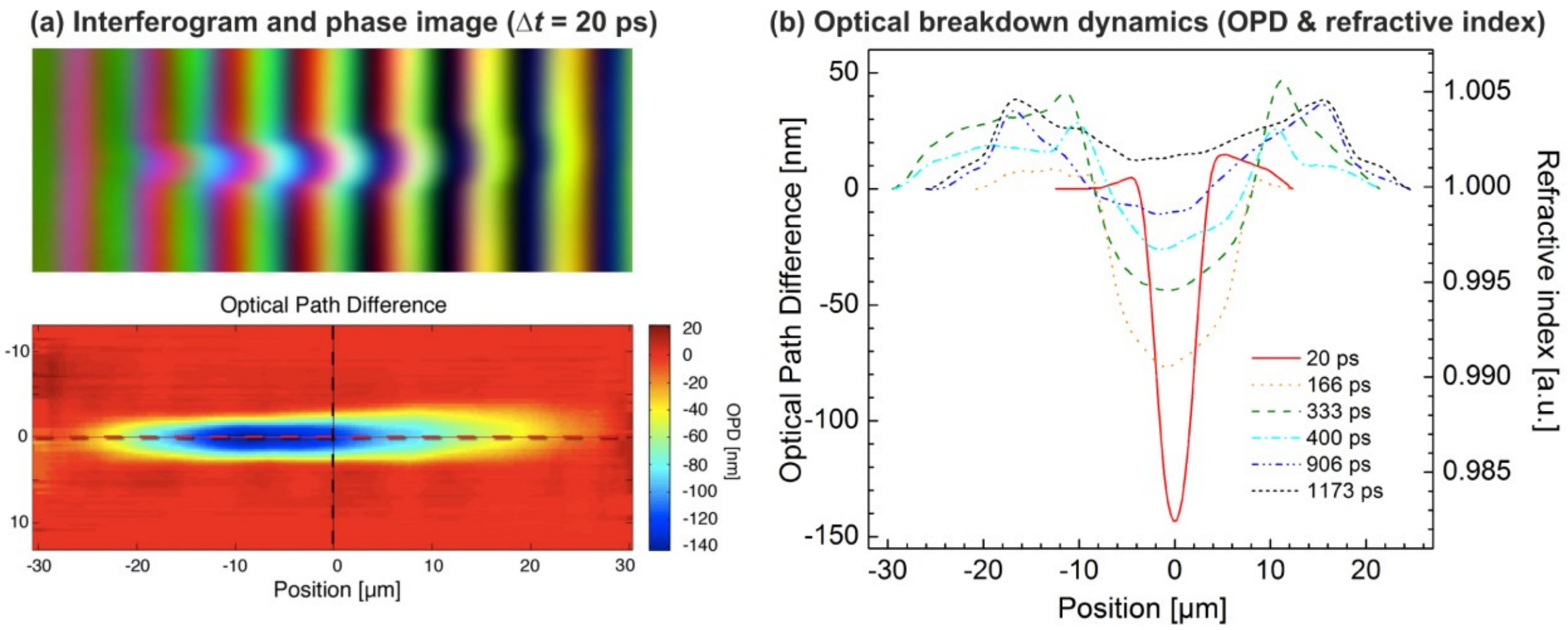

**Fig. 13.32: (a)** Interferogram obtained by ultrafast pump-probe WLIM (top) and corresponding calculated optical path difference (OPD) image (bottom) of air breakdown induced by a single pulse of Ti:sapphire fs-laser pulse after $\Delta t = 20$ ps. ($\lambda_{pump} = 800$ nm, $\tau_{pump} = 130$ fs, $E_p = 115$ µJ, $\lambda_{probe} = 380 - 750$ nm, $\theta_{probe} = 90°$) [Mingareev, 2010]. **(b)** Temporal evolution of the lateral OPD (left ordinate) and refractive index (right ordinate) of an air breakdown as evaluated via lateral cross-sections for six different delays $\Delta t = 20$, 166, 333, 400, 906, and 1173 ps. (Reprinted from [Mingareev, 2010], I. Mingareev et al., Time-resolved white light interferometry for ultrafast metrology, AIP Conf. Proc. **1278**, 891 – 901 (2010), with the permission of AIP Publishing)

In the top panel Figure 13.32a exemplifies an fs-WLIM interferogram recorded at a probe pulse delay time of $\Delta t = 20$ ps after the impact of the pump pulse [$\lambda_{pump} = 800$ nm, $\tau_{pump} = 130$ ps, $E_p = 115$ µJ, $\theta_{probe} = 90°$]. The bottom panel displays the OPDs calculated from the deformation of the fringes in the interferogram according to the procedure described in Mingareev et al. [Mingareev, 2010]. Maximum negative OPDs of up to –140 nm are observed in the near-focal region having an axial extend of 30 µm and a lateral diameter of ~6 µm. Figure 13.32b provides lateral cross-sections extracted at six different delay times between $\Delta t = 20$ ps and 1.173 ns. At early delay times ($\Delta t = 0.20$ ps, red solid curve) the air in the focal area gets ionized by the pump pulse, resulting in a refractive index of the air-plasma being reduced by a few percent compared to the unaffected air ($n = 1.0$). According to the Drude-model, the refractive index change $\Delta n$ is then proportional to the density of free electrons in the air plasma. The laser-induced plasma area subsequently expands with time, manifesting in an increased width of the laser-modified zone after several hundreds of picoseconds (see the dashed and dotted curves). Consecutively, the air surrounding the laser-induced plasma becomes compressed, resulting in a locally increased gas density and an increase of the refractive index and the OPD [Mingareev, 2010].

## 3.4 Phase-Contrast Microscopy

Photographic equipment, the human eye, and typical optoelectronic cameras are only sensitive to intensity variations of the incident light. Therefore, without special arrangements, changes in the optical phase of the detected light are usually not accessible. Nevertheless, optical phase changes often convey relevant information.

A special type of optical microscopy that converts local changes (shifts) of the optical phase upon light propagation through non-absorbing objects in transparent samples via interference effects into local brightness variations is *phase-contrast microscopy* (PCM). The method was developed by Frits Zernicke in the early 1930s and has been awarded the Nobel Prize in 1953 [Zernicke, 1955]. The basic principle of PCM is to make phase changes visible by separating object-scattered light from the illuminating background light and then to manipulate these two contributions differently. Technically, this is realized by using a condenser with a ring-shaped annulus along with special microscope objectives, where a quarter-wave coating (creating an additional 90° optical phase shift) and a gray filter attenuation is added in an adapted ring-shaped area. The beam path of PCM is illustrated in Figure 13.33a (reproduced from [Davidson, 2002]).

An annular "phase ring" is placed at the front focal plane of the condenser lens, conjugate to the objective rear focal plane. The illuminating light (yellow) passes the phase ring and is then focused by the condenser lens as a hollow cone of light onto the transparent sample (specimen). Some of the illuminating light is then scattered/diffracted by the transparent (phase) objects in the sample. The non-deviated remaining light arrives at the rear focal plane of the objective again in the shape of an annular ring of light. It provides the background illumination. The light scattered through the transparent objects in the sample (deviated light) is angularly spread and undergoes an extra phase-shift compared to the background light. It is assumed to be 90° in this example. In PCM, the image contrast is then formed and increased in two ways: First, by generating destructive or constructive interference between the object-scattered light and the background illumination light, and, secondly, by reducing the amount of background light that reaches the image plane for enhancing the modulation amplitude upon interference.

For obtaining the first, the background light is selectively phase-shifted by 90° via passing it through a ring-shaped quarter-wave ($\lambda/4$) phase plate deposited on the rear surface microscope objective lens. This adds a 90° phase difference between the background and the object-scattered light allowing them to destructively interfere in the image plane, here. This results in a decreased brightness of these object-related areas compared to the background (just as for an absorbing object in the sample). This PCM-mode is referred to as "positive phase-contrast".

Finally, for obtaining the above-mentioned selective attenuation of the background light an additional ring-shaped gray filter is coated on the microscope objective. It reduces the intensity by appr. 70 to 90%. Some of the object-scattered light will be phase-shifted and attenuated by the ring-shaped coatings too, but, overall, to a much smaller extent than the background light, which only illuminates the phase-shift and gray filter rings. If the total optical phase-shift caused by the scattering transparent object is deviating from the 90° assumed here, the resulting phase-contrast will be reduced in the images according to the less effective destructive interference.

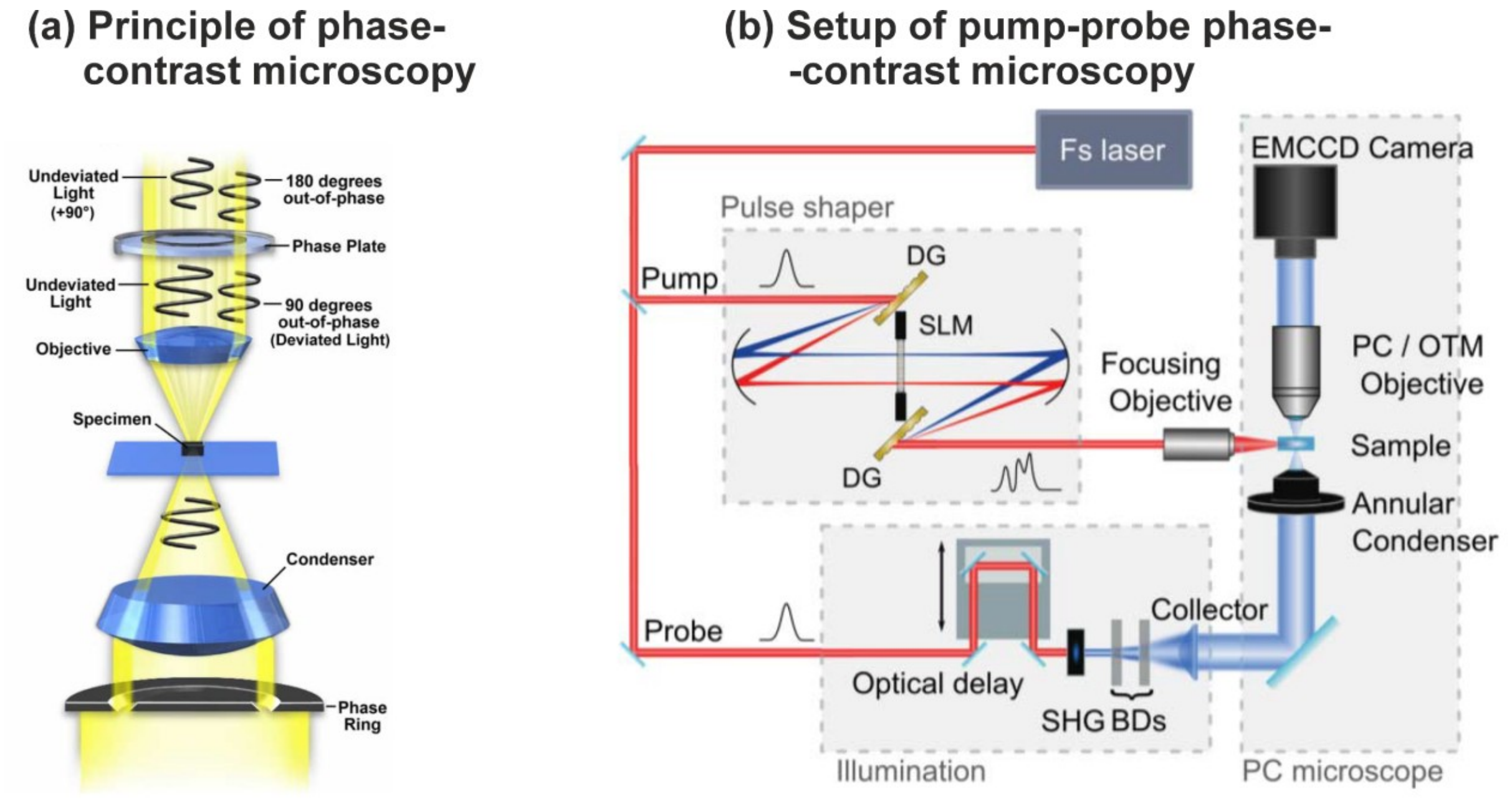

**Fig. 13.33: (a)** Principle of "positive" phase-contrast microscopy (PCM) [Davidson, 2002]. The illuminating light is incident from the bottom (Reprinted from [Davidson, 2002], M.W. Davidson et al., Optical microscopy. pp. 1106 – 1141, in J.P. Horniak (Ed.), *Encyclopedia of Imaging Science and Technology*, Vol. 2, by permission from John Wiley and Sons, Copyright (2002). **(b)** Experimental setup of pump-probe phase-contrast microscopy [Mermillod, 2011]. Abbreviations (alphabetically): BD – beam diffuser, DG – diffraction grating, EMCCD – electron multiplying charge-coupled device, OTM – optical transmission microscopy, PC – phase-contrast, SHG – second harmonic generation, SLM – spatial light modulator. (Reprinted from [Mermillod, 2011], A. Mermillod-Blondin et al., Time-resolved imaging of laser-induced refractive index changes in transparent media, Rev. Sci. Instrum. **82**, 033703 (2011), with the permission of AIP Publishing)

Femtosecond time-resolved pump-probe phase-contrast microscopy was developed at the *Max-Born-Institute for Nonlinear Optics and Short Pulse Spectroscopy* (MBI) in Berlin, Germany. The purpose was to reveal the transient evolution of the fs-laser-induced refractive index changes in transparent glasses and crystals that were shown to be useful for flexibly writing optical waveguides in the bulk of the material. A special focus was placed also on the formation of nanovoids [Mermillod, 2009], the effect of temporally shaped ultrashort laser pulses and pulse sequences [Mermillod, 2011], the visualization of laser-induced photo-acoustic phenomena [Mauclair, 2016], and the measurement of thermo-physical constants in glasses [Bonse, 2018]. A review on the technology of fs-laser direct writing of waveguides in dielectrics can be found in [Mermillod Chapter, 2023].

Figure 13.33b shows an experimental implementation of pump-probe phase-contrast microscopy [Mermillod, 2011]. The pump pulses are split-off from a Ti:sapphire fs-laser beam and pass through a temporal pulse shaping unit before being tightly focused a few hundreds of micrometers behind the surface into the bulk of a transparent and polished sample. For focusing, a large working distance microscope objective (50×, NA = 0.45) is used. The pulse shaper is based on a *spatial light modulator* (SLM) that is capable to manipulate (shift) the spectral phases of the broadband fs-laser laser pulses [Weiner, 2000]. In this way, bandwidth-limited fs-laser pulses can be converted into ultrashort double- or multi-pulse burst sequences featuring flexible pulse-energy ratios. This allows to temporally distribute the spatio-temporal energy deposition into the material for tailoring its material excitation, subsequent relaxation, and final structural state. The probe pulses pass through an optical delay line, are frequency-doubled in an SHG-crystal, and enter a pair of beam diffusers (BDs) that reduce the spatial beam coherence, while homogenizing the spatial distribution. After collimation, the probe beam enters the PCM that is coupled to an electron multiplying charge-coupled device (EMCCD) camera and synchronized with the laser system. An additional bandpass filter centered at 400 nm wavelength suppressed parasite light emitted by the irradiated sample or scattered pump beam radiation (filter not shown in the scheme). For improving the signal-to-noise ratio and reducing the impact of speckles resulting from the used of coherent probe radiation, multiple images can be accumulated at a fixed delay time $\Delta t$, while laterally displacing the sample by well-adjusted motorized translation stages to a fresh spot.

Just by moving the annular condenser module in or out, the imaging mode can be readily changed between phase-contrast microscopy and optical transmission microscopy (OTM), allowing to acquire similar micrographs of the laser-induced probe light absorbing free-electron plasma. For additional details on the spatial and temporal resolution of the fs-pump-probe PCM setup, image data handling, as well as results on the impact of double pump pulse excitation in fused silica, the reader is referred to [Mermillod, 2011].

Mermillod et al. used fs-time-resolved PCM to study local transient refractive index changes in the bulk of fused silica samples in the delay range up to $\Delta t$ = 10 ns [Mermillod, 2009]. The imaging was performed at high resolution perpendicular to the pump beam propagation direction in the region close to the geometrical focus of a single, tightly focused pump laser pulse ($\lambda_{pump}$ = 800 nm, $\tau$ = 100 fs, $\theta_{pump}$ = 0°, $\lambda_{probe}$ = 400 nm, $\theta_{probe}$ = 90°). The energy of the pump laser pulses was chosen at $E_p$ = 0.22 µJ, i.e., around the critical power for self-focusing (~2.8 MW here) and close to the threshold of permanent material modification. "Positive" phase contrast microscope objectives were used in the experimental setup. Hence, local positive refractive index changes $\Delta n$ appear as regions darker that the background level. The results of these measurements are provided in Figure 13.34, where 100 images were accumulated in each frame. The temporal resolution was better than 0.6 ps in this case. For comparison, a micrograph visualizing the permanent material modification along the optical axis (OA, dashed line) is shown (labelled "$\Delta t \to \infty$"). It indicates the well-known positive refractive index change in fs-laser irradiated fused silica over a length of ~10 µm along the OA, having a sub-micrometric radial diameter. This region is interrupted by an elliptical region with a large negative refractive index change. It has been interpreted as a nanovoid locally formed at the region of strongest material excitation surrounded by compacted matter [Mermillod, 2009]. Two vertical dashed lines mark the extent of the spatial region where the void permanently manifests and becomes obvious at delay times $\Delta t$ > 10 ns. Additionally, horizontal cross-sections of the PCM intensity profile (optical phase along OA, represented by the is presented in each micrograph).

Already at early delay times (0.5 ps ≤ $\Delta t$ ≤ 2 ps), an increase of the negative refractive index change can be seen. It can be attributed to fs-laser excited free electrons. After $\Delta t$ = 10 ps (when electron-phonon relaxation became relevant that transferred energy from the electrons to the glass matrix), a peak centered on the location of the future permanent void emerges. It is surrounded by a narrow region of locally increased refractive index. This heat-dominated refractive index pattern dominates for the next 10 ns in the vicinity of the pump laser pulse excited region. Permanent refractive index changes manifest at larger delays.

The thermomechanical void formation launches the emission of *pressure waves* that are emitted from the laser-excited region and exhibits cylindrical and spherical components and can be also visualized via time-resolved PCM on the ns-time scale [Mauclair, 2016].

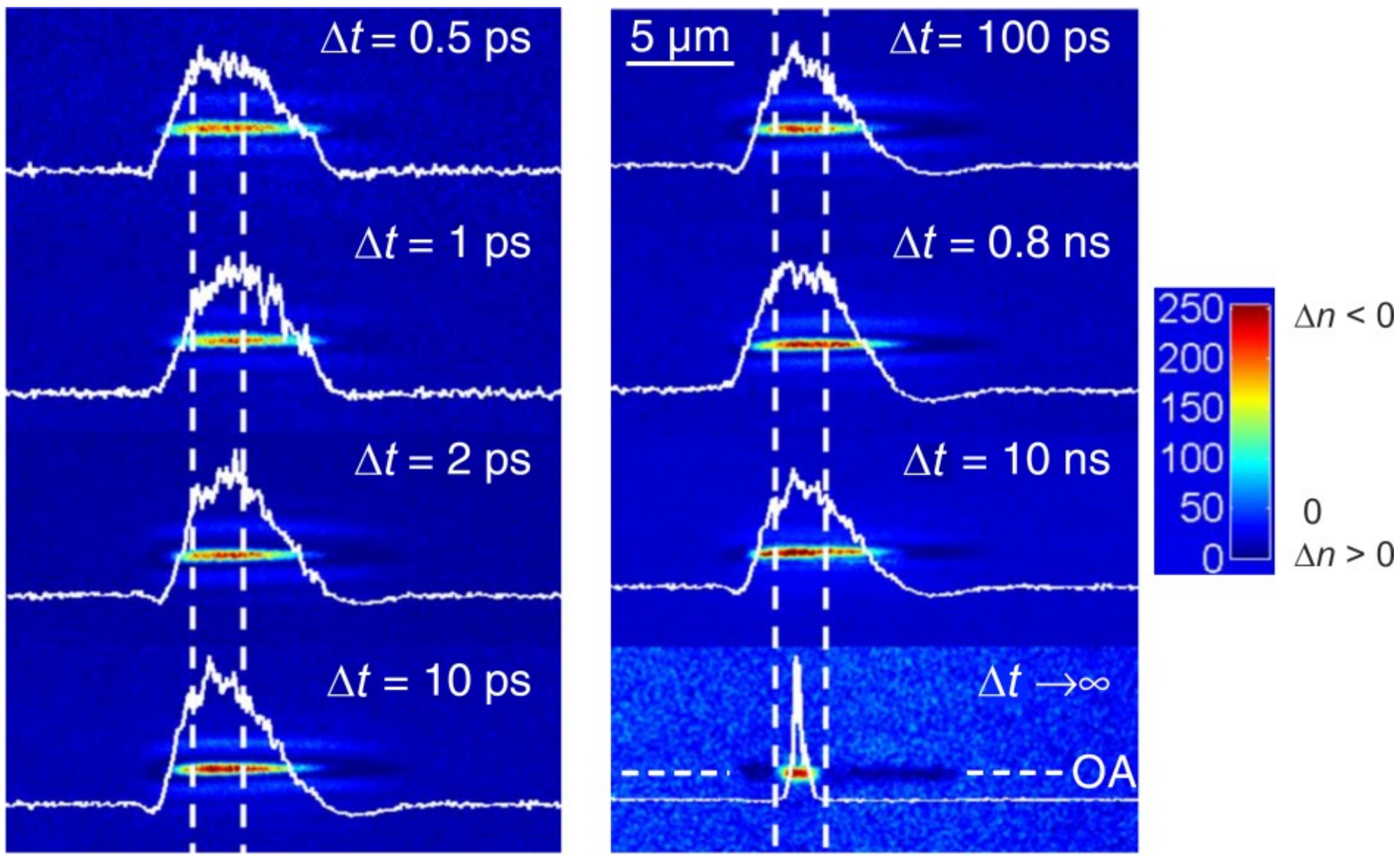

**Fig. 13.34:** Pump-probe side-view phase-contrast microscopy (PCM, corrected from absorption) in fused silica at different delay times $\Delta t$ upon single fs-laser pulse irradiation [Mermillod, 2009]. The laser pulse (pump) is incident from the left [$\lambda_{pump}$ = 800 nm, $\tau_{pump}$ = 100 fs, $E_p$ = 0.22 µJ, $\theta_{pump}$ = 0°, $\lambda_{probe}$ = 400 nm, $\tau_{probe}$ = 100 fs, $\theta_{probe}$ = 90°, 100 image accumulations per frame]. A joint color bar encoding positive and negative refractive index changes ($\Delta n$, arb. units) is provided at the right. OA denotes the optical axis. (Reprinted from [Mermillod, 2009], A. Mermillod-Blondin et al., Dynamics of femtosecond laser induced voidlike structures in fused silica, Appl. Phys. Lett. **94**, 041911 (2009), with the permission of AIP Publishing)

Time-resolved PCM is capable to visualize spatio-temporal thermal transients on longer timescales to reveal how heat flows on the micrometer scale [Bonse, 2018]. The approach relies on the coupling of the refractive index change and the sample temperature through the thermo-optic coefficient d$n$/d$T$. For these experiments, the probe laser was replaced by a ns-pulsed random laser ($\lambda_{probe}$ = 575 nm, $\tau_{probe}$ < 10 ns,) that was synchronized to the Ti:sapphire pump laser pulses and served as the illumination source of the phase-contrast microscope. The random lasers reduce the coherence of the laser radiation and enables a quasi speckle-free full-field imaging [Redding, 2012, Mermillod, 2013]. With that the thermal transients can be studied on a timescale ranging from 10 ns up to 0.1 ms after laser excitation. Beyond providing visualizing transient heat flows, in-depth analysis of the results also allows a quantification of the local thermal diffusivity of the sample [Bonse, 2018]. This powerful capability of PCM is exemplified in Figure 13.35.

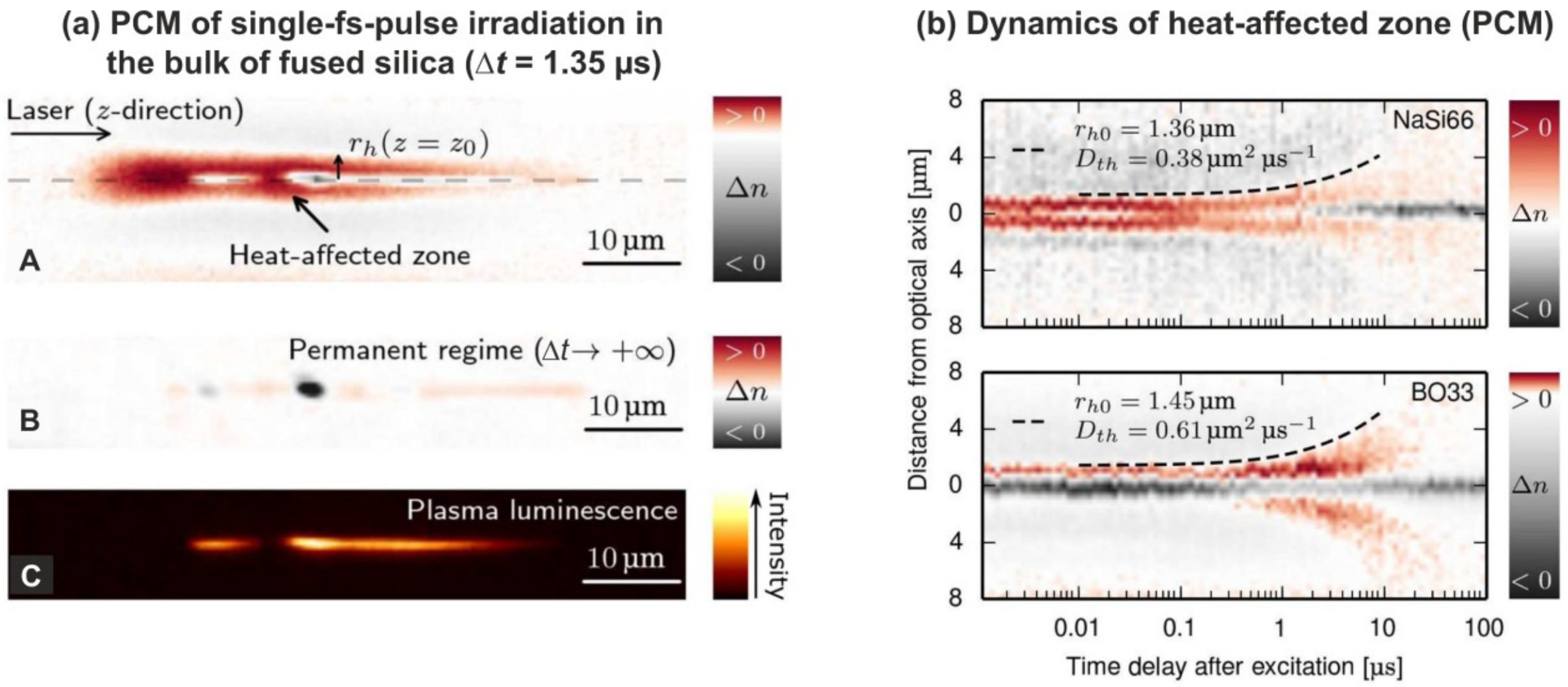

**Fig. 13.35:** Heat flow dynamics in different glasses upon single fs-laser pulse irradiation. The laser pulse (pump) is incident from the left [Bonse, 2018]. **(a)** Pump-probe side-view microscopy in commercial fused silica (SQ1, Sico). Panel A: PCM of refractive index changes recorded at $\Delta t = 1.35$ ns. The laser-heated radius $r_h$ at the position $z = z_0$ is marked by an arrow. Panel B: PCM of permanent refractive index changes. Panel C: Laser-induced plasma luminescence around $\lambda = 575 \pm 15$ nm in absence of the pump beam [$\lambda_{pump} = 790$ nm, $\tau_{pump} = 70$ fs, $E_p = 2.5$ µJ, $\theta_{pump} = 0°$, $\lambda_{probe} = 575$ nm, $\tau_{probe} < 10$ ns, $\theta_{probe} = 90°$]. **(b)** Temporal evolution of the radius $r_h$ of the heat-affected zone in two other glasses as obtained by PCM. Top panel: custom made alkaline earth silicate glass (NaSi66, made of 66 mol% $SiO_2$ and 34 mol% $Na_2O$), bottom panel: commercial borosilicate glass (BO33, Schott). The dashed lines represent least-squares-fits to the experimental data allowing to determine the thermal diffusivity $D_{th}$; $r_{h0}$ is the initial radius of the heat source. For details see the text. (Reprinted from [Bonse, 2018], J. Bonse et al., Time-resolved microscopy of fs-laser-induced heat flows in glasses, Appl. Phys. A **124**, 60, 2018, Springer Nature)

A typical result of such a measurement recorded at $\Delta t = 1.35$ ns after the impact of a single Ti:sapphire laser pulse ($\lambda_{pump} = 790$ nm, $\tau_{pump} = 70$ fs, $E_p = 2.5$ µJ, $\theta_{pump} = 0°$) focused into the bulk of fused silica is presented in Figure 13.35a [Bonse, 2018]. Panel A represents the pump-probe side-view PCM image and visualizes the refractive index change in the heat-affected zone around the focal region. Positive refractive index changes $\Delta n > 0$ (as generated by heat in fused silica) are encoded in red colors. The radius $r_h$ that heat-affected zone at the focal position $z = z_0$ is indicated by an arrow. The corresponding permanent refractive index changes are provided in panel B. Overall, the PCM image reveals a similar morphology as previously shown in Figure 13.34 ($\Delta t \to \infty$), i.e., a small refractive index increase on the OA, interrupted by a void with $\Delta n < 0$ appearing dark here. These regions of permanent material modification also show a characteristic laser-induced plasma luminescence that is detected in absence of the probe beam and is falling into the spectral transmission window of the bandpass filter (±15 nm centered around $\lambda_{probe}$) placed in front of the EMCCD camera for suppressing parasitic light from the probe beam path, see panel C.

Figure 13.35b compiles the results of the temporal evolution of the heated radius as observed in PCM and evaluated at the focal position $z = z_0$. Analyzing and solving the differential equation describing diffusive heat transport under the constraints given here, one obtains a square-root-scaling of the heated radius with evolving delay time $\Delta t$, i.e., $r_h(\Delta t) = (r_{h0}^2 + 4 \cdot D_{th} \cdot \Delta t)^{1/2}$, with $r_{h0}$ being the initial heat radius (immediately after transfer from the laser-excited electric system and the glass matrix via electron-phonon coupling) and $D_{th}$ being the *thermal diffusivity*. [Bonse, 2018]. The latter is related to the thermal conductivity $\kappa$, the mass density $\rho$, and the specific heat $c_p$ via $D_{th} = \kappa/(\rho \cdot c_p)$. The dashed lines in the two panels of Figure 13.35b represent least-squares fit of the measured radius $r_h$ of the HAZ to that equation, resulting in fit parameters of $r_{h0} \sim 1.4$ µm (very similar for both glasses) and thermal diffusivities of $D_{th} = (0.37 \pm 0.05)$ µm$^2$/µs (NaSi66) and $(0.64 \pm 0.09)$ µm$^2$/µs (BO33), respectively. These values excellently agree with values of 0.39 and 0.63 µm$^2$/µs expected from the literature [Bonse, 2018]. Thus, ns-time-resolved PCM represents a suitable method for the measurement of $D_{th}$ in the bulk of transparent materials.

## 3.5 Nomarski Microscopy

The success of PCM has led to the development of a number of subsequent phase-imaging methods. Georges Nomarski patented in 1952 a method that is nowadays known as *differential interference contrast* (DIC) microscopy [Nomarski, 1952]. In DIC microscopy, an additional Wollaston-prism and a linear polarizer are inserted in the imaging beam path of a brightfield microscope [Davidson, 2002]. The Wollaston-prism splits the polarized beam into an extraordinary and an ordinary beam, which are polarized perpendicular to each other and spatially displaced (sheared) at the sample plane. After reflection from the sample (or transmission through another Wollaston-prism), the beams are recombined again before observation. The interference of the two parts at recombination is sensitive to their optical path difference. Thus, the method is particularly sensitive to changes of the refractive index. Adding an adjustable offset phase by moving the Wollaston-prism laterally to the optical beam path, the interference at zero optical path difference in the sample can be adjusted. The image contrast is proportional to the path length gradient along the shear direction, providing the appearance of a 3D physical surface relief corresponding to the variation of optical density of the sample. Thus, the DIC images are emphasizing gradients of the optical phase and are not providing a topographically accurate image.

A femtosecond time-resolved pump-probe microscopy setup was realized by Horn et al. using a Ti:sapphire laser system for probing the fs- and ps-laser-induced volume modifications, the development of laser-induced sound waves, and cracking in the bulk of borosilicate glass (BK7, Schott) at delay times $\Delta t$ in the few tens of ps to some tens of ns range [Horn Diss, 2003 / Horn, 2006].

## 3.6 Polarization Microscopy

Another imaging technique based on the change of the polarization state of light is *polarization microscopy* (PM). In PM, the sample of interest in typically placed between a pair of crossed linear polarizers ("polarizer" and "analyzer") with a high extinction ratio and embedded in a brightfield transmission microscopy scheme. If the sample does not change the linear polarization state being present after passing through the first polarizer, the crossed analyzer completely blocks the transmitted radiation. If, however, parts of the sample locally depolarize the linear state, e.g. via birefringence induced by mechanical stress, some depolarized light can pass the second analyzer and becomes visible on a detector. In imaging mode this typically then manifests as a fringe pattern. Through this, PM has become the standard method for visualizing birefringence effects in (optical) glasses. The combination of a light source with two crossed linear polarizers is called “plane polariscope”. When inserting two additional quarter-wave plates for creating circularly polarized light interacting with the sample, the device is referred to as “circular polariscope”. The latter arrangement has the advantage that an isochromatic fringe pattern is generated. [Wikipedia, Photoelasticity]. As a background-free technique, PM is very sensitive and is perfectly suitable to reveal stress-related effects in laser-processed transparent samples.

In the processing of transparent glasses and crystal with ultrashort laser pulses it was found that irradiation at the surface or in the bulk under specific irradiation conditions can lead to the generation of micro-cracks [Moon, 2009 / Tochio, 2012 / Sakakura, 2015], that may be even controlled by properties of the laser beam [Hendricks LANE, 2016 / Dudutis, 2016]. Soon, the enormous potential of this non-ablative “self-cleaving” mechanism was recognized, leaving behind clean and smooth interfaces, while simultaneously allowing an energy efficient up-scaling of glass sheet processing via sequential polarization-controlled stitching, e.g., for singulation of display glasses. The clarification of the involved mechanisms and the enormous industrial potential has further boosted the development of time-resolved pump-probe polarization microscopy around the second half of the past decade [Hendricks SPIE, 2016 / Kumkar, 2016 / Sakakura, 2017 / Jenne, 2018 / Koritsoglu, 2024].

Jenne et al. implemented a fs-time-resolved pump-probe polarization microscope that based on a circular polariscope, see Figure 13.36 [Jenne, 2018]. Two amplifier modules were seeded by the same fs-laser oscillator but are capable to pick their seed pulses independently with preselected, discrete pump-probe delay as multiples of 20 ns. The pump laser was operated with pulse repetition rates of to 400 kHz at a center wavelength of $\lambda_{pump}$ = 1030 nm. An electronically controlled compressor unit allowed the continuous adjustment of the pump pulse duration $\tau_{pump}$ in the range between 0.3 ps to 20 ps. A subsequent spatial light modulator offered the possibility to alter the beam shape of the pump pulses. The latter were then focused by a microscope objective (NA = 0.4, focal length 10 mm) into a transparent glass sample. The probe laser pulses exhibited a duration of $\tau_{probe} \sim 200$ fs, were frequency-doubled ($\lambda_{probe}$ = 515 nm), variably delayed by a motorized optical delay line ($\Delta t$ up to 20 ns) and used to illuminate the interaction area transversely to the laser processing beam path of the pump beam ($\theta_{pump} = 0°$). The setup allowed to generate delays $\Delta t$ up to several milliseconds and to control them with sub-picosecond resolution. An additional bandpass filter with a spectral width of $\Delta\lambda_{probe} = \pm 1$ nm placed in front of a CMOS camera was used to reduce scattered or other parasitic light. The setup could be operated either in PM mode or in OTM mode. In OTM mode, “optical depth” ($\delta$) images were calculated via $\delta(\Delta t) = \ln(I_0/I_S(\Delta t)) = \int \alpha(l,\Delta t)\mathrm{d}l$ from the transient shadowgraphs ($I_S(\Delta t)$) and the corresponding background images ($I_0$). According to the Lambert-Beer law, the optical depth represents the integral over all local absorption coefficients $a(\Delta t)$ for the probe beam along its line of sight through the initially transparent sample. Hence, such optical depth micrographs exhibit a reduced probe beam intensity, when the probe radiation is absorbed or scattered at laser-excited electrons, defects, or via other transient states. In PM mode, the background images were subtracted, and the intensity values were calibrated against a standard of known optical retardance.

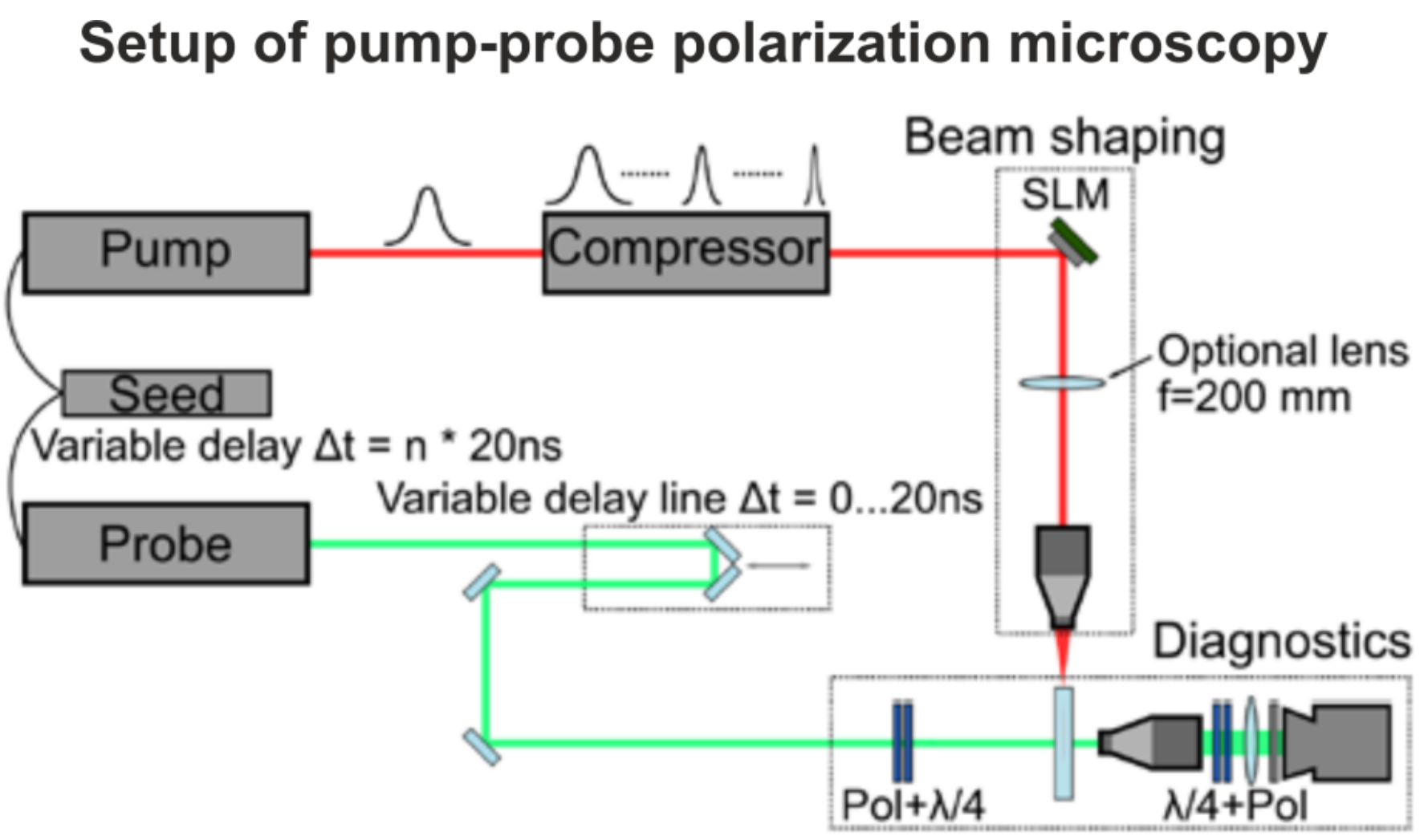

**Fig. 13.36:** Scheme of the experimental pump-probe polarization microscopy setup. Depending on the desired beam profile, an optional lens can be inserted in the beam shaping unit to realize a 4f-arrangement [Jenne, 2018]. The diagnostics relies on a circular polariscope. (Reproduced from Ref. [Jenne, 2018] with permission from Japan Laser Processing Society)

Figure 13.37 shows a collage of PM and OTM micrographs, assembling the transient optical retardance in the left half of each frame, while displaying the corresponding transient optical depth in the right half of the frame. All images were recorded at a fixed delay $\Delta t$ = 10 ns after the exposure of a non-strengthened Gorilla® Glass sample (Corning) by a fixed number (#$N$) of two-pulse pump pulse bursts (single-pulse duration $\tau_{pump}$ = 1 ps, temporal burst pulse separation 60 ns, burst repetition rate 400 kHz) to the same spot (#$N$ = 1, 5, 10, 50, and 100). For comparison, additionally micrographs of the permanent ($\Delta t \to \infty$) material modifications are provided. The total energy in each two-pulse pulse burst was $E_{burst}$ = 13 µJ, evenly distributed between the two individual sub-pulses. Note the 10× enhanced scale of the PM images (retardance) in Figure 13.37 (#1) to (#10).

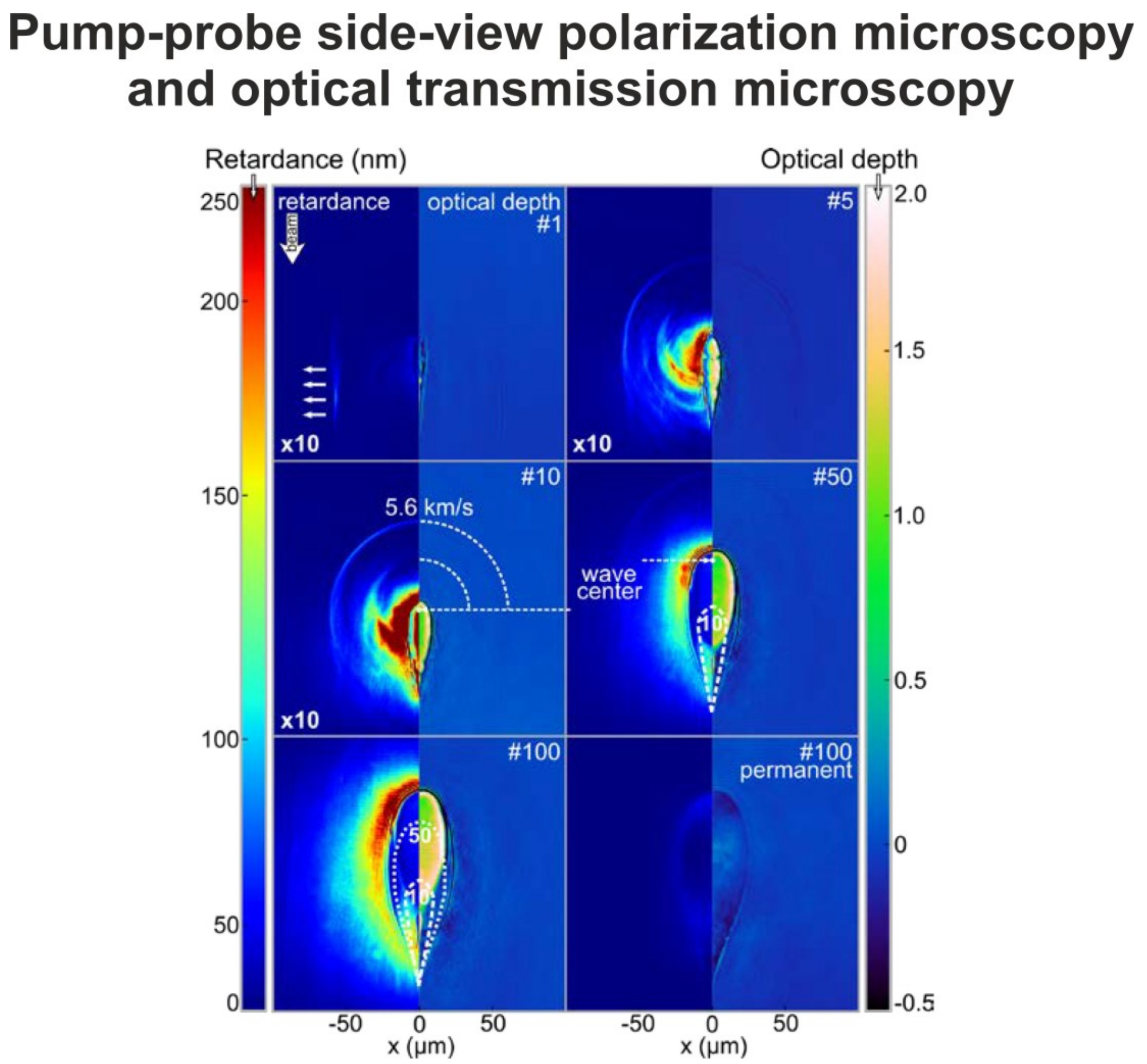

**Fig. 13.37:** Collage of pump-probe side-view micrographs of a two-pulse-burst measurement series (60 ns interval) with $\tau_{pump}$ = 1 ps pulse duration, $\Delta t$ = 10 ns after the last pulse, at a repetition rate of 400 kHz [$\lambda_{pump}$ = 1030 nm, $E_{burst}$ = (6.5 + 6.5) µJ, $\theta_{pump}$ = 0°, $\lambda_{probe}$ = 515 nm, $\tau_{probe}$ = 300 fs, $\theta_{probe}$ = 90°] [Jenne, 2018]. The right-hand side of each frame shows the OTM image and the number of pulse bursts. The left-hand side displays the corresponding measured optical retardance. The laser pulse-bursts (pump) are incident from the top. The color scale of the polarization measurements of #1 to #10 pulse trains are enhanced by a factor of 10 to improve the visibility of the features. (Reproduced from Ref. [Jenne, 2018] with permission from Japan Laser Processing Society)

For irradiation with a single two-pulse burst Figure 13.37 (#1), the absorption region is visible and accompanied by a compressional pressure wave that propagates at a velocity of about 5.6 km/s. This wave represents a longitudinal acoustic wave well [Jenne, 2018]. After irradiation by five and ten two-pulse bursts (#5 and #10), a second pressure wave with reduced velocity becomes visible in the PM measurements. The reduced velocity can be attributed to a transversal acoustic wave, that was also observed in a longitudinal (axial) observation scheme upon double-pulse irradiations [Hendricks SPIE, 2016]. This transversal shear-wave facilitates the cracking of the glass. The direction of the crack can be reproducibly controlled if a spatially elliptical laser beam shape (pump) is used. For such an elliptically shaped beam, there is a tensile stress concentration at the small radii of the ellipse, which determines the crack growth direction [Hendricks LANE, 2016]. Once initiated, the crack-formation is further driven by each of the laser sub-pulses in the burst in an accumulative way, an ablation-free laser process that works well on the tens of ns time scale and enables a continuous “crack-writing” at velocities exceeding several m/s [Hendricks SPIE, 2016 / Hendricks LANE, 2016]. Interestingly, for increasing number of pump pulse irradiations, the virtual emission center of the pressure waves shifts towards the incident beam direction (compare images for #10 and #50). An overall growth of the pump laser modified volume for further pulse bursts can be seen in the frame related to #100. Here, the dotted lines surround the volumes measured in OTM for #10 and #50, respectively. In PM, the highest retardance values appear at the top of the modification, indicating the presence of large compressive stress.

## 3.7 Imaging Reflectometry / Ellipsometry

*Reflectometry* and *ellipsometry* are powerful optical techniques to measure the optical constants of a material by analyzing the polarization dependent reflection of light on basis of the Fresnel formula. The Fresnel reflectivity (and the derived ellipsometric transfer data $\Psi$ and $\Delta$) crucially depends on the irradiation wavelength, the polarization state, and the angle of incidence. Through additional modelling analyses the optical constant can be retrieved, usually expressed via the complex refractive index ($\tilde{n} = n + i\kappa$) or via the complex dielectric permittivity ($\tilde{\varepsilon} = \varepsilon_1 + i\varepsilon_2$), with the relation $\tilde{\varepsilon} = \tilde{n}^2$.

Ultrashort pulsed pump-probe ellipsometry was already implemented in 1974 by Auston et al. for studying the laser-induced carrier dynamics in semiconductors (silicon) at excitation levels below the materials damage threshold [Auston, 1974]. During the second half of 1980's the subject of laser annealing gained industrial attraction and laser-induced melting and solidification of silicon and germanium was addressed by Jellison and co-workers using ns-time resolved ellipsometry, quantifying the optical constants of the molten phase in the visible and near ultraviolet spectral range [Jellison, 1985 / Jellison, 1987].

Around 2003 a north American and a Japanese have extended reflectometry into a pump-probe scheme that allowed to retrieve transient optical constants during femtosecond pulsed laser-induced melting and ablation [Roeser, 2003 / Yoneda, 2003]. Further major steps to fs-time-resolved ellipsometry under laser processing conditions were made a decade later by Rapp et al., studying the fs-laser-induced changes of the complex refractive index metals, such as molybdenum [Rapp, 2016], aluminum, and steel upon single-pulse laser ablation [Winter, 2020]. Ultrafast imaging ellipsometry was implemented by Pflug et al. in 2018 for the investigation of fs-laser irradiated gold films [Pflug, 2018 / Horn, 2022]. Pump-probe spectroscopic ellipsometry was realized by Shikne and Yoneda [Shikne, 2015] for studying liquid metal surfaces, and later by Gutiérrez et al. for analyzing optical phase change media [Gutiérrez, 2022].

Figure 13.38 sketches the principle and experimental setup of time-resolved pump-probe imaging ellipsometry [Horn, 2022]. The pump beam is focused by a lens (FL) to the sample surface under normal incidence, inducing a transient material modification. The probe beam is linearly polarized by a rotatable polarizer (P), typically operated under an angle $\varphi = 45°$. After passing through a telescope the probe beam is imaged onto the sample surface for homogenous illumination over the pump laser excited region. The reflected probe radiation is collimated by a microscope objective (O) and passing through an optical bandpass filter (BPF) that suppresses diffusely reflected pump beam radiation as well as optical emissions from the laser-induced ablation. A motorized compensator (C) and analyzer (A) combination allows to perform rotating compensator ellipsometry [Horn, 2022]. The transmitted light is then imaged by a tube lens (T) onto the chip of a CCD-camera that is synchronized with the probing laser pulses.

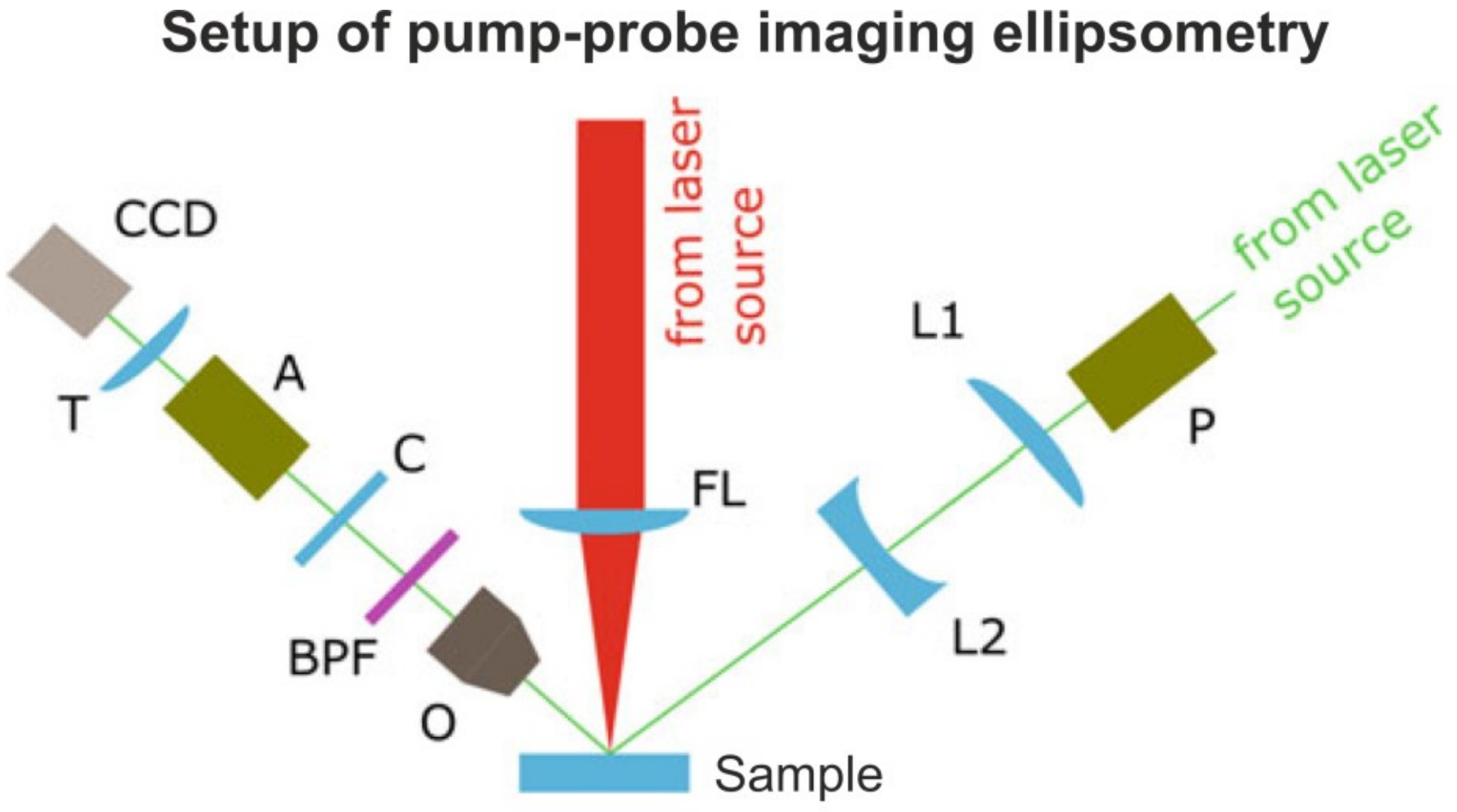

**Fig. 13.38:** Scheme of an ultrafast pump-probe imaging ellipsometer setup [Horn, 2022]. The pump beam is drawn in red, while the probe beam is visualized in green. Abbreviations: A – analyzer; BPF – bandpass filter; C – compensator (rotatable, motorized); CCD – charge-coupled device (camera); FL – focusing lens; L1, L2 – lenses; O – microscope objective; P – linear polarizer; T – tube lens. (Reprinted from [Horn, 2022], A. Horn, *The Physics of Lase Radiation–Matter Interaction*, 2022, Springer Nature)

Using that setup the ablation of a thin gold films upon irradiation with single Ti:sapphire fs-laser pulses ($\lambda_{pump}$ = 800 nm, $\tau_{pump}$ = 35 fs) focused to a beam diameter of ~50 µm was investigated at a fluence level $\phi$ = 0.5 J/cm$^2$ above the melting threshold but below the ablation threshold. The system was probed at a wavelength of $\lambda_{probe}$ = 550 nm and a probe pulse duration of $\tau_{probe} \approx$ 50 fs. Figure 13.39 presents the corresponding results for the spatially resolved variation of complex refractive index ($n$, $\kappa$) as a function of the delay time $\Delta t$ in the left column or alternatively for the corresponding complex dielectric permittivity ($\varepsilon_1$, $\varepsilon_2$) in the middle column [Horn, 2022]. In the right column, the laser-induced normalized surface reflectivity change $\Delta R/R$ is shown. In the top panel it was calculated from the optical constants ($n$, $\kappa$) displayed in the left column, while in the bottom panel it was directly measured for comparison. Both datasets of $\Delta R/R$ agree very well.

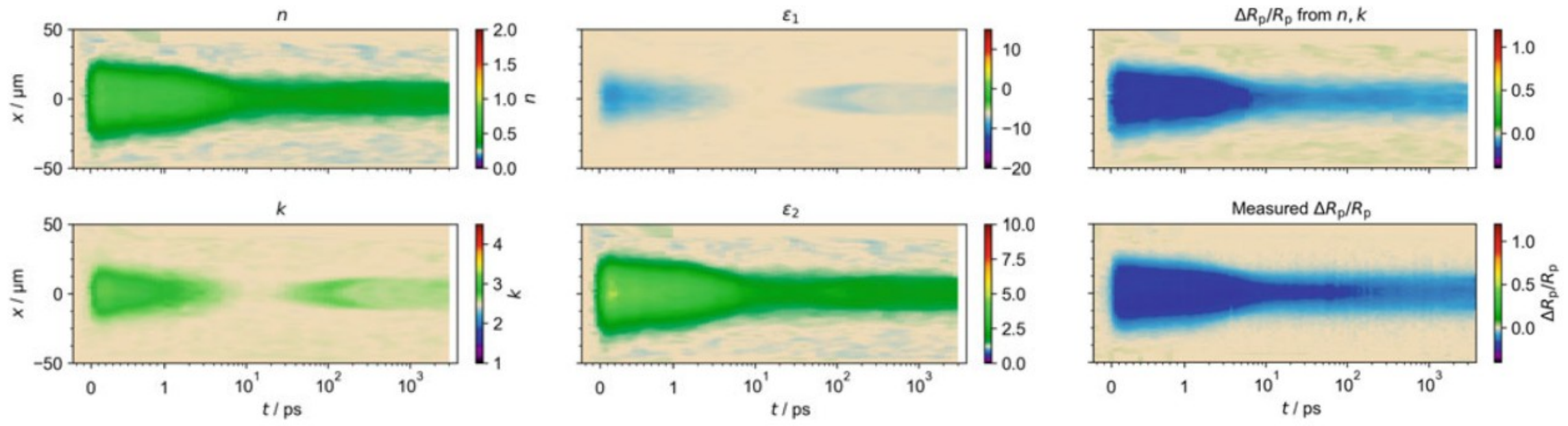

**Fig. 13.39:** Optical properties of a thin gold film after the irradiation by a single fs-pump laser pulse [Horn, 2022]. Left column: refractive index $n$ and extinction coefficient $\kappa$. Middle column: corresponding real part ($\varepsilon_1$) and imaginary part ($\varepsilon_2$) of the dielectric permittivity. Right column: normalized surface reflectivity change $\Delta R/R$ as calculated from $n$ and $\kappa$ and directly measured for comparison. (Reprinted from [Horn, 2022], A. Horn, *The Physics of Lase Radiation–Matter Interaction*, 2022, Springer Nature)

The dynamics of the complex refractive index allows to render some conclusions here that go beyond the ones that can be drawn just from the transient reflectivity measurements: interestingly, the extinction coefficient $\kappa$ shows a characteristic minimum for delay times between ~10 and ~30 ps, see Figure 13.39 (left column, bottom panel), while the refractive index $n$ or the surface reflectivity change $\Delta R/R$ do not. The extinction coefficient $k$ rules the optical absorption coefficient ($\alpha = 4 \cdot \pi \cdot \kappa/\lambda_{pump}$) and the deposited energy that is subsequently shared among the electron system of the gold film. The decrease of $k$ starting immediately after the optical excitation reflects the cooling of the electron system via *electron-phonon coupling* and starts to be thermalized with the lattice of the solid after ~10 ps (see also Sect. 2.2 in Chap. 1, Nolte et al.). The subsequent increase of $\kappa$ for larger delay times can be attributed to the further heating of the metal film and its expansion [Horn, 2022].

# 4 Digital Holography

The development of powerful two-dimensional optoelectronic sensors opens up new and powerful possibilities in microscopy and ultrafast optical imaging (see Sect. 6 below), The method of *Holography* generally takes benefit of the phase information of a coherent optical wave field that allows to store and reconstruct information on an object of interest typically by superimposing a "reference wave" (not interacting with the object) with an "object wave" (interacting with the object) on a photosensitive material or sensor. The phrase derives from the ancient Greek words "holos" (for "whole") and "gráphein" (for "writing" or "drawing"). In the initial form of holography according to Gábor [Gábor, 1948 / Gábor, 1949] (awarded the Nobel Prize in 1971), the information about the object recorded (stored) in the hologram is reconstructed in optical manner. *Digital holography* (DH) differs from classic "analog" holography in that the recorded hologram is not optically reconstructed, but instead mathematically reconstructed (retrieved) on the computer.

*Digital holographic microscopy* (DHM) distinguishes itself from other microscopy methods by not directly recording the projected image of the object. Instead, the light wave front information originating from the object is digitally recorded on a 2D optoelectronic sensor, from which the object image is numerically retrieved. Thus, the image forming lens used in traditional microscopy has been replaced by a computer algorithm.

Currently, the strict classification of microscopic and holographic techniques (and also their distinction) is fading away. It became already a common sense that techniques that store digital information, which can be used to retrieve simultaneously images of optical intensity and phase changes, are nowadays termed DH, DHM, or *digital holographic interferometry* (DHI) [Guo, 2021]. Such a digitally stored information can be either a classical hologram, or an interferogram (see Sect. 3.3 on IFM above), or a diffractogram, or even more complex optical patterns quantitatively encoding the phase and amplitude information of light [Zeng, 2023]. The actual intensity/phase information is then reconstructed on the computer from the fringe pattern using specific mathematical algorithms [Takeda, 1982 / Reid, 1996 / Saville, 2022], often based on the Fourier or Fresnel integral transformations, and thus enables the object to be analyzed in detail in 3D. Note that DH is not restricted to pump-probe approaches but fully compatible with them.

As an example in the context of laser processing, Centurion et al. used a pump-probe diffractogram recorded from a single fs-laser-pulse induced breakdown plasma in air to retrieve the spatial electron density distribution [Centurion, 2004]. Pangovski et al. developed a digital holographic system to image the refractive index distributions associated with the ablation plumes that are emerging in a train of multiple ps-laser pulses interacting with aluminum, titanium, copper, and brass [Pangovski, 2016] (providing results qualitatively very similar to the case of shadowgraphy presented in Figure 13.28 above).

Figure 13.40 provides an example of pump-probe time-resolved digital holography of the propagation and interaction of a fs-pulse beam tightly focused (NA = 0.18) into the bulk of a fused silica sample [Momgaudis, 2022]. The figure assembles a sequence of six images of the optical phase shift, taken at different sub-picosecond pump-probe delay times $\Delta t$ of –0.4 ps, 0 ps, +0.25 ps, +0.30 ps, +0.35 ps, and 0.5 ps, respectively.

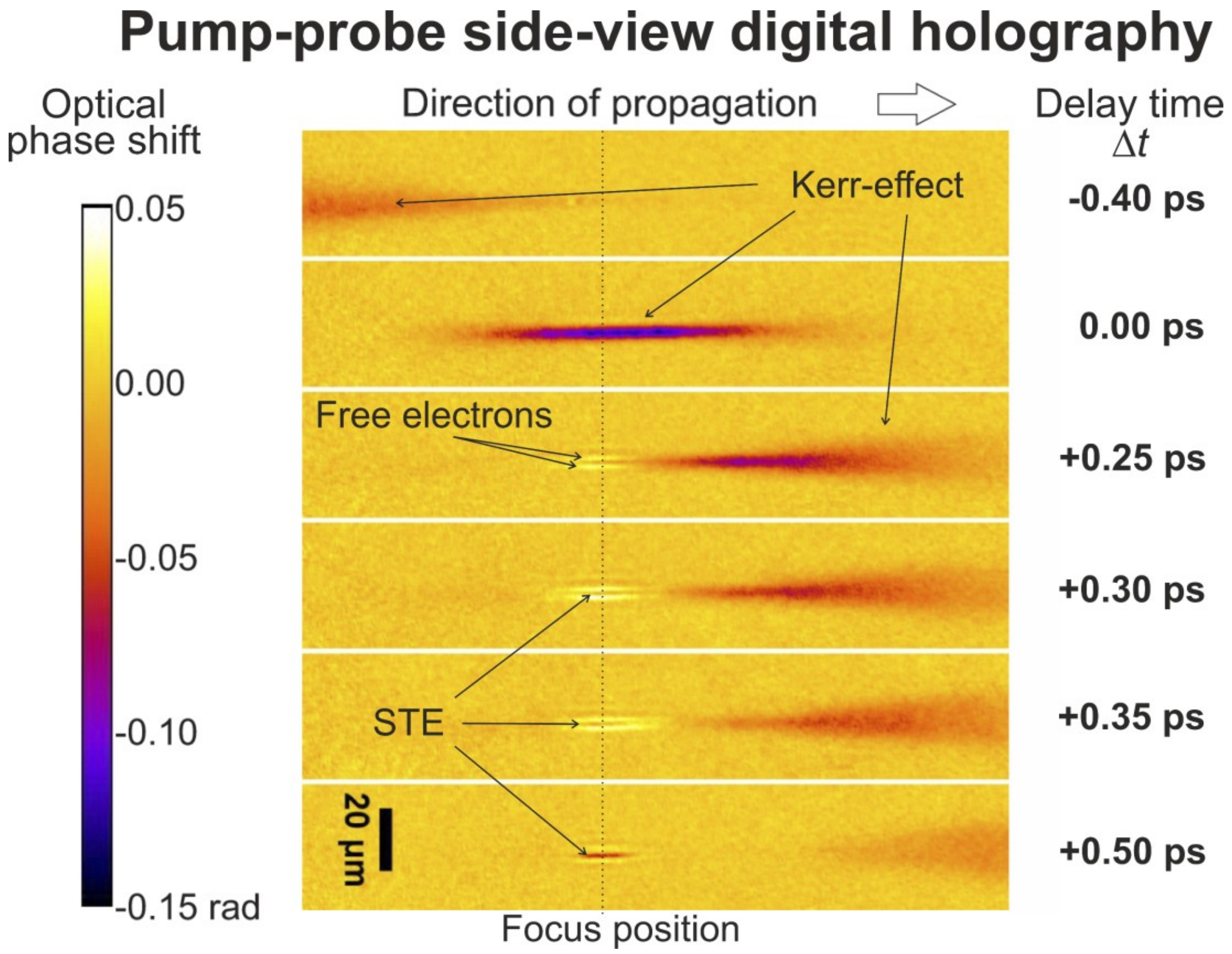

**Fig. 13.40:** Pump-probe side-view digital holography (DH) of tightly focused fs-laser pulse propagating in fused silica through the focal region [Momgaudis, 2022]. The retrieved images of the optical phase shift correspond to different delays $\Delta t$ ranging between –0.40 ps and +0.50 ps. $\Delta t = 0$ marks the arrival of the pump pulse at the focus position. The laser pulse (pump) is incident from the left [$\lambda_{pump} = 1030$ nm, $\tau_{pump} = 360$ fs, $E_p = 0.46$ µJ, $\theta_{pump} = 0°$, $\lambda_{probe} = 550$ nm, $\tau_{probe} = 35$ fs, $\theta_{probe} = 90°$]. (Adapted with permission from [Momgaudis, 2022] © Optica Publishing Group)

The image sequence starts at a negative delay $\Delta t$ = –0.4 ps, visualizing a snapshot of the propagation of the focused light pulse (left part of the image). The high local laser intensities result in the increase of the refractive index due to optical Kerr-effect ($\Delta n \sim n_2 \cdot I(x, y, z, t)$), manifesting as a negative phase shift. When the laser pulse approaches the focal point at $\Delta t = 0$ ps, the negative phase shift due to the Kerr nonlinearity is even larger at the increased local intensities and nonlinear absorption and avalanche ionization set on. This results in the creation of a free electron plasma in the focal region that is locally decreasing the refractive index and, thus, becomes visible as positive phase change in the image frame at $\Delta t$ = +0.25 ps, while the non-absorbed part of the fs-laser pulse leaves the focal region towards the right-hand side of the frame. Such free electrons in the conduction band of fused silica have a short lifetime and scatter within ~170 fs into *self-trapped excitons* (STE) [Petite, 1996 / Sun, 2005], see Chap. 1 (Nolte et al.). At the given probe laser wavelength of $\lambda_{probe}$ = 550 nm, the STE formation has been attributed to the negative phase shift that is visible on the optical axis within the free electron plasma close to the geometrical focus point [Momgaudis, 2022]. For even longer positive delays of $\Delta t$ > +0.3 ps, +0.35 ps, and +0.5 ps the near-focal positive phase shift caused by the free-electron plasma gradually decreases, while the divergent fs-laser pulse leaves the field of view at the speed of light.

# 5 Ultrafast Pump-Probe Tomography

*Tomography* is a method to obtain a 3D-dataset from a series of 2D-projections of an object, typically recorded at different observation angles. Bergner et al. extended the idea of pump-probe shadowgraphy into an ultrafast time-resolved tomography scheme [Bergner OPEX, 2018 / Bergner Diss, 2019]. The experimental setup is sketched in Figure 13.41a. For their experiments the authors implemented advanced ultrafast beam shaping by means of a *spatial light modulator* (SLM), arranged in an 4f-imaging configuration. Via tailored manipulation of the spectral phase of the ultrashort pump pulses the SLM enabled the generation of different Bessel-Gaussian spatial beam shapes of a desired pulse duration inside a dielectric glass sample ($\lambda_{pump}$ = 1026 nm, $\tau_{pump}$ = 7.5 ps). Moreover, the SLM allowed them to locally vary the angle of incidence and propagation of the pump pulses within the glass sample. A frequency-doubled probe beam ($\lambda_{probe}$ = 513 nm, $\tau_{probe}$ = 200 fs) with a fixed illumination direction recorded then side-view shadowgraphic maps of the optical extinction coefficient of locally rotated pump beam induced material modifications, i.e., from different observation angles with respect to the object [Bergner OPEX, 2018].

Three-dimensional tomograms of the ps-laser-induced optical extinction change $\kappa(x, y, z)$ were subsequently retrieved from the pump-probe image sequence by using a sinogram representation of the individual shadowgraphs along with inverse Radon transforms. For additional details on the data retrieval the reader is referred to [Bergner OPEX, 2018 / Bergner Diss, 2019].

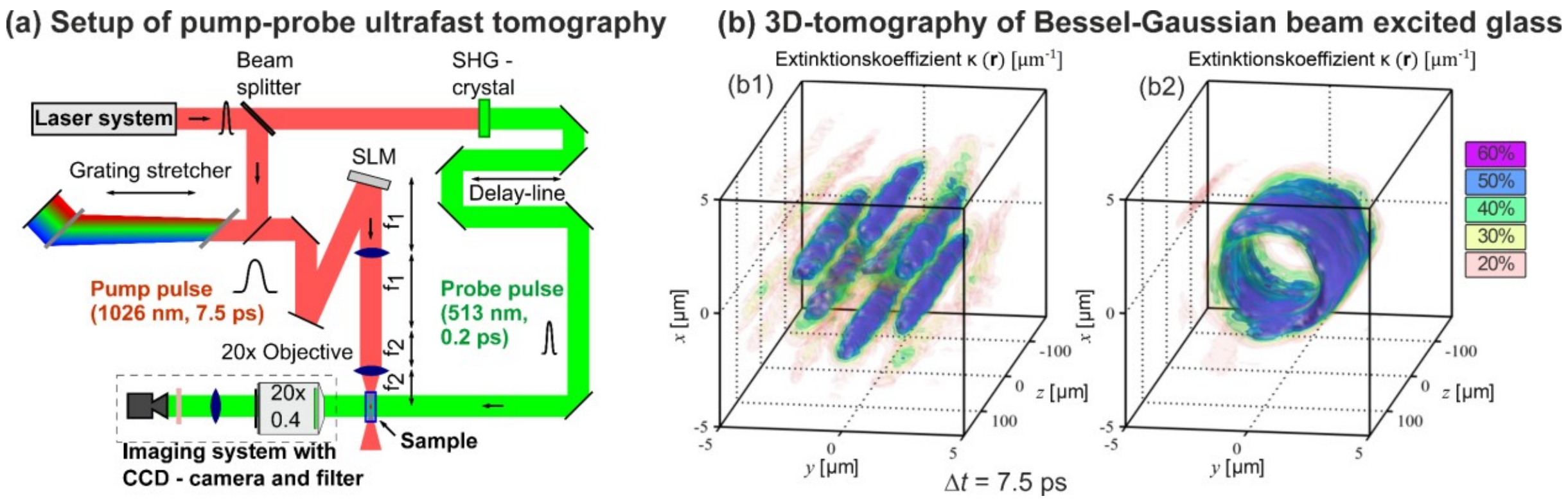

**Fig. 13.41: (a)** Scheme of the experimental setup for pump-probe tomographic microscopy in transparent samples [Bergner OPEX, 2018]. Laser pulses of $\tau_{pump}$ = 7.5 ps duration at $\lambda_{pump}$ = 1026 nm serve as pump beam, whereas pulses of $\tau_{probe}$ = 200 fs duration at $\lambda_{probe}$ = 513 nm are used to spatially and temporally probe the pump beam shaped focal region inside the glass volume. The spatial light modulator (SLM) based beam shaper generates temporally stretched Bessel-Gaussian beams from the pump pulses and ensures the desired beam orientation. **(b)** Iso-surface representation of the reconstructed extinction distribution $\kappa(x, y, z)$ at $\lambda_{probe}$ = 513 nm wavelength after a delay time of $\Delta t$ = +7.5 ps by a Bessel-Gaussian beam ($E_{pump}$ = 74 µJ) of petal-like third-order (panel b1) and pure third-order (panel b2) focused into the Gorilla™ glass sample. The colored iso-contours represent five different absorption thresholds relative to $\kappa_{max}$ = 0.4 µm$^{-1}$ (panel b1) and $\kappa_{max}$ = 0.1 µm$^{-1}$ (panel b2), respectively. (Figure (a) provided and reprinted with permission and courtesy of Klaus Bergner. Figures (b1,b2) reprinted from [Bergner Diss, 2019], with permission and courtesy of Klaus Bergner)

Figure 13.41b exemplifies the capabilities of the ultrafast pump-probe tomography. Two 3D tomograms of the optical extinction coefficients of single ps-laser excited Gorilla™ glass (Corning) are presented for two different Bessel-Gaussian pump beam shapes (pulse energy $E_{pump}$ = 74 µJ), recorded immediately after the laser pulse at a delay time of $\Delta t$ = 7.5 ps. The left sub-panel (b1) is related to the laser excitation with a third-order, petal-like Bessel-Gaussian beam, while the right sub-panel (b2) is associated to a pure third-order Bessel-Gaussian beam [Bergner Diss, 2019].

The ultrafast tomography approach of Bergner et al. discussed here can be employed to any gaseous, liquid, or transparent solid-state matter interacting with ultrashort pulsed laser radiation.

# 6 Recent Ultrafast Optical Imaging Approaches

During the last decade enormous technological efforts were made in different fields of ultrafast imaging, being different from the classical optical microscopy techniques discussed in Sect. 3. The progress of these new approaches in the context of laser processing is briefly outlined for ultrafast photography (Sect. 6.1), light field tomography (Sect. 6.2), optical ptychography (Sect. 6.3), and coherent diffractive imaging (Sect. 6.4). Most of these methods were implemented first for imaging of macroscopic objects. With further increasing detector resolution and imaging magnification, laser-matter interaction processes were identified as very useful rapid transient test objects for benchmarking the imaging capabilities of the individual approaches.

## 6.1 Photography

Visualizing even the ultrafast propagation of "*light-in-flight*" (LiF) has attracted the attention of many scientists [Abramson, 1978 / Denisyuk, 1992]. In 1978 this was realized by Abramson in a classical Gabor-holographic approach to capture a motion picture of light as it propagated and reflected by a mirror on a holographic plate [Abramson, 1978]. More than 30 years later, a new digital camera-based direct imaging implementation was realized at the *Massachusetts Institute of Technology* (MIT) [Velten, 2013 / Faccio, 2018]. This research has then initiated a major boost in the development of ultrafast optical imaging techniques that were driven by several research groups in north America, China, and Japan. Not all of these ultrafast optical techniques represent pump-probe approaches. Several of them are also based on *framing photography* (see the review article [Yao, 2024]).

In these ultrafast photography (UP) techniques specific terms must be distinguished, even for a single-shot (e.g., pump pulse induced) imaging event: The *temporal resolution* must not be directly related to the (inverse) *frame rate*. The latter describes how often the device can capture (probe) image frames per time interval (quantified in frames per second, fps). An upper limit of the frame rate is given by the inverse temporal resolution, but real values of frame rate are usually significantly smaller. Moreover, often the frame rate is specified only over a limited number of image frames, referred to as *sequence depth*. Note that some of the ultrafast cameras can provide extremely high frame rates but only for a limited sequence depth. An example was discussed above already in the context of the *Echelon* pump-probe technique (see Sect. 1.3), where the number of optically replicated probe pulses determines the sequence depth and the inter-probe pulse delay rules the frame rate, while the temporal resolution is given by the duration of the individual probe pulses.

Another important step on that route was the development of the *Compressed Ultrafast Photography* (CUP) technique in 2014, which was capable to acquire non-repetitive time-evolving events at rates up to $10^{11}$ frames per second and with a temporal resolution of tens of picoseconds [Gao, 2014]. In CUP the spatial information is encoded with a pseudo-random binary pattern that is conveyed with a digital micro-mirror device (DMD), in combination with a shearing operation in the temporal domain that is performed by using a streak camera. The encoded, sheared ($x$, $y$, $t$) information is recorded as a single snapshot on the streak camera's back-end CCD camera. The image frames are reconstructed via post-processing then.

Since then, many specific different variants of ultrafast optical imaging were realized and are comprehensively described in several review articles [Mikami, 2016 / Faccio, 2018 / Liang Optica, 2018 / Guo, 2019 / Wang, 2021 / Zeng, 2023 / Yao, 2024 / Wei RM, 2025]. Most of these ultrafast photography approaches are capable of imaging spatially macroscopic sceneries, where the phenomena related to the propagation of light can be spatially resolved more easily than on the microscale. An example of the development of a superluminal Mach-cone visualized via CUP was already presented in Figure 13.5 above. Another impressive light-in-flight recoding of the rotation of the electric field upon fs-laser beam propagation was published in 2020 by Zeng et al. employing *Framing Imaging based on Non-Collinear Optical Parametric Amplification* (FINCOPA) [Zeng, 2020]. However, many of the ultrafast optical imaging techniques were also applied for visualizing dynamical effects manifesting in laser processing in spatial domains towards the microscale. Prominent applications are the laser-induced ablation plasma or the laser-induced optical breakdown in ambient air. Table 13.2 provides an overview on the plethora of different currently available ultrafast photography techniques that were already used in the context of laser processing.

One example of fs-laser beam filamentation manifesting in a glass sample upon loose beam focusing recorded by *Compressed Ultrafast Spectral Photography* (CUSP) is presented in more detail in Figure 13.42 [Wang AS, 2023]. The method was developed in the group of Wang at the *California Institute of Technology* (Caltech), USA, and currently sets the world record by enabling in pump-probe single event recordings extremely high frame rates up to $2.19 \times 10^{14}$ per second (fps), with a temporal resolution of ~110 fs and a sequence depth of 230 frames. In CUSP, to encode time via wavelength, a temporally chirped pulse train is generated by passing a single broadband femtosecond light pulse (e.g. of 70 fs duration at a center wavelength $\lambda_0$ = 805 nm) through a distance-adjustable pair of high-reflection plate beamsplitters (creating the pulse train) and a long rod of controlled length made of highly dispersive glass (inducing a desired spectral chirp). CUSP combines spectral encoding, multi-sub-pulse illumination, temporal shearing, and compressed sensing [Wang AS, 2020].

In brief, in CUSP the transient object is first illuminated by a temporally chirped pulse train so that different wavelengths in each sub-pulse carry unique time stamps. As in CUP, the transient event is subsequently imaged and spatially encoded by a *digital micro-mirror device* (DMD). Additionally, a dispersive element (diffraction grating) separates wavelengths in the image plane. Subsequently, a streak camera collects the raw image and distinguishes the sub-pulses via ultrafast shearing in the direction orthogonal to the wavelength dispersion. Finally, a compressed sensing algorithm extracts a sequence of images from one single acquisition [Wang AS, 2023].

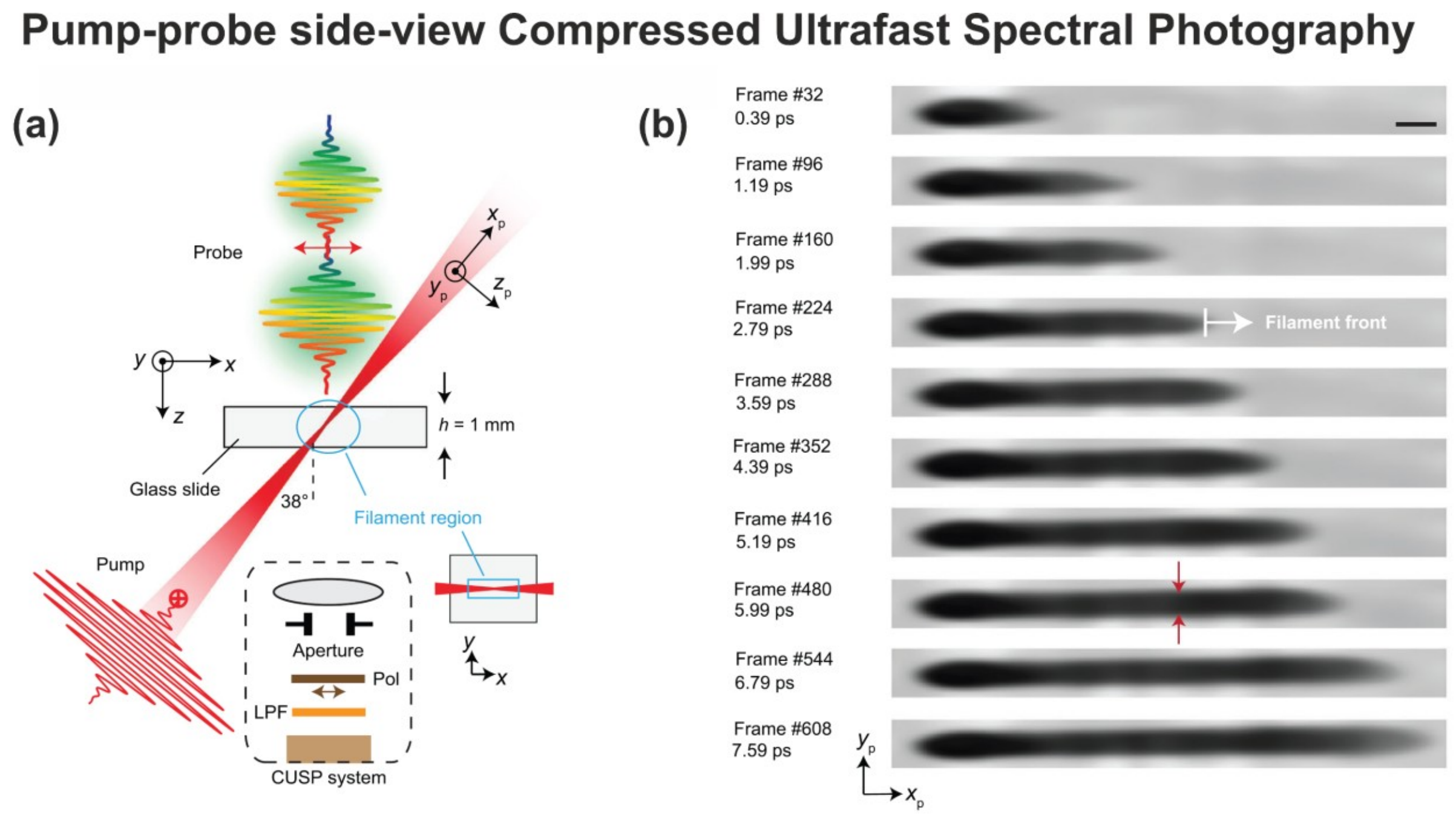

**Fig. 13.42: (a)** Scheme of pump-probe compressed ultrafast spectral photography (CUSP) in a 1 mm thick glass sample [Wang AS, 2023]. Abbreviations: Pol – linear polarizer; LPF – spectral long pass filter. **(b)** Projected side-view CUSP frames recorded at different delay times $\Delta t$ upon single fs-laser (pump) pulse irradiation. The laser pulse (pump) is incident from the left [$\lambda_{pump}$ = 805 nm, $\tau_{pump}$ = 70 fs, $I_{0,pump}$ = 180 TW/cm$^2$, $\theta_{pump}$ = 38°, $\lambda_{probe}$ = 805 nm with –41.63 fs/nm spectral chirp, $\tau_{probe}$ = 1.6 ps, $\theta_{probe}$ = 0°]. The scale bar in (b) represent 50 µm. (Reprinted from [Wang AS, 2023], P. Wang et al., Single-shot reconfigurable femtosecond imaging of ultrafast

optical dynamics. Adv. Sci. **10**, 2207222 (2023), Copyright 2023 under Creative Commons BY 4.0 license. Retrieved from http://dx.doi.org/10.1002/advs.202207222)

To showcase the real-time capabilities of their CUSP setup the authors studied the laser-induced filamentation in glass upon loose laser beam focusing, switching to an imaging mode of $8\times10^{13}$ fps with a sequence depth of 640 frames, allowing a temporal observation window of 8 ps and a temporal resolution of 283 fs. [Wang AS, 2023]. Figure 13.42a presents a scheme of their setup, where a laser pump pulse ($\lambda_{pump}$ = 805 nm, $\tau_{pump}$ = 70 fs) is loosely focused by a 400 mm focal length plano-convex lens (NA = 0.012) at an angle of incidence of $\theta_{pump}$ = 38° into the bulk of the 1 mm thick glass slide, resulting in a peak intensity around 180 TW/cm$^2$. A train of spectrally chirped delayed probe pulses, is then used to for stroboscopic trans-illumination. The fs-laser beam-filamentation process triggered by a single pump pulse is then recorded with the CUSP system as a sequence of shadowgraphs of the laser-induced electron plasma within the glass. A rectangular aperture was used to confine the field of view. Since the pump beam induced filament also propagates into depth ($z$), the depth of field to image the entire filament from its head to the tail was increased by inserting an aperture at the Fourier-plane of the imaging system. Secondary optical plasma emission was removed by a 715 nm spectral long pass filter (LPF). A linear polarizer (Pol) with a transmission direction orthogonal to that of the pump beam was used to suppress the strong surface glare.

Figure 13.42b provides a collage of selected shadowgraphic frames of the pump pulse induced electron plasma taken at ten different pump-probe delays $\Delta t$, individually separated by 800 fs. The laser-induced filament appears dark in regions were pump pulse created a free electron plasma. Since the CUSP raw images are projections into the $x$-$y$ plane, the laser-generated filaments appear shorter than its real physical length in the pump beam propagation direction. Therefore, by using $\theta_{pump}$ = 38° and the glass refractive index of $n$ = 1.51, the CUSP raw images were mapped here to the pump beam coordinate in the $X_p$-$Y_p$ plane to obtain the actual filament geometry. With increasing delay time $\Delta t$, a non-diverging pencil-like filament shape emerged, with a front (head, marked in the frame at $\Delta t$ = 2.79 ps by a white arrow) that propagated at the speed of light towards larger $X_p$-values. Meanwhile, the starting coordinate of the filament remained fixed. Once formed, a constant lateral filament width of ~60 µm was observed. After $\Delta t$ ~ 8 ps, the laser-induced plasma filament then exhibited an aspect ratio of ≈20. The non-diverging propagation characteristics originates from the balancing competition between Kerr-nonlinearity-based self-focusing and plasma defocusing effects, both caused by optical nonlinearities in the glass during the fs-laser beam propagation [Shen, 1975 / Marburger, 1975].

| **Acronym** | **Technique** | **Image acquisition** | **Pump-probe (PP) / Framing (F)** | **Temporal resolution** | **Application** | **Ref.** |
|---|---|---|---|---|---|---|
| **CSMUP** | Chirped Spectral Mapping Ultrafast Photography | reconstruction | F | > 4 ps | Laser ablation | [Yao, 2021] |
| **DMA-CSMUP** | Demosaicing Algorithm - Chirped | reconstruction | F | < 49.1 ps | Laser ablation plasma, Laser | [Yao Science, 2024] |
| | Spectral Mapping Ultrafast Photography | | | | surface ablation, Shockwaves in glass | |
| **SP-CUP** | Stereo-Polarimetric Compressed Ultrafast Photography | reconstruction | F | 41 ps | Laser-induced breakdown in air | [Liang, 2020] |
| **CUSP** | Compressed Ultrafast Spectral Photography | reconstruction | PP/F | 283 fs | Laser-induced beam filamentation plasma in glass | [Wang AS, 2023] |
| **CUST** | Compressed Ultrafast Spectral-Temporal photography | reconstruction | F | 260 | Laser beam propagation | [Lu, 2019] |
| **DRUM** | Diffraction-gated Real-time Ultrahigh-speed Mapping | reconstruction | F | 370 ns | Laser-induced breakdown in water, Bubble dynamics, Laser ablation of biological material | [Liu, 2023] |
| **FINCOPA** | Framing Imaging based on Non-Collinear Optical Parametric Amplification | direct | PP | 50 fs | Laser-induced plasma grating, ultrafast rotating electric field | [Zeng, 2020] |
| **FISI** | Frequency domain Integration Sequential Imaging | direct | PP | 66 fs to 79 fs | Laser-induced damage, Plasma physics, Shockwave interactions | [Zhu, 2022] |
| **FRAME** | Frequency Recognition Algorithm for Multiple Exposures | reconstruction | F | 200 fs | Laser beam propagation | [Ehn, 2017] |
| **M-FTOP** | Multiframe Femtosecond Time-resolved Optical Polarography | direct | ?? | 0.96 ps | Laser pulse propagation in fused silica | [Wang, 2014] |
| **HISAC** | High-Speed Sampling Camera | reconstruction | F | 30 ps | Laser ablation, Laser-induced plasma | [Kodama, 1999] [Fuchs, 2007] |
| **SI-HSSI** | Spatial-multiplexing In-line Holographic Single-shot Sequential Imaging | reconstruction | PP | > 30 fs | Laser-induced breakdown in air | [Huang. 2022] |
| **MA-CS CMOS** | Multiple-Aperture Compressed Sensing CMOS | reconstruction | F | 5 ns | Laser-induced breakdown in air | [Mochizuki, 2016] |
| **MFSI** | Multiframe Femtosecond Stroboscopic Imaging | direct | F | 130 fs / 50 ns | Laser-induced membrane deformation | [Martynowych, 2020] |
| **OCFI** | All-Optical Coaxial Frame Imaging | reconstruction | F | ??? | Laser-induced plasma and electron density | [Chen, 2017] |
| **OPR** | All-Optical Photography with a Raster principle | direct | F | 460 fs | Laser-induced plasma in air and fused silica | [Zhu, 2021] |
| **PUMP** | Polarization-resolved Ultrafast Mapping Photography | reconstruction | F | 850 fs | ITO-film ablation | [Ding, 2023] |
| **SNAP** | Single-shot Non-synchronous Array Photography | direct | F | > 50 fs | Laser-induced plasma | [Sheinman, 2022] |
| **SSISD** | Single-Shot ultrafast Imaging via Spatiotemporal Division | direct | PP | 30 fs | Laser ablation/ionization front | [Yeola, 2018] |
| **STAMP** | Sequentially Timed All-optical Mapping Photography | direct | F | ps | Laser ablation plasma, underwater shock-wave | [Nakagawa, 2014] [Saiki, 2023] |
| **LA-STAMP** | Lens Array - Sequentially Timed All-optical Mapping Photography | direct | F | ps | fs-pulse burst ablation | [Nemoto, 2020] |
| **SF-STAMP** | Spectral Filtering - Sequentially Timed All-optical Mapping Photography | direct | PP | 465 fs | Laser-induced crystalline-to-amorphous phase transition in GST | [Suzuki, 2015] [Suzuki, 2017] |
| **SM-STAMP** | Slicing Mirror - Sequentially Timed All-optical Mapping Photography | direct | F | 7.9 ps | Laser ablation plasma | [Saiki, 2020] |
| **H-STAR** | High-Channel Spectral-Temporal Active Recording | reconstruction | F | < ps | Laser-induced breakdown in air | [Meng, 2024] |
| **STEAM** | Serial Time-Encoded Amplified Microscopy | reconstruction | F | 163 ns | Laser ablation of thin film multi-layer sample | [Goda, 2009] |
| **THPM** | Time-resolved Holographic Polarization Microscopy | reconstruction | PP | 30 ps | Laser-induced damage, phase contrast and amplitude | [Yue, 2017] |
| **TSFM** | Time and Spatial-Frequency Multiplexing | reconstrction | F / PP | 200 fs | Laser ablation of glass, air plasma shadowgraphy | [Moon, 2020] |
| **UFC** | Ultrafast Framing Camera | direct | F | 130 fs | Laser-induced shock waves, shadowgraphy, darkfield imaging | [Dresselhaus-Cooper, 2019] |
| **STS-UFP** | Spatiotemporal Shearing-Based Ultrafast Framing Photography | direct | F | 4.9 ps / 19.7 ps | Laser-induced ablation plasma, shockwaves | [He PR, 2025] |
| **WPMSI** | Wavelength and Polarization Multiplexing Single-shot Imaging | direct | PP | 200 fs | Laser-induced air plasma, laser ablation plasma, Bulk free-electron plasma | [Zhang, 2023] |

**Table. 13.2:** Ultrafast imaging methods that were applied to visualize laser processing. The entries in the left column are alphabetically ordered according to the technique's main acronym

## 6.2 Light Field Tomography

In 2014, Li et al. presented the first ultrafast method for recording optical tomographic movies of objects even evolving at light velocity [Li, 2014]. The authors applied their method to visualize transient refractive index variations (at 400 nm wavelength) induced by the Kerr-effect in the bulk of fused silica glass upon beam propagation of a single focused fs-laser pulse ($\lambda$ = 800 nm, $\tau_p$ = 100 fs) through the focal region ($w_0$ ~ 25 µm radius). In contrast to the ultrafast tomography of Bergner et al. [Bergner OPEX, 2018] presented in Sect. 5 above, the method of Li et al. does not rely on a classical pump-probe approach, realizing the time-dependent measurements through a series of multiple identical irradiation events, recorded individually at different pump-probe delays, but represents a single-shot framing method realizing and deriving five differently delayed probing daughter-pulse from a single ultrashort pulse (a concept already discussed as "Echelon technique" above, see Sect. 1.3). Furthermore, Li et al. probed the refractive index changes not in a side-view geometry but angularly multiplexed (all five probe angles < 10°) in paraxial directions to the strong fs-laser pulse inducing the refractive index change.

In 2021, Feng et al. presented another single-shot ultrafast tomography implementation and coined the term "*Light Field Tomography*" (LFT) as transient imaging strategy that enables highly efficient light field acquisitions within the 4D space-time. They demonstrated a 3D imaging of light-in-flight phenomena with an image sequence depth > 1000, a temporal resolution better than 10 ps, and a *Non-Line-Of-Sight* (NLOS) imaging at a 30 Hz video-rate [Feng, 2021]. In other words, to some extent, the method is even capable to "view at (hidden) objects around the corner".

The central idea of LFT is to reformulate ultrafast photography as a computed tomography problem (see also Sect. 5) by using a set of cylindrical lenses to directly acquire parallel beam projections of the object. Coupled with an electronic streak-camera, LFT can capture the entire 4D spatio-temporal space in a single snapshot, while providing an image resolution of 120×120 pixels at an image sequence depth exceeding 1000. For additional details on the technical and computational implementation, the reader is referred to the original Ref. [Feng, 2021].

The capabilities of ultrafast light field tomography are demonstrated in Figure 13.43, where a recording of the propagation of a single ps-laser pulse ($\lambda = 532$ nm, $\tau_p = 6$ ps) dispersing through a helically wrapped light-diffusing optical fiber is visualized via LFT (a-e) and via conventional photography (f) for comparison [Feng, 2021].

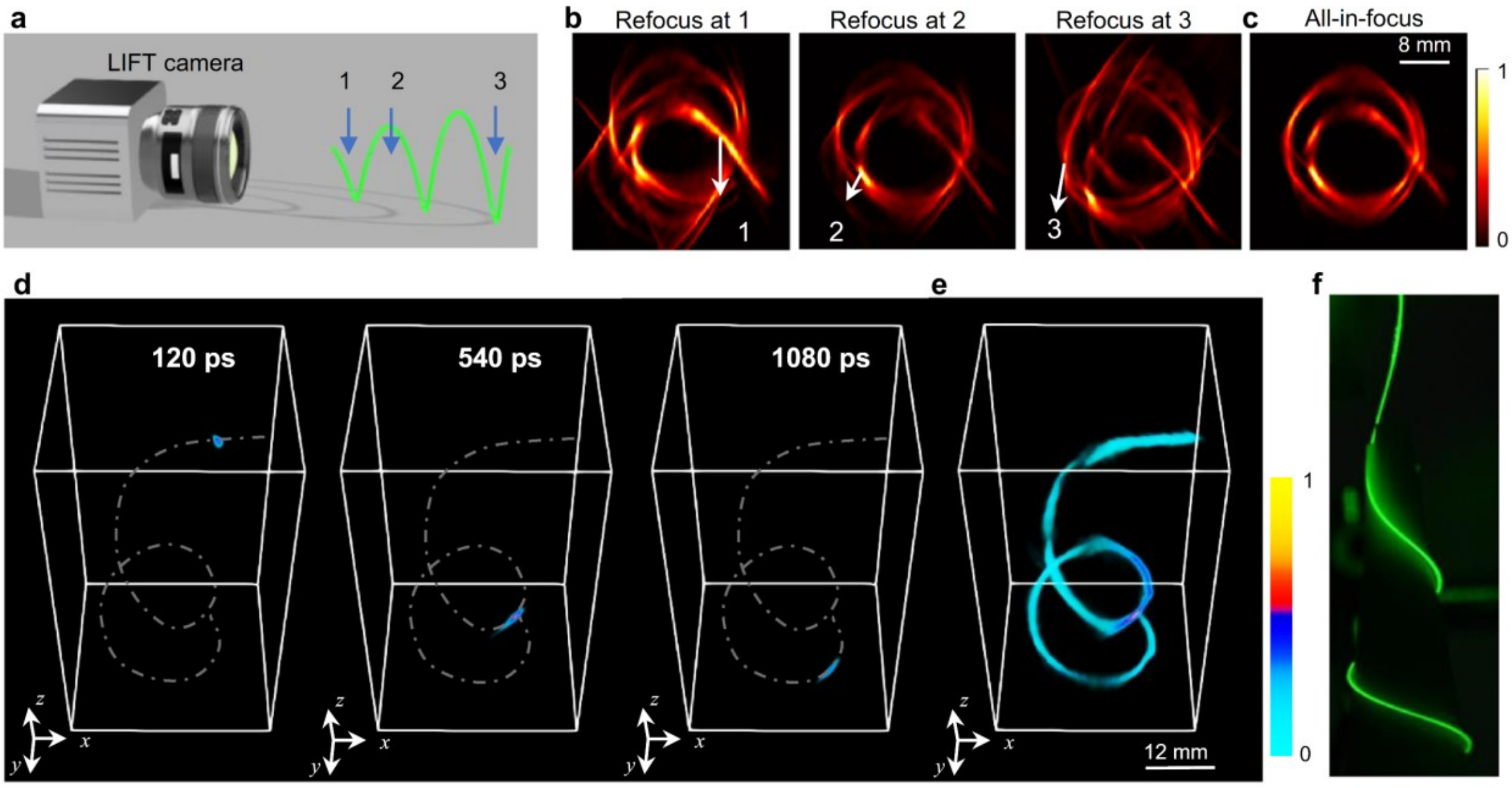

**Fig. 13.43:** (**a**) Experimental setup of transient light field imaging by a LFT camera [Feng, 2021]. The imaged object is a helically wrapped light-diffusing optical fiber that is launched by a single ps-laser pulse ($\lambda = 532$ nm, $\tau_p = 6$ ps). (**b**) Post-capture refocusing LFT images at three different depths. (**c**) Computed all-in-focus time-integrated LFT image. (**d**) 3D LFT images of a ps-laser pulse propagating inside the light-diffusing optical fiber, captured at delay times of $\Delta t = +120$ ps, +540 ps, and +1080 ps, respectively. The helical fiber is artificially drawn by a dashed gray line and overlaid in each 3D frame for visual guidance. (**e**) Time-integrated 3D LFT image of the helical fiber. (**f**) Side-view photograph of the helical fiber. (Reprinted from [Feng, 2021], X. Feng et al., Ultrafast light field tomography for snapshot transient and non-line-of-sight imaging. Nature Comm. **12**, 2179 (2021), Copyright 2021 under Creative Commons BY 4.0 license. Retrieved from https://doi.org/10.1038/s41467-021-22461-0)

Figure 13.43a shows schematically the imaging arrangement of the helically wrapped fiber with a depth range extending over 80 mm. The dynamics of the laser pulse propagation was recorded at a rate of at 0.5 Tera-frames per second with a native temporal resolution of ~3 ps. Figure 13.43b provides the fiber images at three different focal settings of the camera (marked by numbered blue arrows also in sub-figure b), obtained by computationally refocusing the LIFT camera at different depths and integrating time-resolved images along the temporal axis. For each selected focal depth, only a part of the helical fiber remains sharp, as indicated by the white arrows. By employing its post-capture refocusing capability, the LFT computation can also synthesize an all-in-focus image (Figure 13.43c). 3D LFT images of laser pulse propagation inside the helical fiber are visualized in Figure 13.43d at three different delay times $\Delta t$ = +120 ps, +540 ps, and +1080 ps [Feng, 2021]. The LFT-retrieved 3D structure of the fiber provided in Figure 13.43e was obtained by integrating all individual frames of the sequence. It agrees qualitatively well with the photograph presented in Figure 13.43f, convincingly validating capacity of LFT for visualizing extended 3D objects.

# 6.3 Ptychography

*Ptychography* (PG) is a computational method of microscopy [Wikipedia, Ptychography]. The name derives from "ptychē" and "gráphein", two ancient Greek words for "folding" and "writing", respectively. The term was first coined by the German crystallographers Hoppe and Strube in 1970 [Hegerl, 1970], for a method proposed by Hoppe in 1969, aiming to solve the phase problem encountered in electron crystallography [Hoppe, 1969]. In 2004, Faulkner and Rodenburg adopted the iterative phase retrieval framework for ptychographic reconstruction and brought the PG technique to its modern form [Faulkner, 2004].

Ptychography generates images by processing a series of many coherent interference patterns that have been scattered via a locally probing beam from an object of interest and have then been recorded by a 2D detector. Within such an image sequence, the probing beam changes each time its position (or direction) on the object, resulting in distinct coherent scattering patterns. Via mathematical retrieval (e.g., using Fourier (back) transforms) of a series of ptychograms an image of the object can be computed. The recovered phase information allows high-contrast visualization of the object and enables quantitative high-resolution morphological measurements. When only a single ptychogram is recorded at a fixed position at the object, the method is referred to as (conventional) *Coherent Diffraction Imaging* (CDI) [Rodenburg, 2019].

Similar as digital holography (DH, see Sect. 4) PG represents a "lensless imaging" technique. However, in contrast to DH it does not require a second stable reference beam and does not necessarily rely on coherence. The richness of the ptychographic dataset contains information on both, the object and different system components in the setup. Thus, the acquired dataset can be used to jointly recover the object and also the probe beam [Wang OPEX, 2023]. PG requires, however, that the object does not change upon recording the sequence of ptychograms. Thus, in the context of laser processing, PG can only be applied to study sub-damage threshold phenomena. For irradiation conditions exceeding the damage threshold, the CDI technique may be applied (see the following Sect. 6.4).

Barolak et al. combined the concept of optical pump-probe ptychography with recent ultrafast multiplexing photography techniques to establish single-shot *Ultrafast Time-resolved Imaging via Multiplexed Ptychography* (UTIMP) [Barolak, 2024]. The capabilities of the method was demonstrated by dynamic imaging of the conduction band electron population from a two-photon absorption event in a 2 mm thick bulk zinc selenide (ZnSe) window, induced by a single fs-pump laser pulse. The experimental setup is reprinted in Figure 13.44.

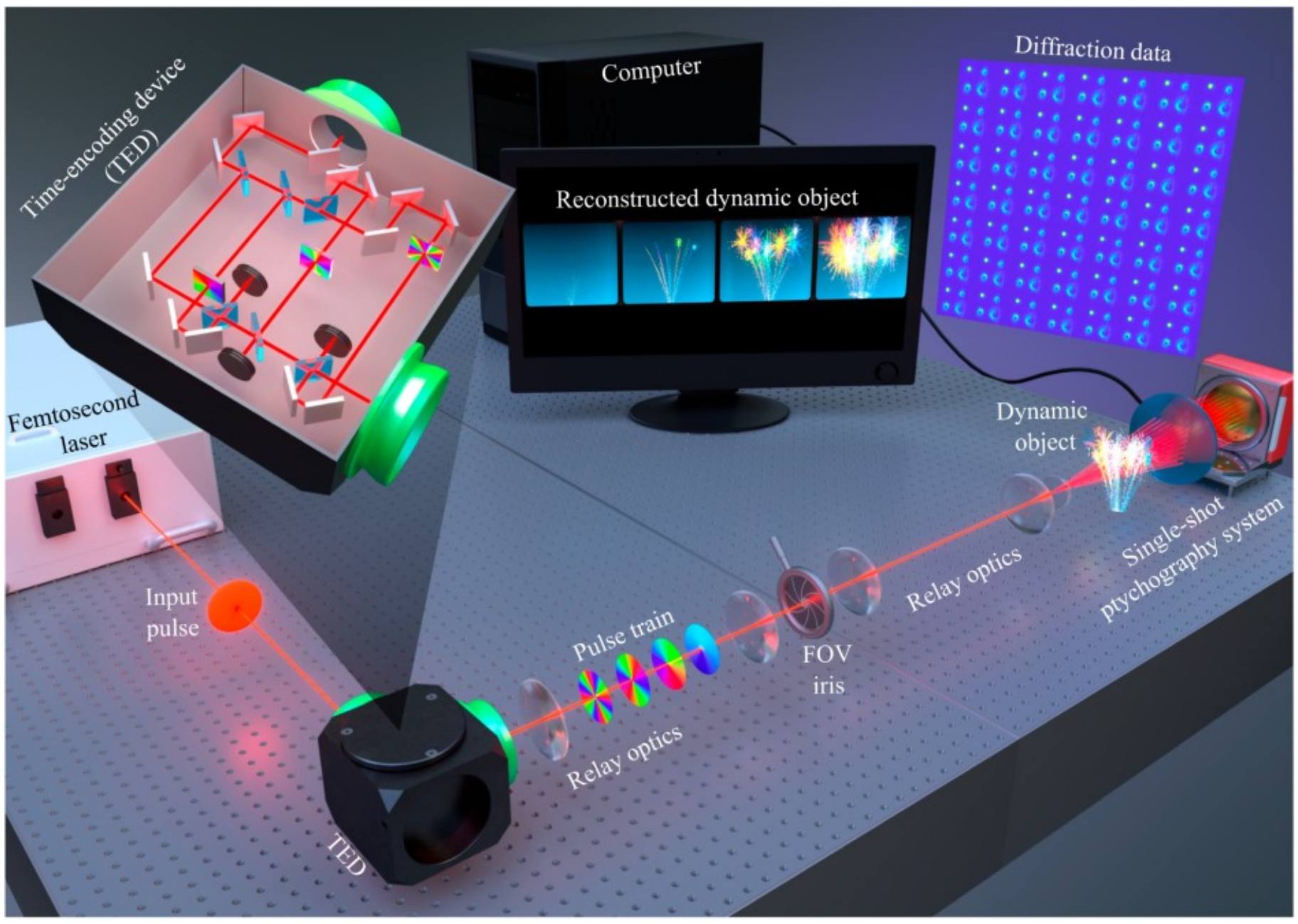

**Fig. 13.44:** Scheme of pump-probe ultrafast time-resolved imaging via multiplexed ptychography (UTIMP) of a dynamics object [Barolak, 2024]. The dynamic object is generated via an ultrashort pump pulse. A fs-laser probe pulse is entering the time-encoding device (TED), where it is split into four temporally separated, phase-stamped sub-pulses. The output pulse train is relay imaged to the single-shot ptychographic (SSP) system. The SSP system collects diffraction data from multiple angular sections of the object. An image of the dynamic object from each probe event is reconstructed from the resulting diffraction data. (Reprinted from [Barolak, 2024], J. Barolak et al., Ultrafast, single-event ptychographic imaging of transient electron dynamics. Ultrafast Sci. **4**, 0058 (2024), Copyright 2024 under Creative Commons BY 4.0 license. Retrieved from https://doi.org/10.34133/ultrafastscience.0058)

The excitation of the sample by a pump pulse ($\lambda_{pump}$ = 800 nm, $\tau_{pump}$ = 35 fs) of a Ti:sapphire fs-laser is creating a dynamic object, i.e., a homogeneously two-photon excited square-shaped area in ZnSe of approximately 1×1 mm$^2$ in size. This dynamic object is probed by a delayed and time-encoded test pulse ($\lambda_{probe}$ = 800 nm, $\tau_{probe}$ < 80 fs). For that the fs-laser probe pulse is first passing through a time-encoding device (TED). The TED is a Michelson-type interferometer that splits a single input pulse into multiple (here four) temporally separated and phase-stamped output pulses. This train of four probe pulses then acts as the illumination source for the dynamics object, where the pulse duration is the "exposure time" and the temporal separation determines the "frame rate". The optical delay stages in each arm of the TED set the timing between pulses in the output pulse train. In this case, the temporal separation was set at 1 ns, with the exception that the last TED output pulse occurred 3 ns after the third pulse. To order the reconstructed images in time, the probe pulses were individually phase-stamped, such that each sub-pulse is different from each other. This was done by adding special optical phase plates into the individual TED arms that are encoding a distinct *orbital angular momentum* (OAM), with a respective "charge" |*l*| of 0, 1, 2, and 4, to the optical phase distribution of the individual sub-pulses of the probe beam train.

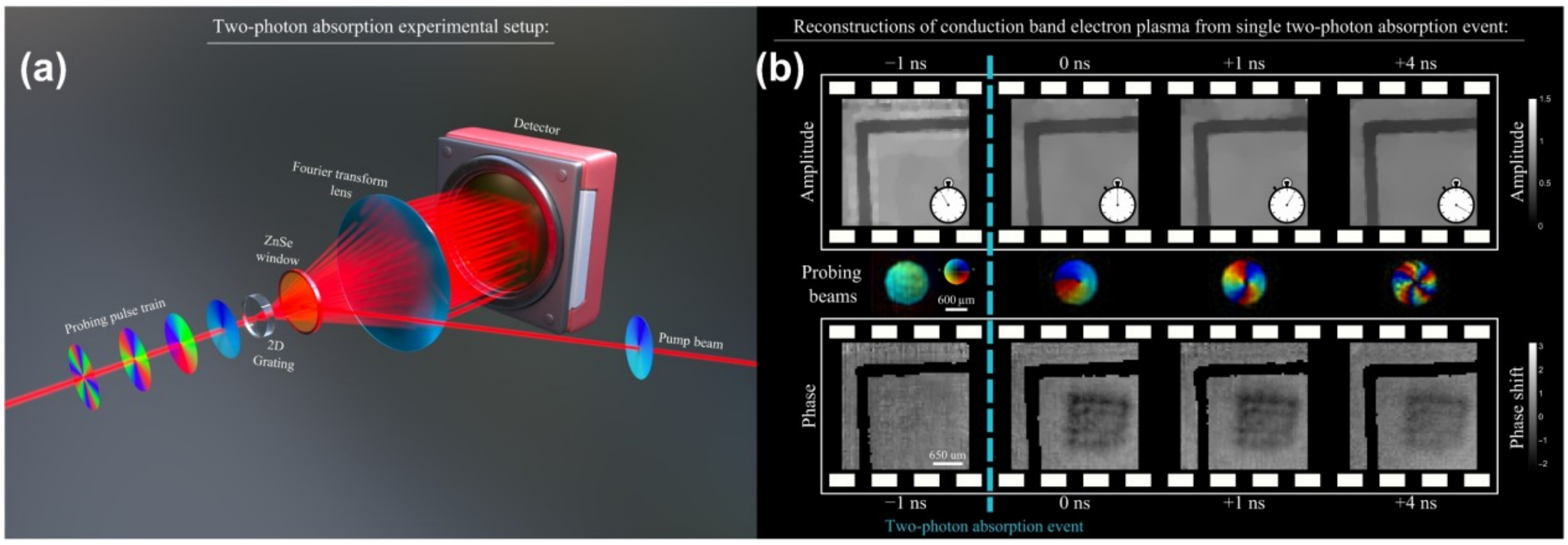

**Fig. 13.45: (a)** Schematic detail of the experimental UTIMP setup for imaging a dynamic object. Here, a single two-photon absorption event in a ZnSe window is generated by a fs-pump pulse between the first and the second probe pulse [Barolak, 2024]. **(b)** Images of retrieved UTIMP object reconstructions at four different delay times $\Delta t$ ranging between –1 ns and +4 ns. The top row provides amplitude images from the 4 reconstructed frames, laser-induced amplitude changes did not manifest under the given excitation conditions. The L shaped pattern is an amplitude fiducial etched into the ZnSe surface marking the region of interest. The bottom row presents the corresponding phase images. The TPA excitation of the ZnSe resembles the spatial pump pulse profile in the phase images around the zero delay and then diminishes again with progressing delay time due to carrier recombination effects. (Reprinted from [Barolak, 2024], J. Barolak et al., Ultrafast, single-event ptychographic imaging of transient electron dynamics. Ultrafast Sci. **4**, 0058 (2024), Copyright 2024 under Creative Commons BY 4.0 license. Retrieved from https://doi.org/10.34133/ultrafastscience.0058)

The output of the TED then was relayed to the input of the single-shot ptychographic (SSP) system, which consists of two sets of two-lens combinations in a 4f-imaging system. The first relay system images the phase plates within the TED interferometer arms to a field-of-view (FOV) iris (see Figure 13.44). The iris was then imaged to the object plane in the SSP system. For breaking up the incident probe radiation into an array of ptychographic probing beamlets, a 2D diffractive grating was inserted that creates a 7×7 array of beamlets, each identical to the input illumination. The dynamic object was then exposed in the divergent beam path to this array of beamlets. A Fourier lens was placed one focal length away from the 2D grating for collimating the array of beamlets. An achromatic doublet lens collected the diffracted beamlets and forms the far-field diffraction pattern on a CCD camera serving as detector (see Figure 13.45a). For numerically retrieving reconstructed amplitude and phase images the RAAR (Relaxed Averaged Alternating Reflections) algorithm was used [Barolak, 2024 / Nashed, 2014].

Figure 13.45b present results of the pump-probe ultrafast ptychography of two-photon laser excited ZnSe samples. The upper row of image frames presents a series of differently delayed temporal snapshots ($\Delta t$ = –1, 0, +1, and +4 ns) of the amplitude of the reconstructed images of the pump laser excited region, while the lower row of frames displays the corresponding set of reconstructed phase images [Barolak, 2024]. At the given laser-excitation conditions (pump), the generated conduction band electrons did not exhibit a significant amplitude contrast, while being probed with infrared light. Contrarily, the optical phase signal shows a pronounced negative laser-induced change, as it can be expected from the presence conduction band electrons.

## 6.4 Coherent Diffraction Imaging

The principle of *Coherent Diffraction Imaging* (CDI), also termed *Coherent Diffraction Microscopy,* is conceptually very similar to that of Ptychography (see Sect. 6.3), with the main difference that the sample is not moved, and a single diffraction pattern is sufficient for CDI – provided that the phase problem can be solved, e.g., by combining the oversampling method with iterative algorithms. The idea of CDI was initially suggested by David Sayre in 1980 [Sayre, 1980 / Miao, 2011]. In 1999, the first experimental demonstration of the technique was reported by Miao et al. [Miao, 1999].

The CDI method typically takes benefit of short wavelength X-rays and, as a lensless imaging method, overcomes far-field diffraction limits of imaging optical elements. The technique became relevant and prominent particularly for coherent high intensity X-ray radiation available from Synchrotrons or, more recently, from Free Electron Lasers that enabled 2D or 3D reconstruction of the image of nanoscale structures, including clusters, nanoparticle, nanotubes, biomolecules, viruses, cells etc. In this context, the question arose, whether the intense and energetic X-rays may affect the diffraction patterns by modifying or destroying the probed object already during the interaction with it? Neutze et al. suggested that, when using intense X-ray pulses being shorter than 10 fs for illuminating a biomolecule, a diffraction pattern may be recorded from the molecule before it is destroyed through the illuminating X-ray pulse [Neutze, 2000]. This has then ruled the principle “diffract before destroy” being key for many investigations.

Chapman et al. demonstrated the feasibility of CDI with an intense, ultrashort FEL pulse at the FLASH facility of DESY [Chapman, 2006]. The authors recorded the diffractogram of a patterned test object by employing a single, intense FEL pulse before the destruction of the object occurred. The successful reconstruction of an image of the test pattern underlined the potential of CDI as an ultrafast imaging technique. About two years later, proof-of-concept time-resolved pump-probe CDI experiments of laser ablation were performed, achieving 10 ps temporal resolution along with 50 nm spatial resolution [Barty, 2008], see Figure 13.46.

In that experiments, a delayed 10 fs XUV-FEL laser pulse (ii, $\lambda_{probe}$ = 13.5 nm) was used to probe under normal incidence ($\theta_{probe}$ = 0°) a patterned sample (iii, see the upper left inset in Figure 13.46) that was pumped by an intense 12.5 ps optical laser pulse (i, $\lambda_{pump}$ = 523 nm, $\phi_{pump}$ ~ 3 J/cm$^2$) in the ablative fluence regime, incident at an angle of $\theta_{pump}$ ~ 45°. As sample, nanometer resolution elephant-shaped patterns were etched by a focused ion beam (FIB) into a 20 nm thick silicon nitride membrane coated with a 100 nm thick iridium film, serving as reference objects. The XUV radiation coherently scattered from the sample was reflected by a 45°-mirror with a graded multilayer coating (iv) and subsequently recorded by a CCD camera (v). In order to prevent damage of the camera by the non-diffracted FEL beam, a hole was drilled in the mirror, preventing its reflection. An additional 100 nm thick zirconium filter covering the CCD chip (not shown) suppressed scattered radiation of optical laser excitation pulse [Barty, 2008].

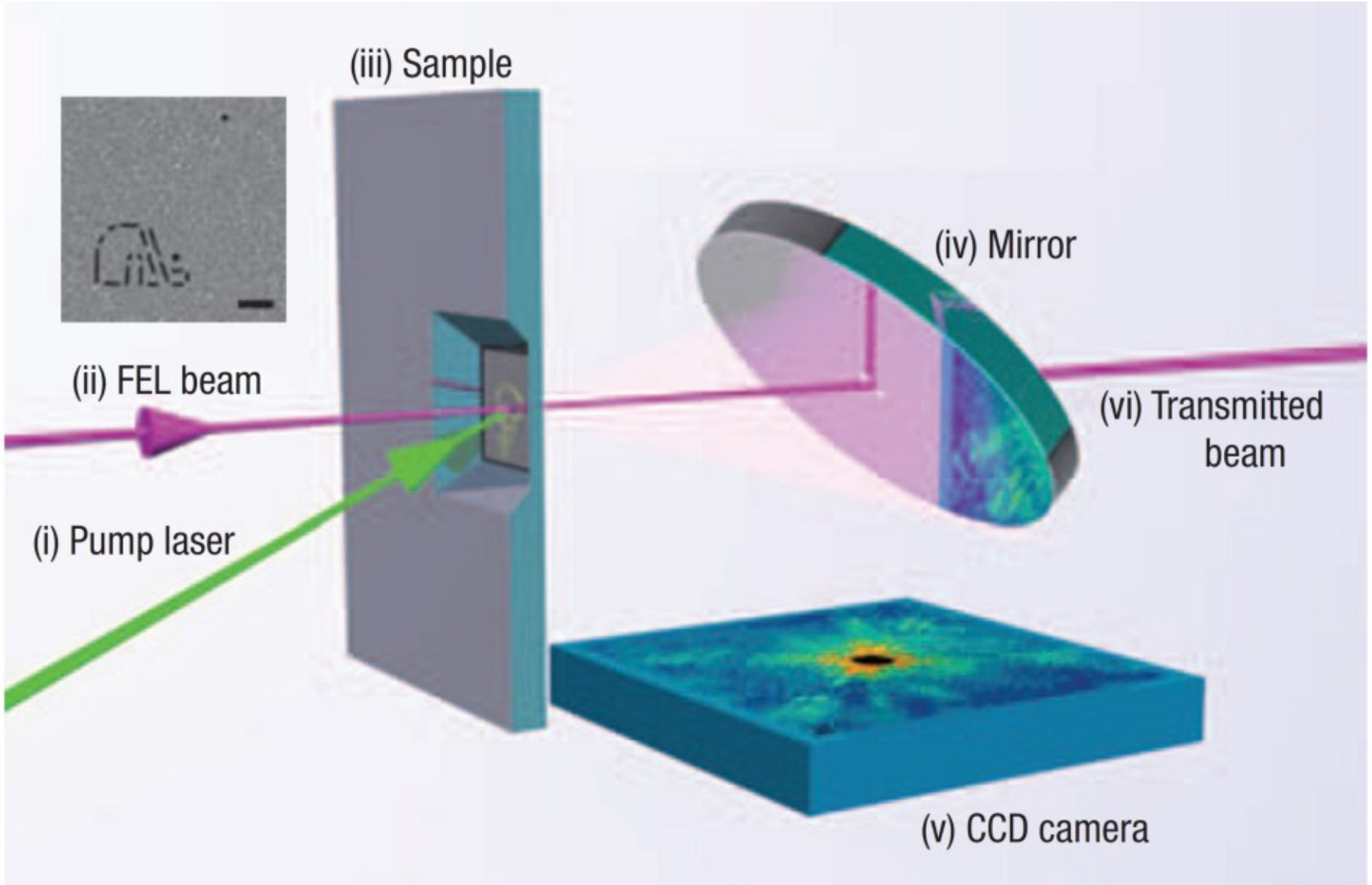

**Fig. 13.46:** Scheme of the non-collinear fs-time-resolved single-pulse coherent diffractive imaging (CDI) performed at the XUV-FEL (FLASH, DESY, Hamburg, Germany) [Barty, 2008]. Single focused ps-laser pump pulses (i, $\lambda_{pump}$ = 523 nm, $\tau_{pump}$ = 12.5 ps) are incident at an angle of ($\theta_{pump} \sim 45°$). Delayed fs-XUV pulses (ii, $\lambda_{probe}$ = 13.5 nm, $\tau_{probe}$ = 10 fs) are probing the laser excited sample (iii) region patterned and coated on a 20 nm thick $Si_3N_4$ membrane under normal incidence ($\theta_{probe} = 0°$). The scattering patterns are reflected by a 45°-mirror (iv) and recorded by a CCD camera (v) that is synchronized with the fs-XUV probe pulses. Scattered optical laser radiation is blocked by a filter (not drawn). The inset in the upper left corned presents an image of the nano-patterned sample surface. (Reprinted from [Barty, 2008], A. Barty et al., Ultrafast single-shot diffraction imaging of nanoscale dynamics, Nature Photon. **2**, 415 – 419, 2008, Springer Nature)

Figure 13.47 compiles some results of the measurements [Barty, 2008]. In panel (a), a top-view SEM micrograph of the nano-patterned reference object is displayed. Panel (b) provides three single-shot diffraction patterns recorded at pump-probe delays $\Delta t$ of –5ps, +10 ps, and +15 ps, respectively. In panel (c), the corresponding CDI images, reconstructed by iterative phase retrieval, are shown underneath the corresponding diffraction patterns.

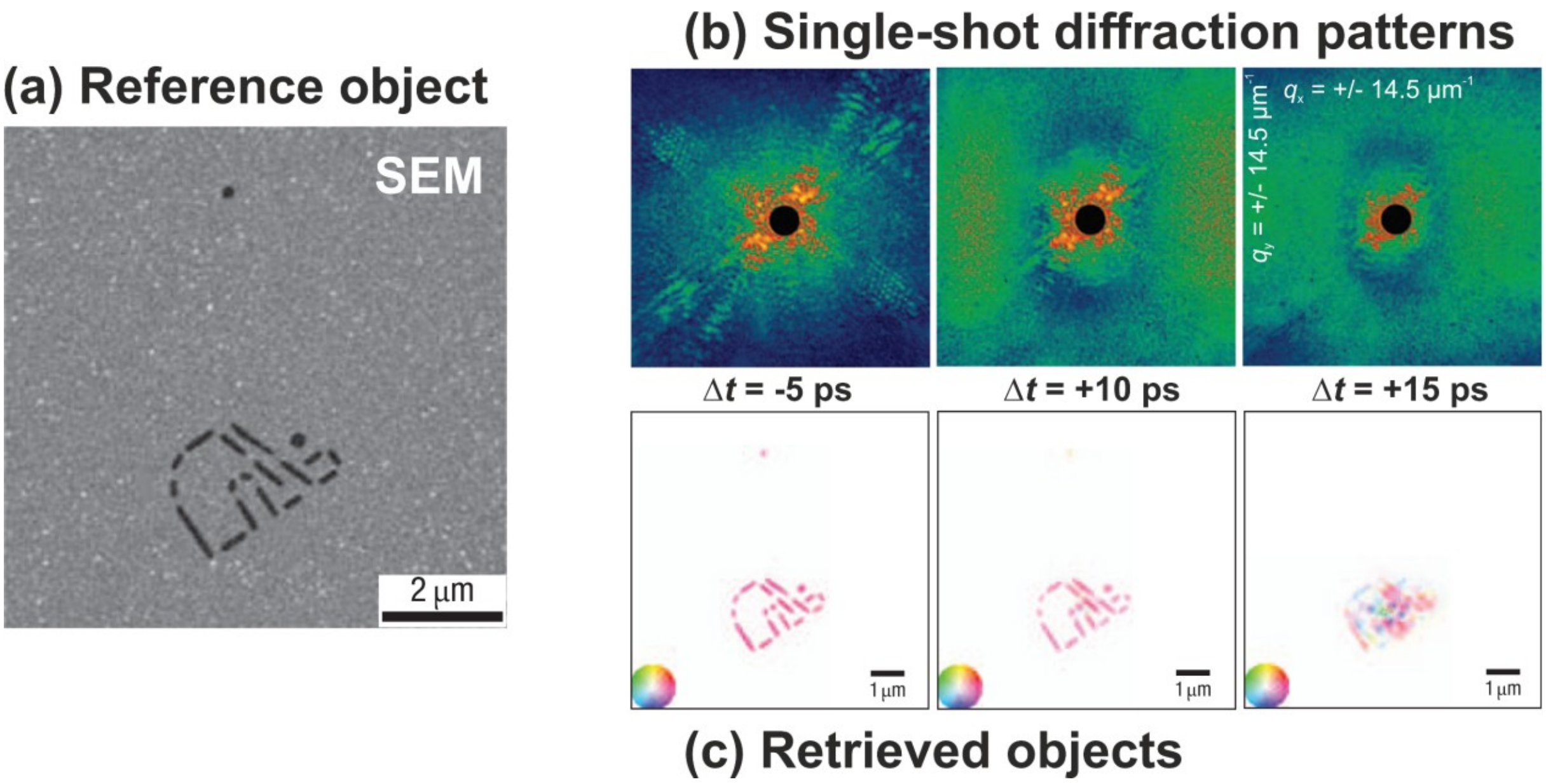

**Fig. 13.47:** (**a**) Top-view SEM image of nanopattern in a 20 nm thick $Si_3N_4$ membrane, subsequently over-coated by a 100 nm thick iridium film and serving as a reference object [Barty, 2008]. (**b**) Measured single-shot XUV-diffraction patterns acquired –5 ps, +10 ps, and +15 ps after the optical laser pulse with the setup shown in Fig. 13.46. The optical pump laser is p-polarized with the linear polarization direction oriented vertically with respect to these images. (**c**) Corresponding images of the object obtained after iterative phase retrieval. The reconstructed complex-valued images are represented by hue and saturation (as indicated in the color wheel). (Reprinted from [Barty, 2008], A. Barty et al., Ultrafast single-shot diffraction imaging of nanoscale dynamics, Nature Photon. **2**, 415 – 419, 2008, Springer Nature)

The retrieved object obtained for $\Delta t = -5$ ps, i.e., shortly before the arrival of the optical laser pulse, demonstrates the successful reconstruction of the image of the reference object. The image of the retrieved object at $\Delta t = +10$ ps confirms that the object has changed very little only during this initial delay span: the edges of the etched structure appear only slightly blurred. However, after a delay of $\Delta t = +15$ ps the reconstructed image indicates the disintegration of the patterned object. Only a part of the pattern can still be recognized since ablation has started on the nanoscale, finally leading to the destruction of the patterned object.

In 2022, Xu et al. presented all-optical single-shot ultrafast multiplexed coherent diffraction imaging (SUM-CDI) applied in the context of laser processing to the visualization of laser beam filamentation and shock wave formation inside K9 glass. The authors quantified a temporal resolution of 10 ns along with a spatial resolution of ~7 µm [Xu PR-B, 2022]. In 2025, He et al. employed a broadband laser wavelength (visible to near-infrared supercontinuum) of temporally chirped picosecond laser pulses as an illumination source for optical pump-probe CDI for studying with 25 frames the surface ablation of glass induced by single, frequency-doubled Ti:sapphire laser pulses ($\lambda_{pump} = 400$ nn, $\tau_p = 35$ fs), with a temporal resolution ranging from several hundred femtoseconds to a few picoseconds [He LPR, 2025].

# 7 Outlook and Future Developments

In this final section of the chapter a (personal) assessment of the current trends and actual limitations in ultrafast optical probing is provided, along with a brief outlook and a discussion of potential future developments.

## 7.1 Shorter Pulse Durations

All optical pump-probe spot probing or imaging experiments are already mature and well established for pulse durations down to ~100 fs. At the near-infrared Ti:sapphire laser wavelength of $\lambda_{probe} = 800$ nm and for a temporally Gaussian pulse, this duration corresponds to a spectral bandwidth of $\Delta\nu$ ~9.4 nm. Shorter laser pulses exhibit a broader spectrum ($\tau_p \cdot \Delta\nu \geq 0.441$, for transform-limited temporally Gaussian pulses) and then care must be taken to keep the probe pulse duration short, particularly for laser wavelengths in the visible and ultraviolet spectral range. This often requires extra efforts regarding the bandwidth of coatings, dispersion compensation, pulse broadening upon focusing, costs, etc. Moreover, the condition of optical wave propagation imposes a fundamental lower pulse duration limit at $\tau_{min} = \lambda/c_0$ since the optical pulses must at least have one single optical cycle. At $\lambda_{probe}$ = 800 nm, the minimum theoretical pulse duration is then about $\tau_{min}$ (800 nm) = 2.7 fs. In the ultraviolet spectral range at $\lambda$ = 400 nm the value is twice smaller at ~1.3 fs. Thus, attosecond laser pulse can only be generated at even shorter wavelengths. In view of these difficulties is not expected that significant further efforts will be made to significantly shorten the probe pulse durations in all-optical pump-probe experiments.

## 7.2 Other Wavelengths

In *far-field* optics, the spatial resolution of optical instruments is ruled by the optical *diffraction limit*, providing a lower boundary at ~ $\lambda/(2\cdot NA)$, with NA being the numerical aperture of the limiting optical element. Hence, the resolution is typically larger than half the probe laser wavelength. Assuming a probe wavelength in the visible spectral range (e.g. $\lambda_{probe}$ = 500 nm), then objects smaller than 250 nm cannot be resolved anymore. Many of the all-optical pump-probe microscopy approaches presented above have almost reached the fundamental resolution limit of optical devices, for example when being successfully imaging even sub-wavelength LIPSS or near-field diffraction phenomena in BF-R mode.

Alternatively, *near-field* optical scanning probe microscopy concepts may be employed for obtaining higher spatial resolution. While near-field optical structuring was already stablished in laser processing for fabrication of structural feature sizes down to ~10 nm only far below the optical diffraction limit (see Chap. 1, Nolte et al. and Ref. [Leiderer, 2023]), high-resolution spatio-temporal pump-probe investigations have not yet been established as useful tools for probing laser processing. The two currently established near-field imaging concepts are either conventional *Scanning Near-field Optical Microscopy* (SNOM) or the variant of *Apertureless SNOM*. In the latter technique the electromagnetic light field is strongly enhanced in the local vicinity of a sharp metallic tip that is placed close to the object of interest. Commercial fs-pump-probe SNOM devices are already available.

With that time-resolved SNOM technique the dynamics of thermally- or laser-induced insulator-to-metal transition (IMT) in $VO_2$ thin films was explored on the sub-nanosecond timescale [Sternbach, 2021]. In that experiments sample was excited by an IR optical laser pulse ($\lambda_{pump}$ = 1023 nm, $\tau_{pump}$ = 280 fs) and near-field-probed through the metallized tip of an atomic force microscope that was delay-controlled illuminated with a frequency-shifted probe pulse ($\lambda_{probe}$ = 10 µm). The evanescent field confined to the local vicinity of the tip apex (~20 nm spatial resolution) was subsequently scattered to a detector. Nano-Fourier transform infrared spectroscopy (nano-FTIR) or pseudo-heterodyne detection methods were used to extract the complex frequency dependent near-field amplitude and phase of the scattered light [Sternbach, 2021].

However, both SNOM techniques rely on a back-feeded contact sensitive scanning of a local probe across a surface, where the electronic feedback mechanism to keep the probe tip at a controlled distance from the surface is too slow to compensate for very fast changes (transient laser heating, mass density, etc.) manifesting during localized laser ablation. Thus, large area scanning optical near-field imaging will always be a very time consuming (serial) approach that is still of limited feasibility in a pump-probe approach during laser processing.

## Shorter Wavelengths

For spatially resolving objects with a size of less than ~250 nm, probe beam wavelengths below $\lambda_{probe}$ = 500 nm must be employed, soon leaving the optical spectral range towards the ultraviolet (UV), extended UV (XUV), and X-ray spectral ranges. For UV wavelengths below ~200 nm, the ambient air is absorbing the electromagnetic radiation and, thus, enforces the experiments to be performed in vacuum or special gas environments. With ultrashort pulsed radiation at X-ray wavelengths the nanometric mesoscale and even molecular and atomic scales become accessible, allowing to study fundamental physical, chemical, and biological processes [Pfeifer, 2006 / Seddon, 2017] and matter under extreme conditions [Pascarelli, 2023].

## Ultrafast X-ray Sources

During the last decades, enormous progress was made in the development of ultrafast X-ray sources [Giulietti, 1998 / Seddon, 2017 / Schoenlein, 2019], particularly with respect to *4th generation light sources* (4GLS) and *X-ray Free Electron Lasers* (XFELs) [Winick, 1998 / Geloni, 2017 / Kovalchuk, 2022] that exceed the performance of the *3rd generation light sources* (storage rings) in the spectral peak brightness by up to 10 orders of magnitude [Altarelli, 2004], see also Figure 13.48 [Tschentscher, 2004 / Kovalchuk, 2022].

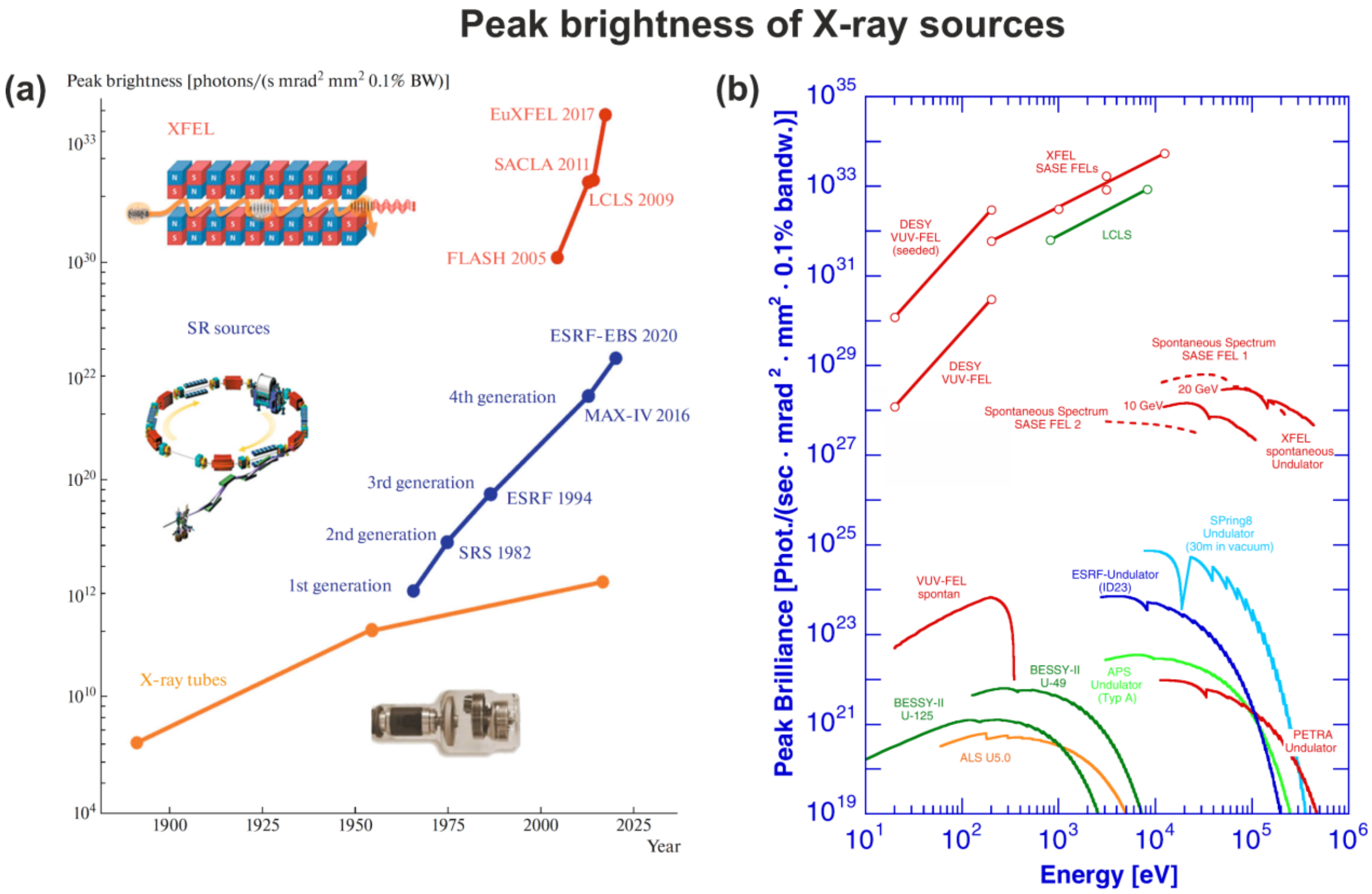

**Fig. 13.48:** (**a**) Timeline of the development and peak brightness of X-ray sources. (Reprinted from [Kovalchuk, 2022], M.V. Kovalchuk et al., European X-ray Free-Electron Laser, Crystallogr. Rep. **67**, 631 – 675, 2022,

Springer Nature) (**b**) Peak brightness of 3rd and 4th generation light sources as function of the emitted photon energy in 2004. (Reprinted from [Tschentscher, 2004], Chem. Phys., **299**, T. Tschentscher, Investigation of ultrafast processes using X-ray free-electron laser radiation, 271 – 276, Copyright (2004), with permission from Elsevier)

With respect to the "scientific evergreen" of Laser-induced Periodic Surface Structures (LIPSS), major contributions can be expected for further exploring their formation mechanism and dynamics on ultrafast meso- and atomic scales that are not directly accessible by optical means. An example for an optical-pump / XUV-probe time-resolved coherent scattering experiment for uncovering the dynamics of optical sub-wavelength surface structure formation in the ablative laser processing regime was already presented and discussed in detail above in Sect. 2.5. These XUV experiments are currently extended into the X-ray spectral range by performing *fs-time-resolved Small Angle X-ray Scattering* (fs-SAXS) and *fs-time-resolved Grazing-Incidence Small Angle X-ray Scattering* (fs-GISAXS) on sub-wavelength high spatial frequency LIPSS (HSFL), see the Perspective article [Bonse & Sokolowski-Tinten, 2024], and [Nakatsutsumi, 2025].

### Secondary Sources

The enormous developments in the field of laser technology during the last decade have enabled the availability and commercialization of ultrafast laser sources at the kW average output power level. This came particularly along with a boost in the pulse repetition frequency of energetic laser pulses into the MHz range. These laser sources are currently being used to develop so-called *Secondary Sources* (SS) [Thoss, 2023 / Thoss, 2024], where the primary extremely intense laser radiation generates fluxes of secondary emissions, e.g. in the form of energetic particles (protons, neutrons, electrons) or photons in the extreme-ultraviolet (EUV) down to X-rays spectral range. A prominent billion-$-market example of a secondary source is the laser-driven generation of EUV radiation for photolithography in ASML's latest generation electronic microprocessor fabrication facilities. While the photon fluxes of secondary sources usually do not reach that of 4GLS or XFELs, SS are ideally tabletop and their construction, operation, and maintenance costs are lower by some orders of magnitude. Thus, secondary sources may be of high interest for extending ultrafast pump-probe techniques particular in application areas between laboratory and large-scale research facilities.

### Longer Wavelength

Recently, Dong et al. proposed and demonstrate a single-shot ultrafast THz imaging system [Dong, 2023], where the authors exploited the electro-optic sampling (EOS) technique [Wu, 1995] for ultrafast THz radiation detection, using an optical probe beam multiplexed in both the time and the spatial-frequency domains. With that optical pump - THz probe imaging system, the authors recorded the ultrafast THz signature of Ti:sapphire laser ($\lambda_{pump}$ = 800 nm, $\tau_p$ =150 fs) generated carriers in a 500 µm thick silicon wafer [Dong, 2023] at excitation levels below the materials damage threshold. However. Given the wavelength of THz radiation in the 0.1 to 1 mm range, *in-situ* far-field observation of THz radiation during laser processing will remain rather on the macroscopic spatial scale.

## 7.3 New Ultrafast Cameras

The scientific and technological field of ultrafast photography is currently very rapidly emerging and is providing a plethora of new imaging techniques (see Sect. 6). The field emerged from digitally capturing light-in-flight phenomena, where larger objects (lower magnifications) are beneficial. Currently, UP continuously conquers the microscale, where the already well-established microscopy techniques (see Sect. 3) started from, while being extended in pump-probe approaches toward the ultrafast timescales. Rapid transient laser processing phenomena were recognized as useful objects for UP, in order to test the ultrafast camera devices. However, significant new scientific findings on laser processing were not accomplished yet by the new emerging lines of ultrafast photography. Nevertheless, enormous frame rates (inverse proportional to the frame interval times) were reached already by several imaging techniques (CUP, FRAME, STAMP, FINCOPA, and CUSP), see the timeline of the development of cameras and imaging methods presented in Figure 13.49.

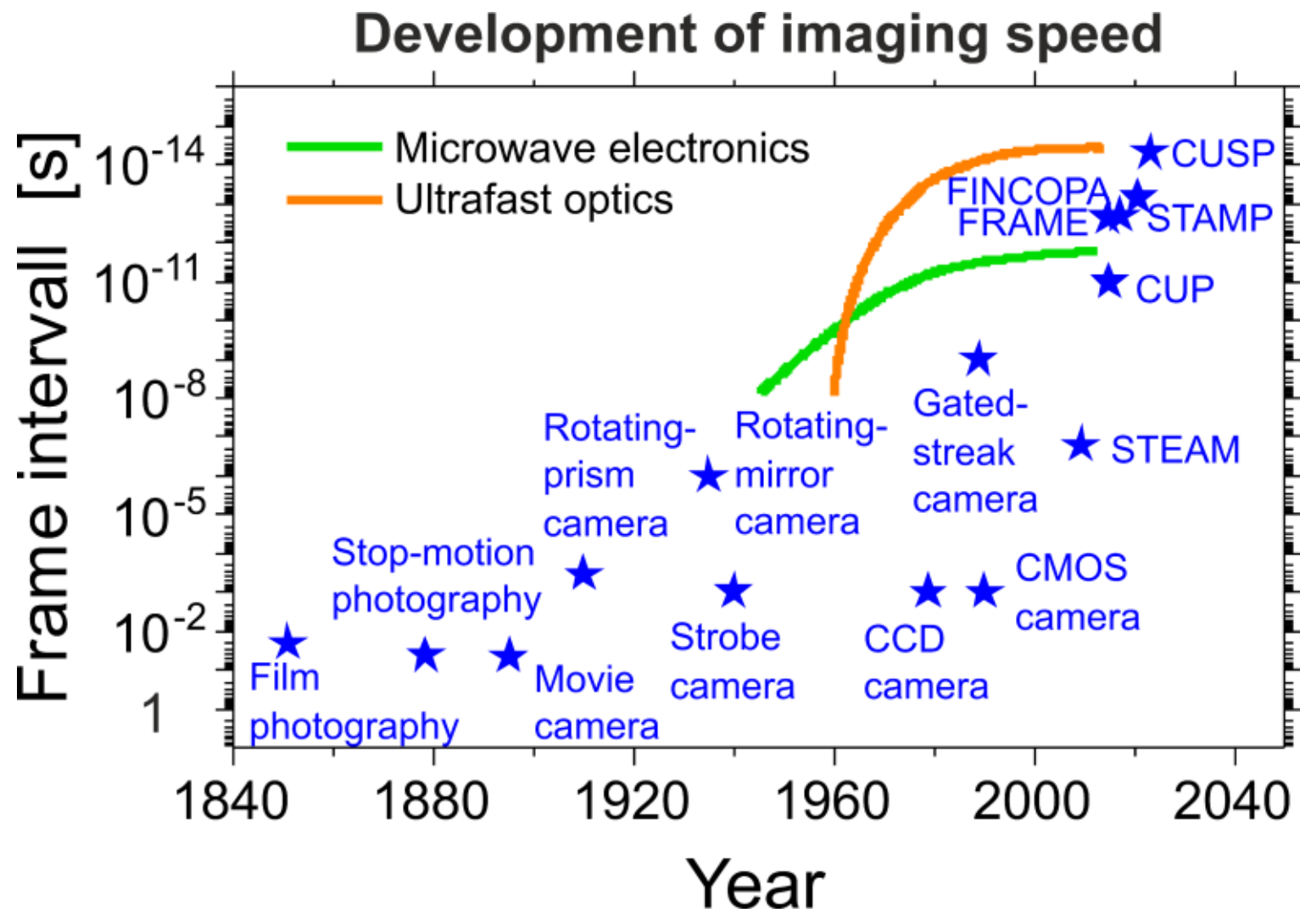

**Fig. 13.49:** Timeline of the development of the frame interval time of cameras and imaging methods. For abbreviations refer to Table 13.2. Data merged from [Pogue, 2014 / Nakagawa, 2015] and extended with selected imaging methods from Table 13.2. The green and orange solid lines represent limiting curves from [Krausz, 2009]

However, with the further improvements in 2D optoelectronic sensor performance, smart multiplexing information encoding, and fast data storage systems, it can be expected that the UP will become commercial, speeds up further, approaches spatially the optical diffraction limit and can provide new insights into the laser processing.

## 7.4 Extended Pump-Probe Schemes

As demonstrated in this chapter, all-optical pump-probe methods have been explored already in detail and in many different imaging modes in the context of laser processing. Additional potential may be raised if the “all-optical” approach is extended towards a “secondary probing” technique, as briefly outlined in the following.

### Secondary Probing

In such a *Secondary Probing* (SP) time-resolved pump-probe approach (see the scheme provided in Figure 13.50), the optical pump pulse triggers a transient (laser processing) event in the irradiated material. The following short wavelength probe pulse explicitly interrogates the system state via the creation of a secondary emission that is uniquely representing the exact state of the system, e.g., in the form of electrons, other particles, photons, etc. that may be then further analyzed – typically in a spectroscopic manner. Currently, such techniques are still explored through complex scientific experiments in research laboratories and at large scale facilities, as they require highly specialized personnel.

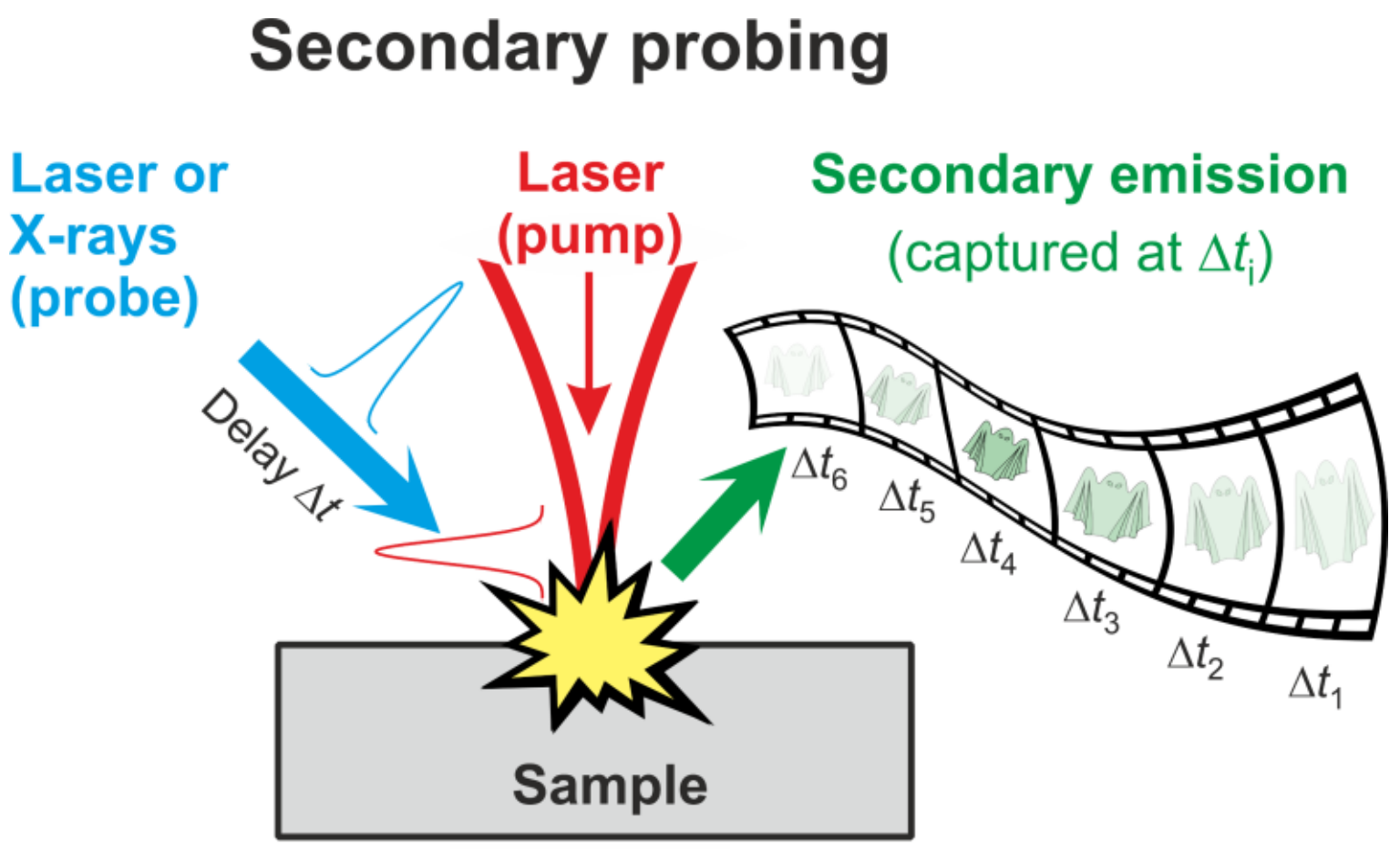

**Fig. 13.50:** Time-resolved recording of a secondary emission in a pump-probe scheme

An example of such a pump-probe approach was presented in 2004 by Glover et al., who took benefit of a synchrotron source and implemented a ps-time-resolved core-level X-ray photoelectron spectroscopy for studying the fs-laser ablation of silicon in a spot-probing geometry [Glover, 2004]. An ultrashort Ti:sapphire laser pump pulse ($\lambda_{pump}$ = 800 nm, $\tau_p$ = 200 fs) ablated the surface of a silicon wafer in a vacuum environment using multiple pulses at high laser fluences between 4 and 12 J/cm$^2$. A delayed synchrotron X-ray pulse ($\lambda_{probe}$ = 3.1 nm, $\tau_{probe}$ = 80 ps) caused the secondary emission of photoelectrons that were collected by a hemispherical analyzer. Via spectroscopically analyzing the Si 2p core-level photoelectrons, they found that material is ejected predominantly in a condensed (liquid) phase and that the subsequent solidification occurring rapidly within less than 50 ps. Moreover, a relative absence of vapor particles in the ejecta (estimated at <10%) was observed, indicating that semiconductor micro-particles do not arise from condensation of a dilute vapor but can be associated with rapid vacuum expansion fragments of the ejected material. These findings are fully in line with MD simulations (see Chap. 1, Nolte et al.).

Other examples of time-resolved photoelectron spectroscopy (tr-PES), two-photon photo emission (2PPE), and even time-resolved photoemission electron microscopy (tr-PEEM) as ultrafast secondary probing approach at material excitation levels below the damage threshold are found in the recent review article [Aeschlimann, 2025]. Also, ultrafast time-resolved scanning tunneling microscopy (tr-STM) [Nguyen, 2021 / Gross, 2023] can represents such a secondary probe approach. The transfer of these techniques and *in-situ* applicability to laser processing is to be proven.

*High Harmonic Generation* (HHG) created by the coherent superposition of a pump and probe fields provides a unique optical signature of strong-field, sub-cycle processes. This has, for instance, been shown by Jürgens et al., who used the light-cycle-driven and phase-controlled generation of low-order harmonics in the visible to ultraviolet spectral range for probing the strong laser-field-induced free-electron plasma formation in transparent solids [Jürgens, 2020 / Jürgens, 2022]. For recording a time-resolved spectrogram, the authors focused a mid-infrared pump pulse ($\lambda_{pump}$ = 2.1 µm, $\tau_{pump}$ = 45 fs) along with a weak, time-delayed near-infrared probe laser pulse ($\lambda_{probe}$ = 790 nm, $\tau_{probe}$ = 45 fs) into the bulk of a fused silica sample, using a cross-polarized, almost collinear geometry. Through numerical simulations and in combination with a measurement of the total plasma density, the authors were able to identify the (relative) contribution of two competing ionization mechanisms, i.e. *strong-field ionization* (SFI) and *electron-impact ionization*. The authors were able to render the important conclusion that SFI-induced signals manifest already well below the threshold for a permanent modification of the material and, thus, provide the potential for an in-situ detection of the onset of optical damage already before catastrophic damage manifests permanently [Jürgens, 2022].

## Acknowledgements

The author acknowledges the long-lasting and fruitful collaborations in the field of time-resolved experiments with Alexandre Mermillod-Blondin, Arkadi Rosenfeld, Cyril Mauclair, Daniel Puerto, Guillaume Bachelier, Jan Siegel, Javier Solis, Klaus Sokolowski-Tinten, Mark Wiggins, Motoaki Nakatsutsumi, Razvan Stoian, Sandra Höhm (listed alphabetically). Alexandre Mermillod-Blondin gave valuable detailed feedback for improving this chapter. Klaus Sokolowski-Tinten and Sebastian Kraft provided valuable hints to literature. The author thanks Samuel W. Teitelbaum for producing Figure 13.8b and authorizing its reprint in this chapter and Keith Nelson for help. The author thanks Christopher B. Schaffer for producing Figures 13.15a,b and authorizing the reprint in this chapter. The author thanks Vasily V. Temnov for producing Figure 13.29b and authorizing its reprint in this chapter. The author thanks Klaus Bergner for providing and producing Figures 13.41a,b and authorizing the reprint in this chapter. Sebastian Kraft and Guido Mann helped with the proof reading.